# First M87 Event Horizon Telescope Results. V. Physical Origin of the Asymmetric Ring

The Event Horizon Telescope Collaboration
(See the end matter for the full list of authors.)

## Abstract

The Event Horizon Telescope (EHT) has mapped the central compact radio source of the elliptical galaxy M87 at 1.3 mm with unprecedented angular resolution. Here we consider the physical implications of the asymmetric ring seen in the 2017 EHT data. To this end, we construct a large library of models based on general relativistic magnetohydrodynamic (GRMHD) simulations and synthetic images produced by general relativistic ray tracing. We compare the observed visibilities with this library and confirm that the asymmetric ring is consistent with earlier predictions of strong gravitational lensing of synchrotron emission from a hot plasma orbiting near the black hole event horizon. The ring radius and ring asymmetry depend on black hole mass and spin, respectively, and both are therefore expected to be stable when observed in future EHT campaigns. Overall, the observed image is consistent with expectations for the shadow of a spinning Kerr black hole as predicted by general relativity. If the black hole spin and M87's large scale jet are aligned, then the black hole spin vector is pointed away from Earth. Models in our library of non-spinning black holes are inconsistent with the observations as they do not produce sufficiently powerful jets. At the same time, in those models that produce a sufficiently powerful jet, the latter is powered by extraction of black hole spin energy through mechanisms akin to the Blandford-Znajek process. We briefly consider alternatives to a black hole for the central compact object. Analysis of existing EHT polarization data and data taken simultaneously at other wavelengths will soon enable new tests of the GRMHD models, as will future EHT campaigns at 230 and 345 GHz.

## 1. Introduction

In 1918 the galaxy Messier 87 (M87) was observed by Curtis and found to have “a curious straight ray ... apparently connected with the nucleus by a thin line of matter” (Curtis 1918, p. 31). Curtis’s ray is now known to be a jet, extending from sub-pc to several kpc scales, and can be observed across the electromagnetic spectrum, from the radio through $\gamma$-rays. Very long baseline interferometry (VLBI) observations that zoom in on the nucleus, probing progressively smaller angular scales at progressively higher frequencies up to 86 GHz by the Global mm-VLBI Array (GMVA; e.g., Hada et al. 2016; Boccardi et al. 2017; Kim et al. 2018; Walker et al. 2018), have revealed that the jet emerges from a central core. Models of the stellar velocity distribution imply a mass for the central core $M \approx 6.2 \times 10^9\, M_\odot$ at a distance of 16.9 Mpc (Gebhardt et al. 2011); models of arcsecond-scale emission lines from ionized gas imply a mass that is lower by about a factor of two (Walsh et al. 2013).

The conventional model for the central object in M87 is a black hole surrounded by a geometrically thick, optically thin, disk accretion flow (e.g., Ichimaru 1977; Rees et al. 1982; Narayan & Yi 1994, 1995; Reynolds et al. 1996). The radiative power of the accretion flow ultimately derives from the gravitational binding energy of the inflowing plasma. There is no consensus model for jet launching, but the two main scenarios are that the jet is a magnetically dominated flow that is ultimately powered by tapping the rotational energy of the black hole (Blandford & Znajek 1977) and that the jet is a magnetically collimated wind from the surrounding accretion disk (Blandford & Payne 1982; Lynden-Bell 2006).

VLBI observations of M87 at frequencies $\gtrsim$230 GHz with the Event Horizon Telescope (EHT) can resolve angular scales of tens of $\mu$as, comparable to the scale of the event horizon (Doeleman et al. 2012; Akiyama et al. 2015; EHT Collaboration et al. 2019a, 2019b, 2019c, hereafter Paper I, II, and III). They therefore have the power to probe the nature of the central object and to test models for jet launching. In addition, EHT observations can constrain the key physical parameters of the system, including the black hole mass and spin, accretion rate, and magnetic flux trapped by accreting plasma in the black hole.

In this Letter we adopt the working hypothesis that the central object is a black hole described by the Kerr metric, with mass $M$ and dimensionless spin $a_*$, $-1 < a_* < 1$. Here $a_* \equiv Jc/GM^2$, where $J$, $G$, and $c$ are, respectively, the black hole angular momentum, gravitational constant, and speed of light. In our convention $a_* < 0$ implies that the angular momentum of the accretion flow and that of the black hole are anti-aligned. Using general relativistic magnetohydrodynamic (GRMHD) models for the accretion flow and synthetic images of these simulations produced by general relativistic radiative transfer calculations, we test whether or not the results of the 2017 EHT observing campaign (hereafter EHT2017) are consistent with the black hole hypothesis.

This Letter is organized as follows. In Section 2 we review salient features of the observations and provide order-of-magnitude estimates for the physical conditions in the source. In Section 3 we describe the numerical models. In Section 4 we outline our procedure for comparing the models to the data in a way that accounts for model variability. In Section 5 we show that many of the models cannot be rejected based on EHT data alone.

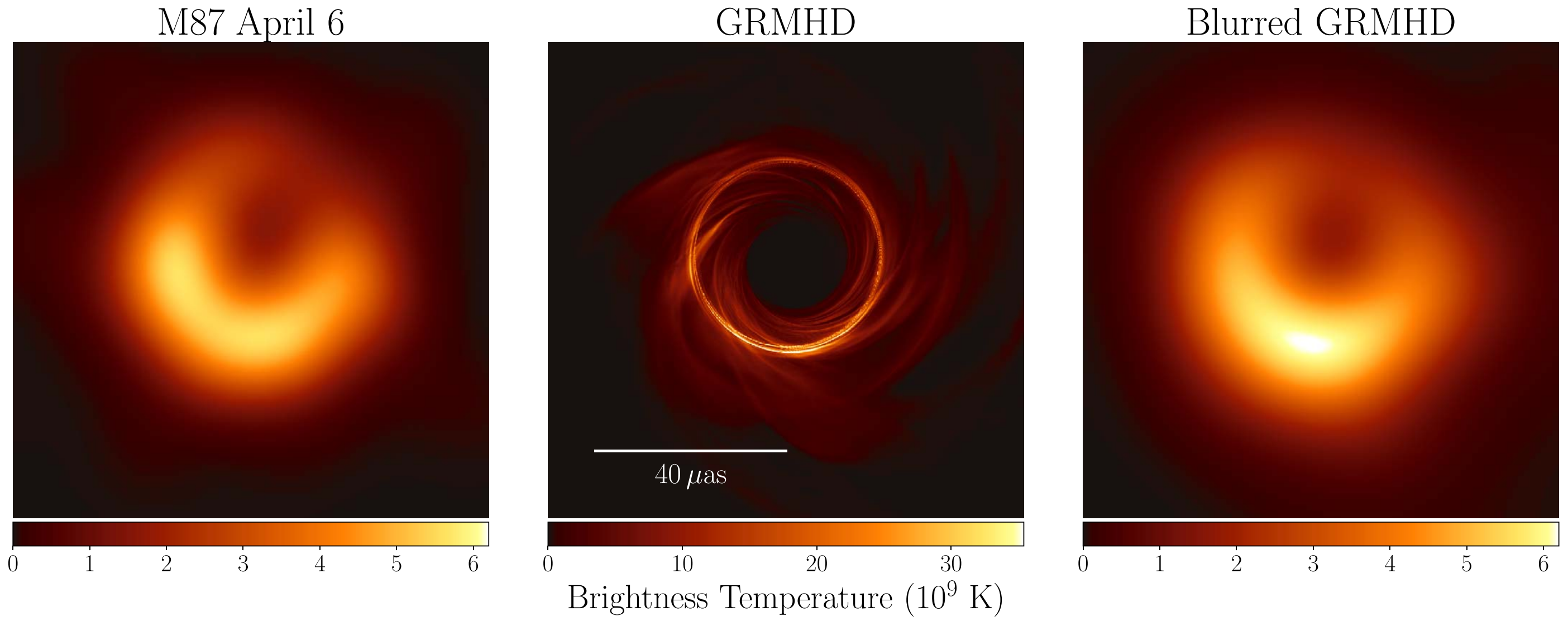

**Figure 1.** Left panel: an EHT2017 image of M87 from Paper IV of this series (see their Figure 15). Middle panel: a simulated image based on a GRMHD model. Right panel: the model image convolved with a 20 $\mu$as FWHM Gaussian beam. Although the most evident features of the model and data are similar, fine features in the model are not resolved by EHT.

In Section 6 we combine EHT data with other constraints on the radiative efficiency, X-ray luminosity, and jet power and show that the latter constraint eliminates all $a_* = 0$ models. In Section 7 we discuss limitations of our models and also briefly discuss alternatives to Kerr black hole models. In Section 8 we summarize our results and discuss how further analysis of existing EHT data, future EHT data, and multiwavelength companion observations will sharpen constraints on the models.

## 2. Review and Estimates

In EHT Collaboration et al. (2019d; hereafter Paper IV) we present images generated from EHT2017 data (for details on the array, 2017 observing campaign, correlation, and calibration, see Paper II and Paper III). A representative image is reproduced in the left panel of Figure 1.

Four features of the image in the left panel of Figure 1 play an important role in our analysis: (1) the ring-like geometry, (2) the peak brightness temperature, (3) the total flux density, and (4) the asymmetry of the ring. We now consider each in turn.

(1) The compact source shows a bright ring with a central dark area without significant extended components. This bears a remarkable similarity to the long-predicted structure for optically thin emission from a hot plasma surrounding a black hole (Falcke et al. 2000). The central hole surrounded by a bright ring arises because of strong gravitational lensing (e.g., Hilbert 1917; von Laue 1921; Bardeen 1973; Luminet 1979). The so-called "photon ring" corresponds to lines of sight that pass close to (unstable) photon orbits (see Teo 2003), linger near the photon orbit, and therefore have a long path length through the emitting plasma. These lines of sight will appear comparatively bright if the emitting plasma is optically thin. The central flux depression is the so-called black hole "shadow" (Falcke et al. 2000), and corresponds to lines of sight that terminate on the event horizon. The shadow could be seen in contrast to surrounding emission from the accretion flow or lensed counter-jet in M87 (Broderick & Loeb 2009).

The photon ring is nearly circular for all black hole spins and all inclinations of the black hole spin axis to the line of sight (e.g., Johannsen & Psaltis 2010). For an $a_* = 0$ black hole of mass $M$ and distance $D$, the photon ring angular radius on the sky is

$$\theta_p \equiv \frac{\sqrt{27}\ GM}{c^2 D} = 18.8 \left(\frac{M}{6.2 \times 10^9 M_\odot}\right)\left(\frac{D}{16.9\ \mathrm{Mpc}}\right)^{-1} \mu\mathrm{as}, \quad (1)$$

where we have scaled to the most likely mass from Gebhardt et al. (2011) and a distance of 16.9 Mpc (see also EHT Collaboration et al. 2019e, (hereafter Paper VI; Blakeslee et al. 2009; Bird et al. 2010; Cantiello et al. 2018). The photon ring angular radius for other inclinations and values of $a_*$ differs by at most 13% from Equation (1), and most of this variation occurs at $1 - |a_*| \ll 1$ (e.g., Takahashi 2004; Younsi et al. 2016). Evidently the angular radius of the observed photon ring is approximately $\sim$20 $\mu$as (Figure 1 and Paper IV), which is close to the prediction of the black hole model given in Equation (1).

(2) The observed peak brightness temperature of the ring in Figure 1 is $T_{b,pk} \sim 6 \times 10^9$ K, which is consistent with past EHT mm-VLBI measurements at 230 GHz (Doeleman et al. 2012; Akiyama et al. 2015), and GMVA 3 mm-VLBI measurements of the core region (Kim et al. 2018). Expressed in electron rest-mass ($m_e$) units, $\Theta_{b,pk} \equiv k_B T_{b,pk}/(m_e c^2) \simeq 1$, where $k_B$ is Boltzmann's constant. The true peak brightness temperature of the source is higher if the ring is unresolved by EHT, as is the case for the model image in the center panel of Figure 1.

The 1.3 mm emission from M87 shown in Figure 1 is expected to be generated by the synchrotron process (see Yuan & Narayan 2014, and references therein) and thus depends on the electron distribution function (eDF). If the emitting plasma has a thermal eDF, then it is characterized by an electron temperature $T_e \geqslant T_b$, or $\Theta_e \equiv k_B T_e/(m_e c^2) > 1$, because $\Theta_e > \Theta_{b,pk}$ if the ring is unresolved or optically thin.

Is the observed brightness temperature consistent with what one would expect from phenomenological models of the source? Radiatively inefficient accretion flow models of M87 (Reynolds et al. 1996; Di Matteo et al. 2003) produce mm emission in a geometrically thick donut of plasma around the black hole. The emitting plasma is collisionless: Coulomb scattering is weak at these low densities and high temperatures. Therefore, the electron and ion temperatures need not be the same (e.g., Spitzer 1962). In radiatively inefficient accretion flow models, the ion temperature is slightly less than the ion virial temperature,

$$T_i \sim 0.3\, T_{i,\mathrm{vir}} = 0.3\, m_p c^2 r_\mathrm{g}/(3k_\mathrm{B} r) = 1.1 \times 10^{12} (r_\mathrm{g}/r)\ \mathrm{K}, \quad (2)$$

where $r_\mathrm{g} \equiv GM/c^2$ is the gravitational radius, $r$ is the Boyer–Lindquist or Kerr–Schild radius, and $m_p$ is the proton mass. Most models have an electron temperature $T_e < T_i$ because of electron cooling and preferential heating of the ions by turbulent dissipation (e.g., Yuan & Narayan 2014; Mościbrodzka et al. 2016). If the emission arises at $\sim 5\, r_\mathrm{g}$, then $\Theta_e \simeq 37(T_e/T_i)$, which is then consistent with the observed $\Theta_{b,pk}$ if the source is unresolved or optically thin.

(3) The total flux density in the image at 1.3 mm is $\simeq$0.5 Jy. With a few assumptions we can use this to estimate the electron number density $n_e$ and magnetic field strength $B$ in the source. We adopt a simple, spherical, one-zone model for the source with radius $r \simeq 5\, r_\mathrm{g}$, pressure $n_i k T_i + n_e k T_e = \beta_\mathrm{p} B^2/(8\pi)$ with $\beta_\mathrm{p} \equiv p_\mathrm{gas}/p_\mathrm{mag} \sim 1$, $T_i \simeq 3 T_e$, and temperature $\theta_e \simeq 10\theta_{b,pk}$, which is consistent with the discussion in (2) above. Setting $n_e = n_i$ (i.e., assuming a fully ionized hydrogen plasma), the values of $B$ and $n_e$ required to produce the observed flux density can be found by solving a nonlinear equation (assuming an average angle between the field and line of sight, 60°). The solution can be approximated as a power law:

$$n_e = 2.9 \times 10^4 \left(\frac{r}{5\, r_\mathrm{g}}\right)^{-1.3} \beta_\mathrm{p}^{0.62} \times \left(\frac{T_i}{3T_e}\right)^{-0.47} \left(\frac{\theta_e}{10\theta_{b,pk}}\right)^{-2.4} \mathrm{cm}^{-3}, \quad (3)$$

$$B = 4.9 \left(\frac{r}{5\, r_\mathrm{g}}\right)^{-0.63} \beta_\mathrm{p}^{-0.19} \times \left(\frac{T_i}{3T_e}\right)^{0.14} \left(\frac{\theta_e}{10\theta_{b,pk}}\right)^{-0.71} \mathrm{G} \quad (4)$$

assuming that $M = 6.2 \times 10^9\, M_\odot$ and $D = 16.9$ Mpc, and using the approximate thermal emissivity of Leung et al. (2011). Then the synchrotron optical depth at 1.3 mm is $\sim$0.2. One can now estimate an accretion rate from (3) using

$$\dot{M} = 4\pi r^2 \rho v^r \sim 4\pi (5 r_\mathrm{g})^2\, n_e m_p\, (c/\sqrt{5}) \sim 2.7 \times 10^{-3} M_\odot\ \mathrm{yr}^{-1} \quad (5)$$

assuming spherical symmetry. The Eddington accretion rate is

$$\dot{M}_\mathrm{Edd} = \frac{L_\mathrm{Edd}}{\epsilon c^2} = \frac{2.2}{\epsilon} \left(\frac{M}{10^9\, M_\odot}\right) M_\odot\ \mathrm{yr}^{-1}, \quad (6)$$

where $L_\mathrm{Edd} \equiv 4\pi GMcm_p/\sigma_T$ is the Eddington luminosity ($\sigma_T$ is the Thomson cross section). Setting the efficiency $\epsilon = 0.1$ and $M = 6.2 \times 10^9\, M_\odot$, $\dot{M}_\mathrm{Edd} = 137\, M_\odot\ \mathrm{yr}^{-1}$, and therefore $\dot{M}/\dot{M}_\mathrm{Edd} \sim 2.0 \times 10^{-5}$.

This estimate is similar to but slightly larger than the upper limit inferred from the 230 GHz linear polarization properties of M87 (Kuo et al. 2014).

(4) The ring is brighter in the south than the north. This can be explained by a combination of motion in the source and Doppler beaming. As a simple example we consider a luminous, optically thin ring rotating with speed $v$ and an angular momentum vector inclined at a viewing angle $i > 0°$ to the line of sight. Then the approaching side of the ring is Doppler boosted, and the receding side is Doppler dimmed, producing a surface brightness contrast of order unity if $v$ is relativistic. The approaching side of the large-scale jet in M87 is oriented west–northwest (position angle PA $\approx$ 288°; in Paper VI this is called $\mathrm{PA_{FJ}}$), or to the right and slightly up in the image. Walker et al. (2018) estimated that the angle between the approaching jet and the line of sight is 17°. If the emission is produced by a rotating ring with an angular momentum vector oriented along the jet axis, then the plasma in the south is approaching Earth and the plasma in the north is receding. This implies a clockwise circulation of the plasma in the source, as projected onto the plane of the sky. This sense of rotation is consistent with the sense of rotation in ionized gas at arcsecond scales (Harms et al. 1994; Walsh et al. 2013). Notice that the asymmetry of the ring is consistent with the asymmetry inferred from 43 GHz observations of the brightness ratio between the north and south sides of the jet and counter-jet (Walker et al. 2018).

All of these estimates present a picture of the source that is remarkably consistent with the expectations of the black hole model and with existing GRMHD models (e.g., Dexter et al. 2012; Mościbrodzka et al. 2016). They even suggest a sense of rotation of gas close to the black hole. A quantitative comparison with GRMHD models can reveal more.

## 3. Models

Consistent with the discussion in Section 2, we now adopt the working hypothesis that M87 contains a turbulent, magnetized accretion flow surrounding a Kerr black hole. To test this hypothesis quantitatively against the EHT2017 data we have generated a Simulation Library of 3D time-dependent ideal GRMHD models. To generate this computationally expensive library efficiently and with independent checks on the results, we used several different codes that evolved matching initial conditions using the equations of ideal GRMHD. The codes used include `BHAC` (Porth et al. 2017), `H-AMR` (Liska et al. 2018; K. Chatterjee et al. 2019, in preparation), `iharm` (Gammie et al. 2003), and `KORAL` (Sądowski et al. 2013b, 2014). A comparison of these and other GRMHD codes can be found in O. Porth et al. 2019 (in preparation), which shows that the differences between integrations of a standard accretion model with different codes is smaller than the fluctuations in individual simulations.

From the Simulation Library we have generated a large Image Library of synthetic images. Snapshots of the GRMHD evolutions were produced using the general relativistic ray-tracing (GRRT) schemes `ipole` (Mościbrodzka & Gammie 2018), `RAPTOR` (Bronzwaer et al. 2018), or `BHOSS` (Z. Younsi et al. 2019b, in preparation). A comparison of these and other GRRT codes can be found in Gold et al. (2019), which shows that the differences between codes is small.

In the GRMHD models the bulk of the 1.3 mm emission is produced within $\lesssim 10\, r_\mathrm{g}$ of the black hole, where the models can reach a statistically steady state. It is therefore possible to compute predictive radiative models for this compact component of the source without accurately representing the accretion flow at all radii.

We note that the current state-of-the-art models for M87 are radiation GRMHD models that include radiative feedback and electron-ion thermodynamics (Ryan et al. 2018; Chael et al. 2019). These models are too computationally expensive for a wide survey of parameter space, so that in this Letter we consider only nonradiative GRMHD models with a parameterized treatment of the electron thermodynamics.

### 3.1. Simulation Library

All GRMHD simulations are initialized with a weakly magnetized torus of plasma orbiting in the equatorial plane of the black hole (e.g., De Villiers et al. 2003; Gammie et al. 2003; McKinney & Blandford 2009; Porth et al. 2017). We do not consider tilted models, in which the accretion flow angular momentum is misaligned with the black hole spin. The limitations of this approach are discussed in Section 7.

The initial torus is driven to a turbulent state by instabilities, including the magnetorotational instability (see e.g., Balbus & Hawley 1991). In all cases the outcome contains a moderately magnetized midplane with orbital frequency comparable to the Keplerian orbital frequency, a corona with gas-to-magnetic-pressure ratio $\beta_{\rm p} \equiv p_{\rm gas}/p_{\rm mag} \sim 1$, and a strongly magnetized region over both poles of the black hole with $B^2/\rho c^2 \gg 1$. We refer to the strongly magnetized region as the funnel, and the boundary between the funnel and the corona as the funnel wall (De Villiers et al. 2005; Hawley & Krolik 2006). All models in the library are evolved from $t = 0$ to $t = 10^4\, r_{\rm g} c^{-1}$.

The simulation outcome depends on the initial magnetic field strength and geometry insofar as these affect the magnetic flux through the disk, as discussed below. Once the simulation is initiated the disk transitions to a turbulent state and loses memory of most of the details of the initial conditions. This relaxed turbulent state is found inside a characteristic radius that grows over the course of the simulation. To be confident that we are imaging only those regions that have relaxed, we draw snapshots for comparison with the data from $5 \times 10^3 \leq t/r_{\rm g} c^{-1} \leq 10^4$.

GRMHD models have two key physical parameters. The first is the black hole spin $a_*$, $-1 < a_* < 1$. The second parameter is the absolute magnetic flux $\Phi_{\rm BH}$ crossing one hemisphere of the event horizon (see Tchekhovskoy et al. 2011; O. Porth et al. 2019, in preparation for a definition). It is convenient to recast $\Phi_{\rm BH}$ in dimensionless form $\phi \equiv \Phi_{\rm BH}(\dot{M} r_{\rm g}^2\, c)^{-1/2}$.[110]

The magnetic flux $\phi$ is nonzero because magnetic field is advected into the event horizon by the accretion flow and sustained by currents in the surrounding plasma. At $\phi > \phi_{\rm max} \sim 15$,[111] numerical simulations show that the accumulated magnetic flux erupts, pushes aside the accretion flow, and escapes (Tchekhovskoy et al. 2011; McKinney et al. 2012). Models with $\phi \sim 1$ are conventionally referred to as Standard and Normal Evolution (SANE; Narayan et al. 2012; Sądowski et al (2013a)) models; models with $\phi \sim \phi_{\rm max}$ are conventionally referred to as Magnetically Arrested Disk (MAD; Igumenshchev et al. 2003; Narayan et al. 2003) models.

The Simulation Library contains SANE models with $a_* = -0.94$, $-0.5$, 0, 0.5, 0.75, 0.88, 0.94, 0.97, and 0.98, and MAD models with $a_* = -0.94$, $-0.5$, 0, 0.5, 0.75, and 0.94. The Simulation Library occupies 23 TB of disk space and contains a total of 43 GRMHD simulations, with some repeated at multiple resolutions with multiple codes, with consistent results (O. Porth et al. 2019, in preparation).

### 3.2. Image Library Generation

To produce model images from the simulations for comparison with EHT observations we use GRRT to generate a large number of synthetic images and derived VLBI data products. To make the synthetic images we need to specify the following: (1) the magnetic field, velocity field, and density as a function of position and time; (2) the emission and absorption coefficients as a function of position and time; and (3) the inclination angle between the accretion flow angular momentum vector and the line of sight $i$, the position angle PA, the black hole mass $M$, and the distance $D$ to the observer. In the following we discuss each input in turn. The reader who is only interested in a high-level description of the Image Library may skip ahead to Section 3.3.

(1) GRMHD models provide the absolute velocity field of the plasma flow. Nonradiative GRMHD evolutions are invariant, however, under a rescaling of the density by a factor $\mathscr{M}$. In particular, they are invariant under $\rho \to \mathscr{M}\rho$, field strength $B \to \mathscr{M}^{1/2} B$, and internal energy $u \to \mathscr{M} u$ (the Alfvén speed $B/\rho^{1/2}$ and sound speed $\propto\sqrt{u/\rho}$ are invariant). That is, there is no intrinsic mass scale in a nonradiative model as long as the mass of the accretion flow is negligible in comparison to $M$.[112] We use this freedom to adjust $\mathscr{M}$ so that the average image from a GRMHD model has a 1.3 mm flux density $\approx$0.5 Jy (see Paper IV). Once $\mathscr{M}$ is set, the density, internal energy, and magnetic field are fully specified.

The mass unit $\mathscr{M}$ determines $\dot{M}$. In our ensemble of models $\dot{M}$ ranges from $2 \times 10^{-7} \dot{M}_{\rm Edd}$ to $4 \times 10^{-4} \dot{M}_{\rm Edd}$. Accretion rates vary by model category. The mean accretion rate for MAD models is $\sim 10^{-6} \dot{M}_{\rm Edd}$. For SANE models with $a_* > 0$ it is $\sim 5 \times 10^{-5} \dot{M}_{\rm Edd}$; and for $a_* < 0$ it is $\sim 2 \times 10^{-4} \dot{M}_{\rm Edd}$.

(2) The observed radio spectral energy distributions (SEDs) and the polarization characteristics of the source make clear that the 1.3 mm emission is synchrotron radiation, as is typical for active galactic nuclei (AGNs). Synchrotron absorption and emission coefficients depend on the eDF. In what follows, we adopt a relativistic, thermal model for the eDF (a Maxwell-Jüttner distribution; Jüttner 1911; Rezzolla & Zanotti 2013). We discuss the limitations of this approach in Section 7.

All of our models of M87 are in a sufficiently low-density, high-temperature regime that the plasma is collisionless (see Ryan et al. 2018, for a discussion of Coulomb coupling in M87). Therefore, $T_e$ likely does not equal the ion temperature $T_i$, which is provided by the simulations. We set $T_e$ using the GRMHD density $\rho$, internal energy density $u$, and plasma $\beta_{\rm p}$

[110] $\phi$ is determined by the outcome of the simulation and cannot be trivially predicted from the initial conditions, but by repeated experiment it is possible to manipulate the size of the initial torus and strength and geometry of the initial field to produce a target $\phi$.

[111] In Heaviside units, where a factor of $\sqrt{4\pi}$ is absorbed into the definition of $B$, $\phi_{\rm max} \simeq 15$. In the Gaussian units used in some earlier papers, $\phi_{\rm max} \simeq 50$.

[112] For a black hole accreting at the Eddington rate, the ratio of the accreting mass onto a black hole mass is $\sim 10^{-22}(M/M_\odot)$; in our models mass accretion rate is far below the Eddington rate.

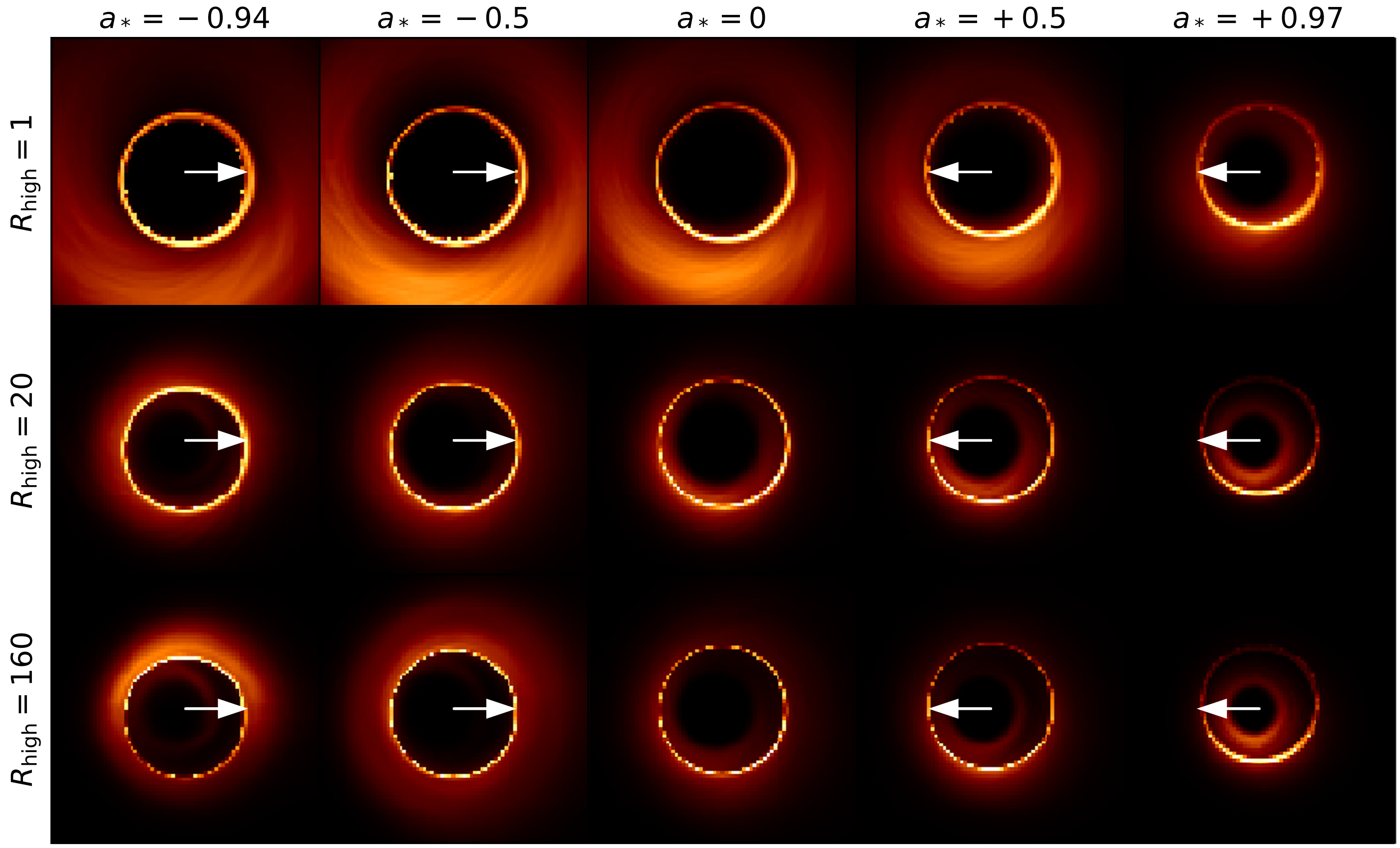

**Figure 2.** Time-averaged 1.3 mm images generated by five SANE GRMHD simulations with varying spin ($a_* = -0.94$ to $a_* = +0.97$ from left to right) and $R_{high}$ ($R_{high} = 1$ to $R_{high} = 160$ from top to bottom; increasing $R_{high}$ corresponds to decreasing electron temperature). The colormap is linear. All models are imaged at $i = 163°$. The jet that is approaching Earth is on the right (west) in all the images. The black hole spin vector projected onto the plane of the sky is marked with an arrow and aligned in the east–west direction. When the arrow is pointing left the black hole rotates in a clockwise direction, and when the arrow is pointing right the black hole rotates in a counterclockwise direction. The field of view for each model image is 80 $\mu$as (half of that used for the image libraries) with resolution equal to $1\mu$as/pixel (20 times finer than the nominal resolution of EHT2017, and the same employed in the library images).

using a simple model:

$$T_e = \frac{2 m_p u}{3 k \rho (2 + R)}, \quad (7)$$

where we have assumed that the plasma is composed of hydrogen, the ions are nonrelativistic, and the electrons are relativistic. Here $R \equiv T_i / T_e$ and

$$R = R_{high} \frac{\beta_p^2}{1 + \beta_p^2} + \frac{1}{1 + \beta_p^2}. \quad (8)$$

This prescription has one parameter, $R_{high}$, and sets $T_e \simeq T_i$ in low $\beta_p$ regions and $T_e \simeq T_i / R_{high}$ in the midplane of the disk. It is adapted from Mościbrodzka et al. (2016) and motivated by models for electron heating in a turbulent, collisionless plasma that preferentially heats the ions for $\beta_p \gtrsim 1$ (e.g., Howes 2010; Kawazura et al. 2018).

(3) We must specify the observer inclination $i$, the orientation of the observer through the position angle PA, the black hole mass $M$, and the distance $D$ to the source. Non-EHT constraints on $i$, PA, and $M$ are considered below; we have generated images at $i = 12°$, 17°, 22°, 158°, 163°, and 168° and a few at $i = 148°$. The position angle (PA) can be changed by simply rotating the image. All features of the models that we have examined, including $\dot{M}$, are insensitive to small changes in $i$. The image morphology does depend on whether $i$ is greater than or less than 90°, as we will show below.

The model images are generated with a $160 \times 160$ $\mu$as field of view and $1 \mu$as pixels, which are small compared to the $\sim$20 $\mu$as nominal resolution of EHT2017. Our analysis is insensitive to changes in the field of view and the pixel scale.

For $M$ we use the most likely value from the stellar absorption-line work, $6.2 \times 10^9 M_\odot$ (Gebhardt et al. 2011). For the distance $D$ we use 16.9 Mpc, which is very close to that employed in Paper VI. The ratio $GM/(c^2 D) = 3.62$ $\mu$as (hereafter $M/D$) determines the angular scale of the images. For some models we have also generated images with $M = 3.5 \times 10^9 M_\odot$ to check that the analysis results are not predetermined by the input black hole mass.

### 3.3. Image Library Summary

The Image Library contains of order 60,000 images. We generate images from 100 to 500 distinct output files from each of the GRMHD models at each of $R_{high} = 1$, 10, 20, 40, 80, and 160. In comparing to the data we adjust the PA by rotation and the total flux and angular scale of the image by simply rescaling images from the standard parameters in the Image Library (see Figure 29 in Paper VI). Tests indicate that comparisons with the data are insensitive to the rescaling procedure unless the angular scaling factor or flux scaling factor is large.[113]

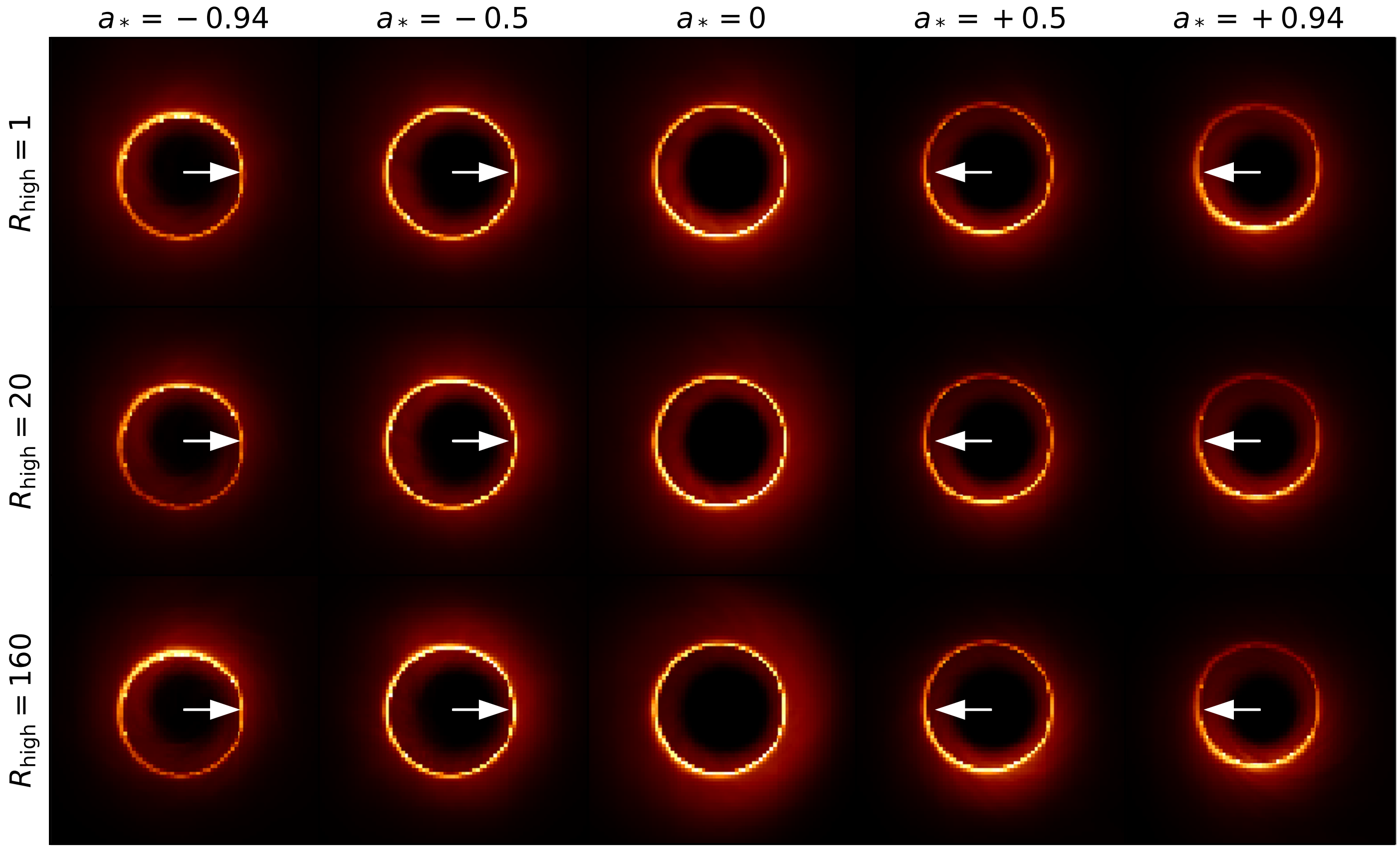

**Figure 3.** Same as in Figure 2 but for selected MAD models.

The comparisons with the data are also insensitive to image resolution.[114]

A representative set of time-averaged images from the Image Library are shown in Figures 2 and 3. From these figures it is clear that varying the parameters $a_*$, $\phi$, and $R_{high}$ can change the width and asymmetry of the photon ring and introduce additional structures exterior and interior to the photon ring.

The location of the emitting plasma is shown in Figure 4, which shows a map of time- and azimuth-averaged emission regions for four representative $a_* > 0$ models. For SANE models, if $R_{high}$ is low (high), emission is concentrated more in the disk (funnel wall), and the bright section of the ring is dominated by the disk (funnel wall).[115] Appendix B shows images generated by considering emission only from particular regions of the flow, and the results are consistent with Figure 4.

Figures 2 and 3 show that for both MAD and SANE models the bright section of the ring, which is generated by Doppler beaming, shifts from the top for negative spin, to a nearly symmetric ring at $a_* = 0$, to the bottom for $a_* > 0$ (except the SANE $R_{high} = 1$ case, where the bright section is always at the bottom when $i > 90°$). That is, the location of the peak flux in the ring is controlled by the black hole spin: it always lies roughly 90 degrees counterclockwise from the projection of the spin vector on the sky. Some of the ring emission originates in the funnel wall at $r \lesssim 8\ r_g$. The rotation of plasma in the funnel wall is in the same sense as plasma in the funnel, which is controlled by the dragging of magnetic field lines by the black hole. The funnel wall thus rotates opposite to the accretion flow if $a_* < 0$. This effect will be studied further in a later publication (Wong et al. 2019). The resulting relationships between disk angular momentum, black hole angular momentum, and observed ring asymmetry are illustrated in Figure 5.

The time-averaged MAD images are almost independent of $R_{high}$ and depend mainly on $a_*$. In MAD models much of the emission arises in regions with $\beta_p \sim 1$, where $R_{high}$ has little influence over the electron temperature, so the insensitivity to $R_{high}$ is natural (see Figure 4). In SANE models emission arises at $\beta_p \sim 10$, so the time-averaged SANE images, by contrast, depend strongly on $R_{high}$. In low $R_{high}$ SANE models, extended emission outside the photon ring, arising near the equatorial plane, is evident at $R_{high} = 1$. In large $R_{high}$ SANE models the inner ring emission arises from the funnel wall, and once again the image looks like a thin ring (see Figure 4).

Figure 6 and the accompanying animation show the evolution of the images, visibility amplitudes, and closure phases over a 5000 $r_g c^{-1} \approx 5$ yr interval in a single simulation for M87. It is evident from the animation that turbulence in the simulations produces large fluctuations in the images, which

113 In particular the distribution of best-fit $M/D$, which is defined in Section 4, have mean and standard deviation of $M/D = 3.552 \pm 0.605\ \mu$as when the images are made with an input $M/D = 3.62\ \mu$as, and $3.564 \pm 0.537\ \mu$as when the images are made with an input $M/D = 2.01\ \mu$as. We have also checked images made with an input 1.3 mm flux ranging from 0.1 to 1.5 Jy and find relative changes in $M/D$ and PA of less than 1%.

114 In particular, doubling the image resolution changes the mean best-fit $M/D$ by 7 nano-arcsec, and the best-fit PA by $\sim 0\overset{\circ}{.}3$.

115 In GRMHD models the jet core is effectively empty and the density is set by numerical "floors." In our radiative transfer calculations emission from regions with $B^2/\rho > 1$ is explicitly set to zero.

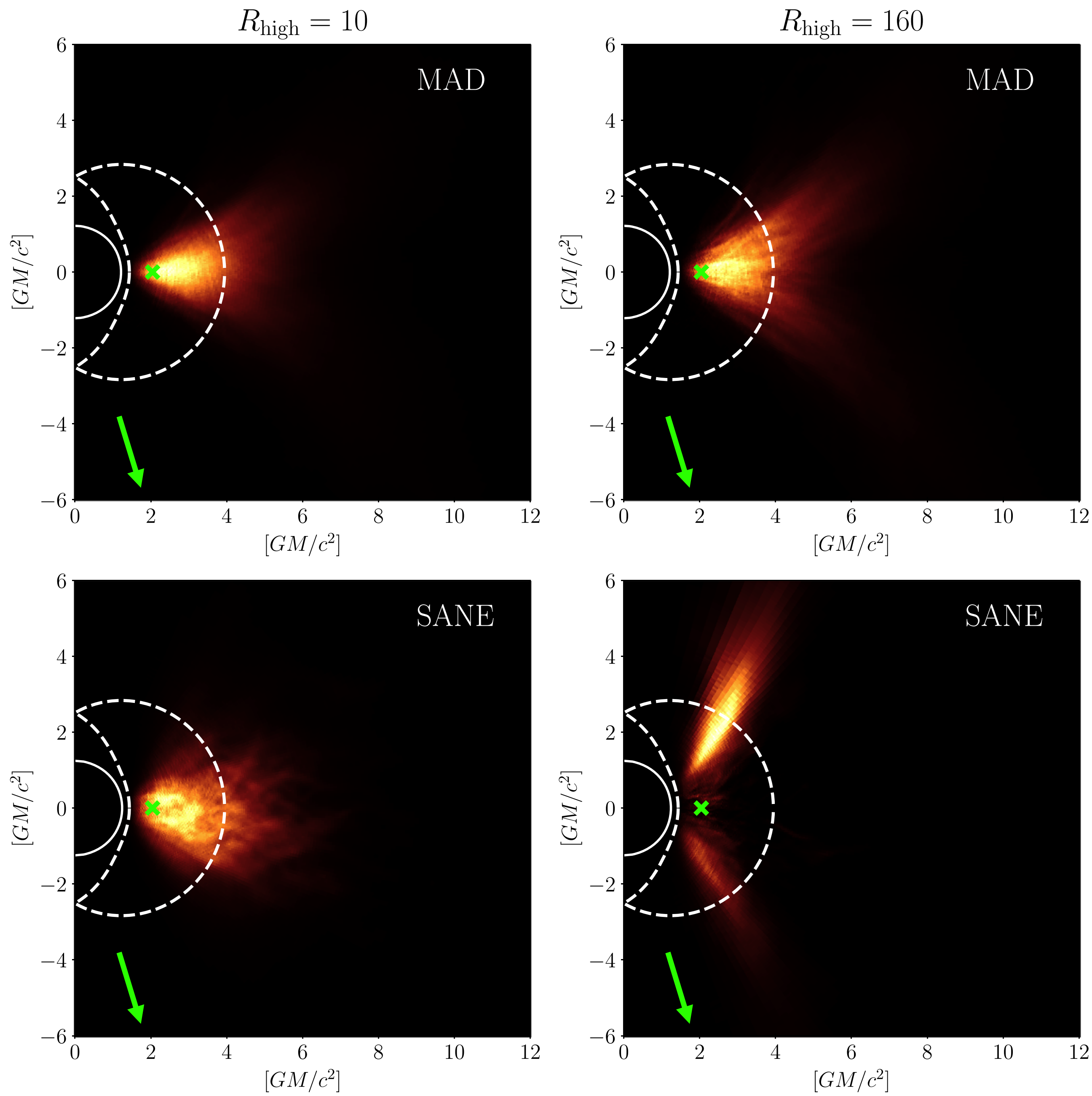

**Figure 4.** Binned location of the point of origin for all photons that make up an image, summed over azimuth, and averaged over all snapshots from the simulation. The colormap is linear. The event horizon is indicated by the solid white semicircle and the black hole spin axis is along the figure vertical axis. This set of four images shows MAD and SANE models with $R_{\rm high} = 10$ and 160, all with $a_* = 0.94$. The region between the dashed curves is the locus of existence of (unstable) photon orbits (Teo 2003). The green cross marks the location of the innermost stable circular orbit (ISCO) in the equatorial plane. In these images the line of sight (marked by an arrow) is located below the midplane and makes a 163° angle with the disk angular momentum, which coincides with the spin axis of the black hole.

imply changes in visibility amplitudes and closure phases that are large compared to measurement errors. The fluctuations are central to our procedure for comparing models with the data, described briefly below and in detail in Paper VI.

The timescale between frames in the animation is $50\, r_{\rm g} c^{-1} \simeq 18$ days, which is long compared to EHT2017 observing campaign. The images are highly correlated on timescales less than the innermost stable circular orbit (ISCO) orbital period, which for $a_* = 0$ is $\simeq 15\, r_{\rm g} c^{-1} \simeq 5$ days, i.e., comparable to the duration of the EHT2017 campaign. If drawn from one of our models, we would expect the EHT2017 data to look like a single snapshot (Figures 6) rather than their time averages (Figures 2 and 3).

## 4. Procedure for Comparison of Models with Data

As described above, each model in the Simulation Library has two dimensionless parameters: black hole spin $a_*$ and magnetic flux $\phi$. Imaging the model from each simulation adds five new parameters: $R_{\rm high}$, $i$, PA, $M$, and $D$, which we set to 16.9 Mpc. After fixing these parameters we draw snapshots from the time evolution at a cadence of 10 to 50 $r_{\rm g} c^{-1}$. We then compare these snapshots to the data.

The simplest comparison computes the $\chi^2_\nu$ (reduced chi square) distance between the data and a snapshot. In the course of computing $\chi^2_\nu$ we vary the image scale $M/D$, flux density $F_\nu$, position angle PA, and the gain at each VLBI station in order to give each image every opportunity to fit the data. The best-fit parameters ($M/D$, $F_\nu$, PA) for each snapshot are found by two pipelines independently: the THEMIS pipeline using a Markov chain Monte Carlo method (A. E. Broderick et al. 2019a, in preparation), and the GENA pipeline using an evolutionary algorithm for multidimensional minimization (Fromm et al. 2019a; C. Fromm et al. 2019b, in preparation; see also Section 4 of Paper VI for details). The best-fit parameters contain information about the source and we use the distribution of best-fit parameters to test the model by asking whether or not they are consistent with existing measurements of $M/D$ and estimates of the jet PA on larger scales.

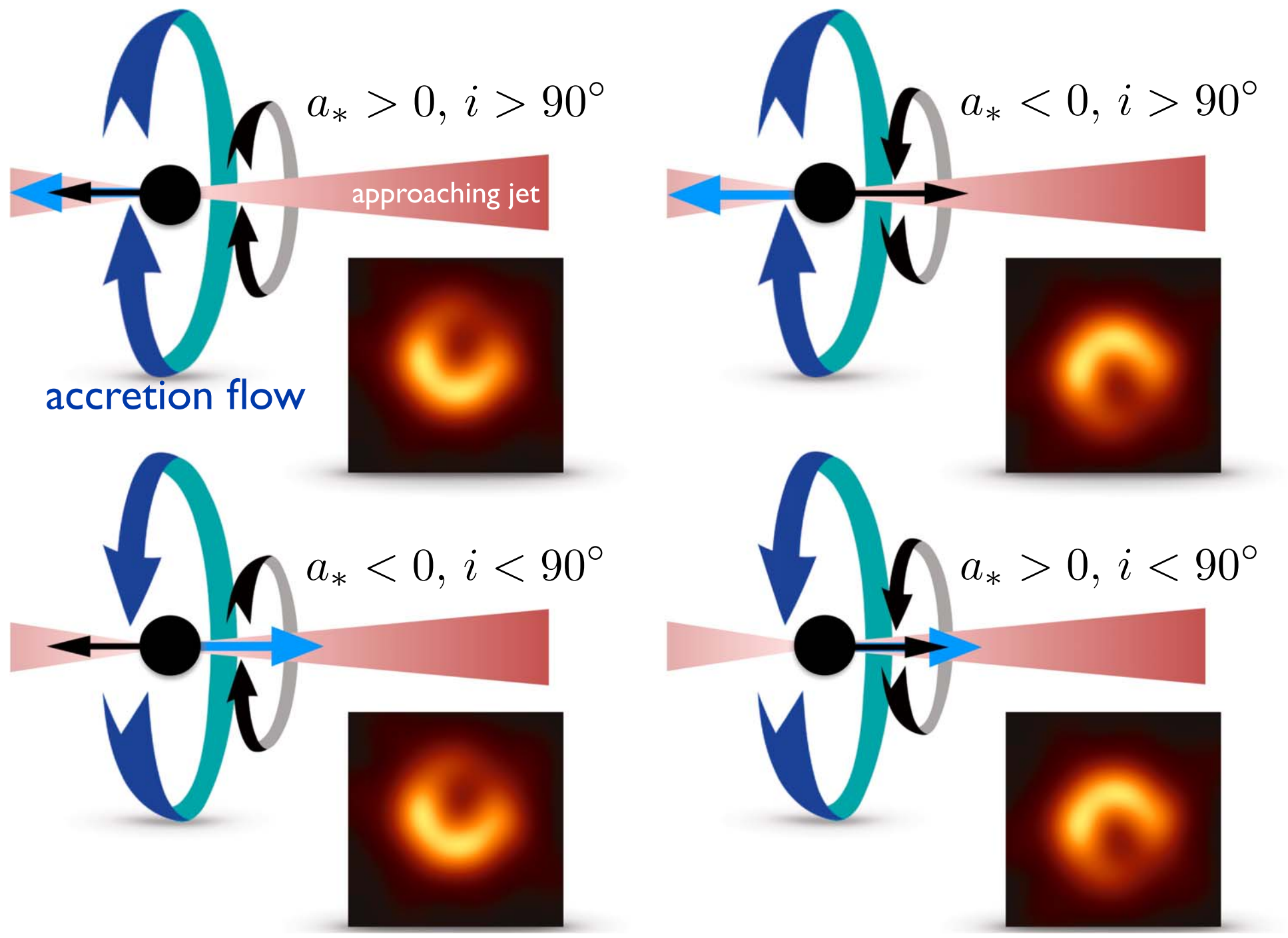

**Figure 5.** Illustration of the effect of black hole and disk angular momentum on ring asymmetry. The asymmetry is produced primarily by Doppler beaming: the bright region corresponds to the approaching side. In GRMHD models that fit the data comparatively well, the asymmetry arises in emission generated in the funnel wall. The sense of rotation of both the jet and funnel wall are controlled by the black hole spin. If the black hole spin axis is aligned with the large-scale jet, which points to the right, then the asymmetry implies that the black hole spin is pointing away from Earth (rotation of the black hole is clockwise as viewed from Earth). The blue ribbon arrow shows the sense of disk rotation, and the black ribbon arrow shows black hole spin. Inclination $i$ is defined as the angle between the disk angular momentum vector and the line of sight.

The $\chi^2_\nu$ comparison alone does not provide a sharp test of the models. Fluctuations in the underlying GRMHD model, combined with the high signal-to-noise ratio for EHT2017 data, imply that individual snapshots are highly unlikely to provide a formally acceptable fit with $\chi^2_\nu \simeq 1$. This is borne out in practice with the minimum $\chi^2_\nu = 1.79$ over the entire set of the more than 60,000 individual images in the Image Library. Nevertheless, it is possible to test if the $\chi^2_\nu$ from the fit to the data is consistent with the underlying model, using "Average Image Scoring" with THEMIS (THEMIS-AIS), as described in detail in Appendix F of Paper VI). THEMIS-AIS measures a $\chi^2_\nu$ distance (on the space of visibility amplitudes and closure phases) between a trial image and the data. In practice we use the average of the images from a given model as the trial image (hence THEMIS-AIS), but other choices are possible. We compute the $\chi^2_\nu$ distance between the trial image and synthetic data produced from each snapshot. The model can then be tested by asking whether the data's $\chi^2_\nu$ is likely to have been drawn from the model's distribution of $\chi^2_\nu$. In particular, we can assign a probability $p$ that the data is drawn from a specific model's distribution.

In this Letter we focus on comparisons with a single data set, the 2017 April 6 high-band data (Paper III). The eight EHT2017 data sets, spanning four days with two bands on each day, are highly correlated. Assessing what correlation is expected in the models is a complicated task that we defer to later publications. The 2017 April 6 data set has the largest number of scans, 284 detections in 25 scans (see Paper III) and is therefore expected to be the most constraining.[116]

## 5. Model Constraints: EHT2017 Alone

The resolved ring-like structure obtained from the EHT2017 data provides an estimate of $M/D$ (discussed in detail in Paper VI) and the jet PA from the immediate environment of the central black hole. As a first test of the models we can ask whether or not these are consistent with what is known from other mass measurements and from the orientation of the large-scale jet.

Figure 7 shows the distributions of best-fit values of $M/D$ for a subset of the models for which spectra and jet power estimates are available (see below). The three lines show the $M/D$ distribution for all snapshots (dotted lines), the best-fit 10% of snapshots (dashed lines), and the best-fit 1% of snapshots (solid lines) within each model. Evidently, as better fits are required, the distribution narrows and peaks close to $M/D \sim 3.6\ \mu$as with a width of about $0.5\mu$as.

The distribution of $M/D$ for the best-fit $<10\%$ of snapshots is qualitatively similar if we include only MAD or SANE models, only models produced by individual codes (BHAC,

[116] Paper I and Paper IV focus instead on the April 11 data set.

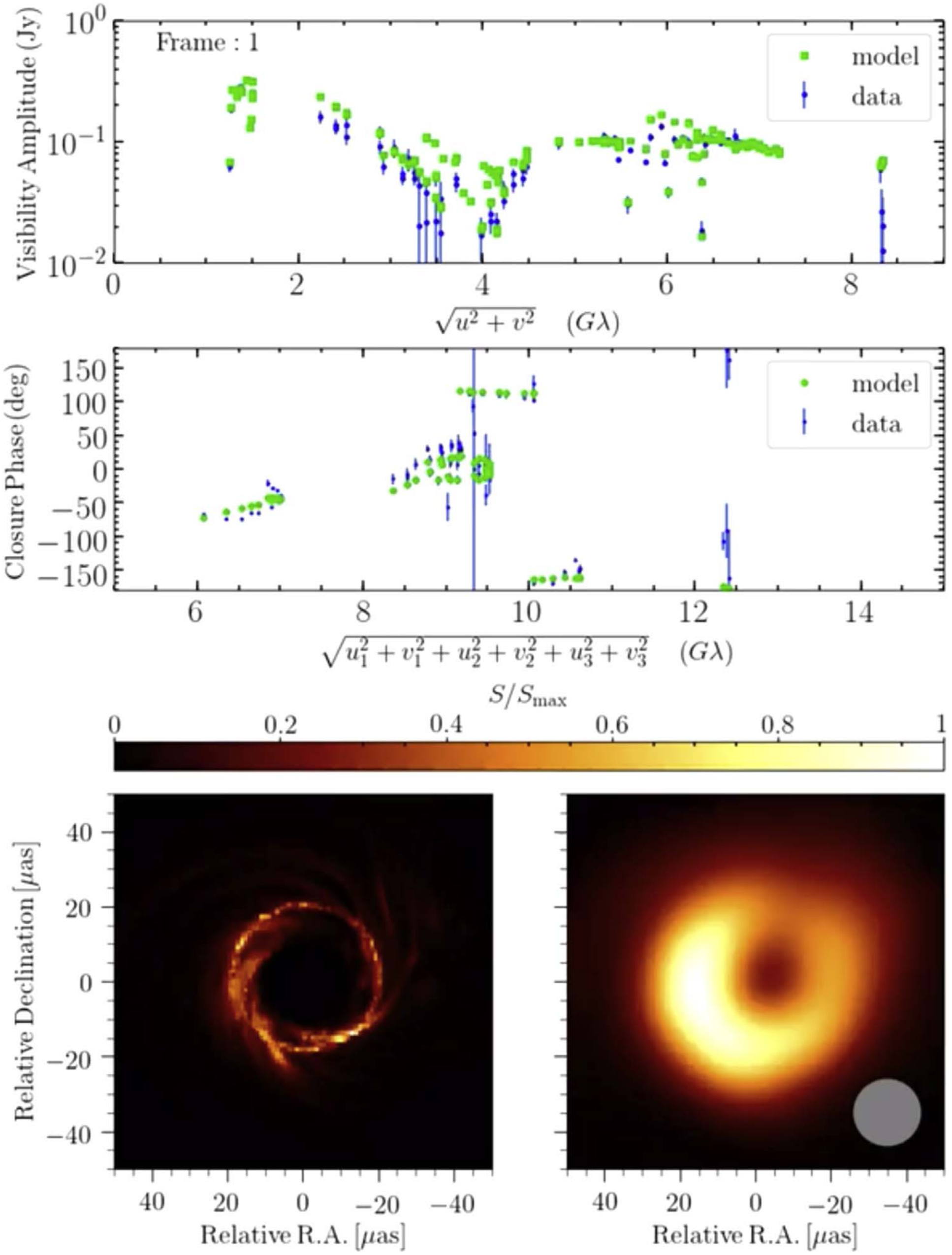

**Figure 6.** Single frame from the accompanying animation. This shows the visibility amplitudes (top), closure phases plotted by Euclidean distance in 6D space (middle), and associated model images at full resolution (lower left) and convolved with the EHT2017 beam (lower right). Data from 2017 April 6 high-band are also shown in the top two plots. The video shows frames 1 through 100 and has a duration of 10 s.

(An animation of this figure is available.)

H-AMR, iharm, or KORAL), or only individual spins. As the thrust of this Letter is to test the models, we simply note that Figure 7 indicates that the models are broadly consistent with earlier mass estimates (see Paper VI for a detailed discussion). This did not have to be the case: the ring radius could have been significantly larger than 3.6 $\mu$as.

We can go somewhat further and ask if any of the individual models favor large or small masses. Figure 8 shows the distributions of best-fit values of $M/D$ for each model (different $a_*$, $R_{\rm high}$, and magnetic flux). Most individual models favor $M/D$ close to 3.6 $\mu$as. The exceptions are $a_* \leqslant 0$ SANE models with $R_{\rm high} = 1$, which produce the bump in the $M/D$ distribution near 2$\mu$as. In these models, the emission is produced at comparatively large radius in the disk (see Figure 2) because the inner edge of the disk (the ISCO) is at a large radius in a counter-rotating disk around a black hole with $|a_*| \sim 1$. For these models, the fitting procedure identifies EHT2017's ring with this outer ring, which forces the photon ring, and therefore $M/D$, to be small. As we will show later, these models can be rejected because they produce weak jets that are inconsistent with existing jet power estimates (see Section 6.3).

Figure 8 also shows that $M/D$ increases with $a_*$ for SANE models. This is due to the appearance of a secondary inner ring inside the main photon ring. The former is associated with emission produced along the wall of the approaching jet.

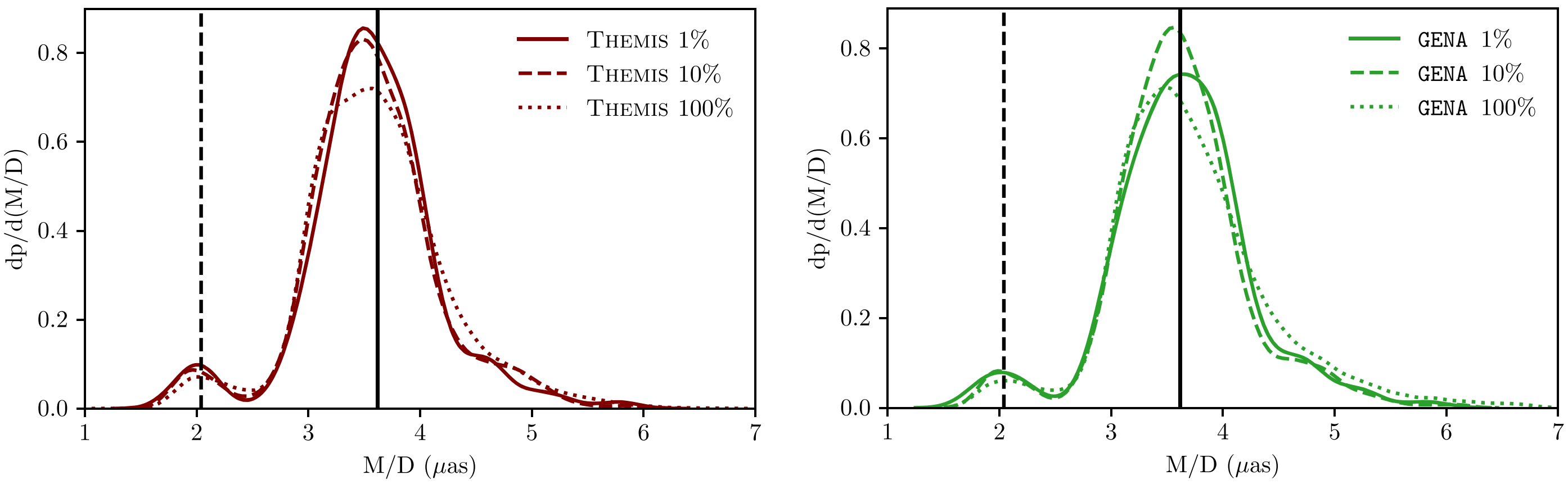

**Figure 7.** Distribution of $M/D$ obtained by fitting Image Library snapshots to the 2017 April 6 data, in $\mu$as, measured independently using the (left panel) THEMIS and (right panel) GENA pipelines with qualitatively similar results. Smooth lines were drawn with a Gaussian kernel density estimator. The three lines show the best-fit 1% within each model (solid); the best-fit 10% within each model (dashed); and all model images (dotted). The vertical lines show $M/D = 2.04$ (dashed) and 3.62 $\mu$as (solid), corresponding to $M = 3.5$ and $6.2 \times 10^9\, M_\odot$. The distribution uses a subset of models for which spectra and jet power estimates are available (see Section 6). Only images with $a_* > 0$, $i > 90°$ and $a_* < 0$, $i < 90°$ (see also the left panel of Figure 5) are considered.

Because the emission is produced in front of the black hole, lensing is weak and it appears at small angular scale. The inner ring is absent in MAD models (see Figure 3), where the bulk of the emission comes from the midplane at all values of $R_{\rm high}$ (Figure 4).

We now ask whether or not the PA of the jet is consistent with the orientation of the jet measured at other wavelengths. On large (∼mas) scales the extended jet component has a PA of approximately 288° (e.g., Walker et al. 2018). On smaller (∼100 $\mu$as) scales the apparent opening angle of the jet is large (e.g., Kim et al. 2018) and the PA is therefore more difficult to measure. Also notice that the jet PA may be time dependent (e.g., Hada et al. 2016; Walker et al. 2018). In our model images the jet is relatively dim at 1.3 mm, and is not easily seen with a linear colormap. The model jet axis is, nonetheless, well defined: jets emerge perpendicular to the disk.

Figure 9 shows the distribution of best-fit PA over the same sample of snapshots from the Image Library used in Figure 7. We divide the snapshots into two groups. The first group has the black hole spin pointed away from Earth ($i > 90°$ and $a_* > 0$, or $i < 90°$ and $a_* < 0$). The spin-away model PA distributions are shown in the top two panels. The second group has the black hole spin pointed toward Earth ($i > 90$ and $a_* < 0$ or $i > 90°$ and $a_* < 0$). These spin-toward model PA distributions are shown in the bottom two panels. The large-scale jet orientation lies on the shoulder of the spin-away distribution (the distribution can be approximated as a Gaussian with, for THEMIS (GENA) mean 209 (203)° and $\sigma_{\rm PA} = 54\,(55)°$; the large-scale jet PA lies $1.5\sigma_{\rm PA}$ from the mean) and is therefore consistent with the spin-away models. On the other hand, the large-scale jet orientation lies off the shoulder of the spin-toward distribution and is inconsistent with the spin-toward models. Evidently models in which the black hole spin is pointing away from Earth are strongly favored.

The width of the spin-away and spin-toward distributions arises naturally in the models from brightness fluctuations in the ring. The distributions are relatively insensitive if split into MAD and SANE categories, although for MAD the averaged PA is $\langle {\rm PA} \rangle = 219°$, $\sigma_{\rm PA} = 46°$, while for SANE $\langle {\rm PA} \rangle = 195°$ and $\sigma_{\rm PA} = 58°$. The $a_* = 0$ and $a_* > 0$ models have similar distributions. Again, EHT2017 data strongly favor one sense of black hole spin: either $|a_*|$ is small, or the spin vector is pointed away from Earth. If the fluctuations are such that the fitted PA for each epoch of observations is drawn from a Gaussian with $\sigma_{\rm PA} \simeq 55°$, then a second epoch will be able to identify the true orientation with accuracy $\sigma_{\rm PA}/\sqrt{2} \simeq 40°$ and the $N$th epoch with accuracy $\sigma_{\rm PA}/\sqrt{N}$. If the fitted PA were drawn from a Gaussian of width $\sigma_{\rm PA} = 54°$ about PA = 288°, as would be expected in a model in which the large-scale jet is aligned normal to the disk, then future epochs have a >90% chance of seeing the peak brightness counterclockwise from its position in EHT2017.

Finally, we can test the models by asking if they are consistent with the data according to THEMIS-AIS, as introduced in Section 4. THEMIS-AIS produces a probability $p$ that the $\chi^2_\nu$ distance between the data and the average of the model images is drawn from the same distribution as the $\chi^2_\nu$ distance between synthetic data created from the model images, and the average of the model images. Table 1 takes these $p$ values and categorizes them by magnetic flux and by spin, aggregating (averaging) results from different codes, $R_{\rm high}$, and $i$. Evidently, most of the models are formally consistent with the data by this test.

One group of models, however, is rejected by THEMIS-AIS: MAD models with $a_* = -0.94$. On average this group has $p = 0.01$, and all models within this group have $p \leqslant 0.04$. Snapshots from MAD models with $a_* = -0.94$ exhibit the highest morphological variability in our ensemble in the sense that the emission breaks up into transient bright clumps. These models are rejected by THEMIS-AIS because none of the snapshots are as similar to the average image as the data. In other words, it is unlikely that EHT2017 would have captured an $a_* = -0.94$ MAD model in a configuration as unperturbed as the data seem to be.

The remainder of the model categories contain at least some models that are consistent with the data according to the average image scoring test. That is, most models are variable and the associated snapshots lie far from the average image. These snapshots are formally inconsistent with the data, but their distance from the average image is consistent with what is expected from the models. Given the uncertainties in the model—and our lack of knowledge of the source prior to EHT2017—it is remarkable that so many of the models are acceptable. This is likely because the source structure is dominated by the photon ring, which is produced by gravitational lensing, and is therefore relatively insensitive to the details of the accretion flow and jet physics. We can further narrow the range of acceptable models, however, using additional constraints.

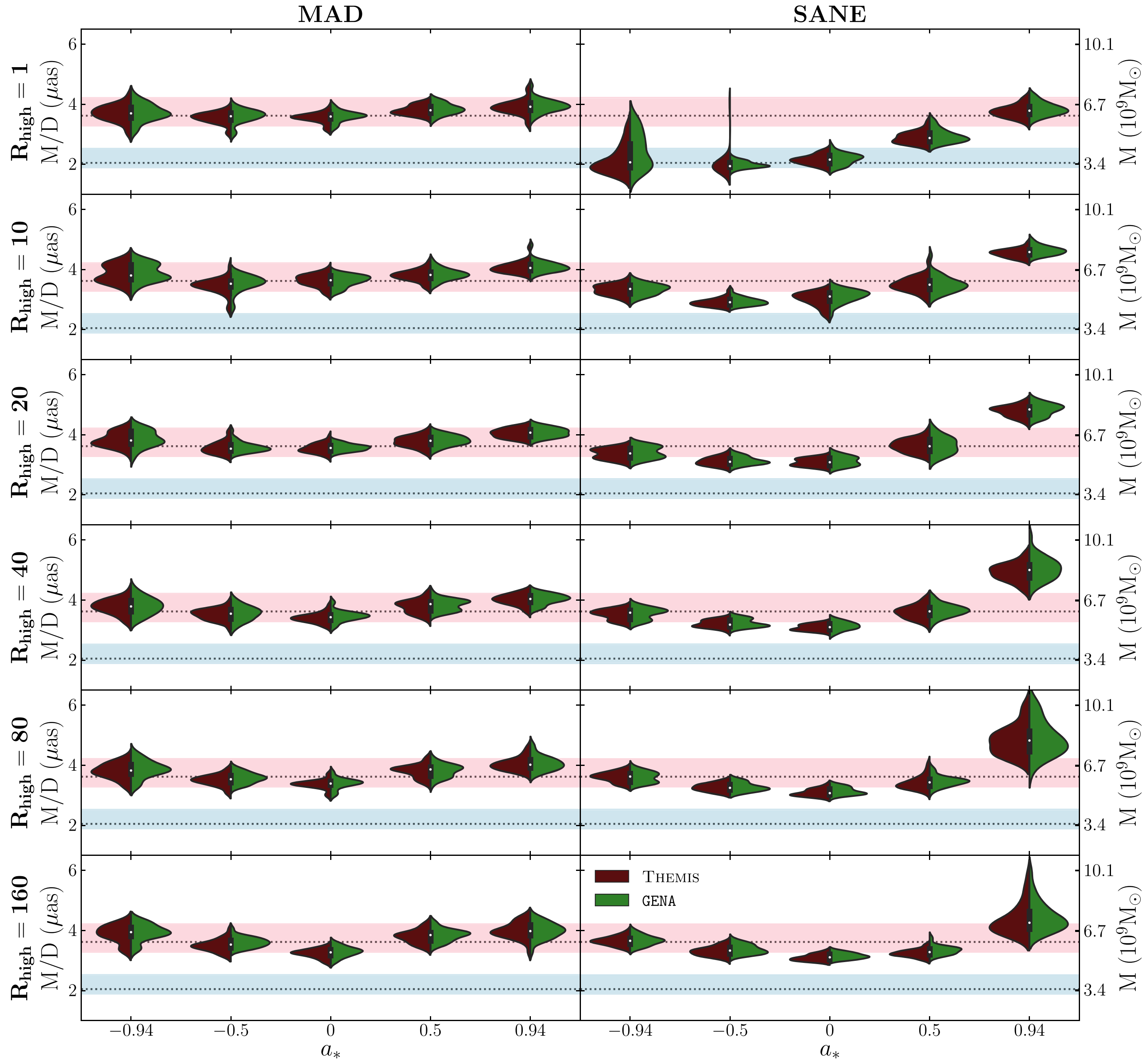

**Figure 8.** Distributions of $M/D$ and black hole mass with $D = 16.9$ Mpc reconstructed from the best-fit 10% of images for MAD (left panel) and SANE (right panel) models ($i = 17^\circ$ for $a_* \leq 0$ and 163° for $a_* > 0$) with different $R_{\mathrm{high}}$ and $a_*$, from the THEMIS (dark red, left), and GENA (dark green, right) pipelines. The white dot and vertical black bar correspond, respectively, to the median and region between the 25th and 75th percentiles for both pipelines combined. The blue and pink horizontal bands show the range of $M/D$ and mass at $D = 16.9$ Mpc estimated from the gas dynamical model (Walsh et al. 2013) and stellar dynamical model (Gebhardt et al. 2011), respectively. Constraints on the models based on average image scoring (THEMIS-AIS) are discussed in Section 5. Constraints based on radiative efficiency, X-ray luminosity, and jet power are discussed in Section 6.

## 6. Model Constraints: EHT2017 Combined with Other Constraints

We can apply three additional arguments to further constrain the source model. (1) The model must be close to radiative equilibrium. (2) The model must be consistent with the observed broadband SED; in particular, it must not overproduce X-rays. (3) The model must produce a sufficiently powerful jet to match the measurements of the jet kinetic energy at large scales. Our discussions in this Section are based on simulation data that is provided in full detail in Appendix A.

### 6.1. Radiative Equilibrium

The model must be close to radiative equilibrium. The GRMHD models in the Simulation Library do not include radiative cooling, nor do they include a detailed prescription for particle energization. In nature the accretion flow and jet are expected to be cooled and heated by a combination of synchrotron and Compton cooling, turbulent dissipation, and Coulomb heating, which transfers energy from the hot ions to the cooler electrons. In our suite of simulations the parameter $R_{\rm high}$ can be thought of as a proxy for the sum of these processes. In a fully self-consistent treatment, some models would rapidly cool and settle to a lower electron temperature (see Mościbrodzka et al. 2011; Ryan et al. 2018; Chael et al. 2019). We crudely test for this by calculating the radiative efficiency $\epsilon \equiv L_{\rm bol}/(\dot{M}c^2)$, where $L_{\rm bol}$ is the bolometric luminosity. If it is larger than the radiative efficiency of a thin, radiatively efficient disk,[117] which depends only on $a_*$ (Novikov & Thorne 1973), then we reject the model as physically inconsistent.

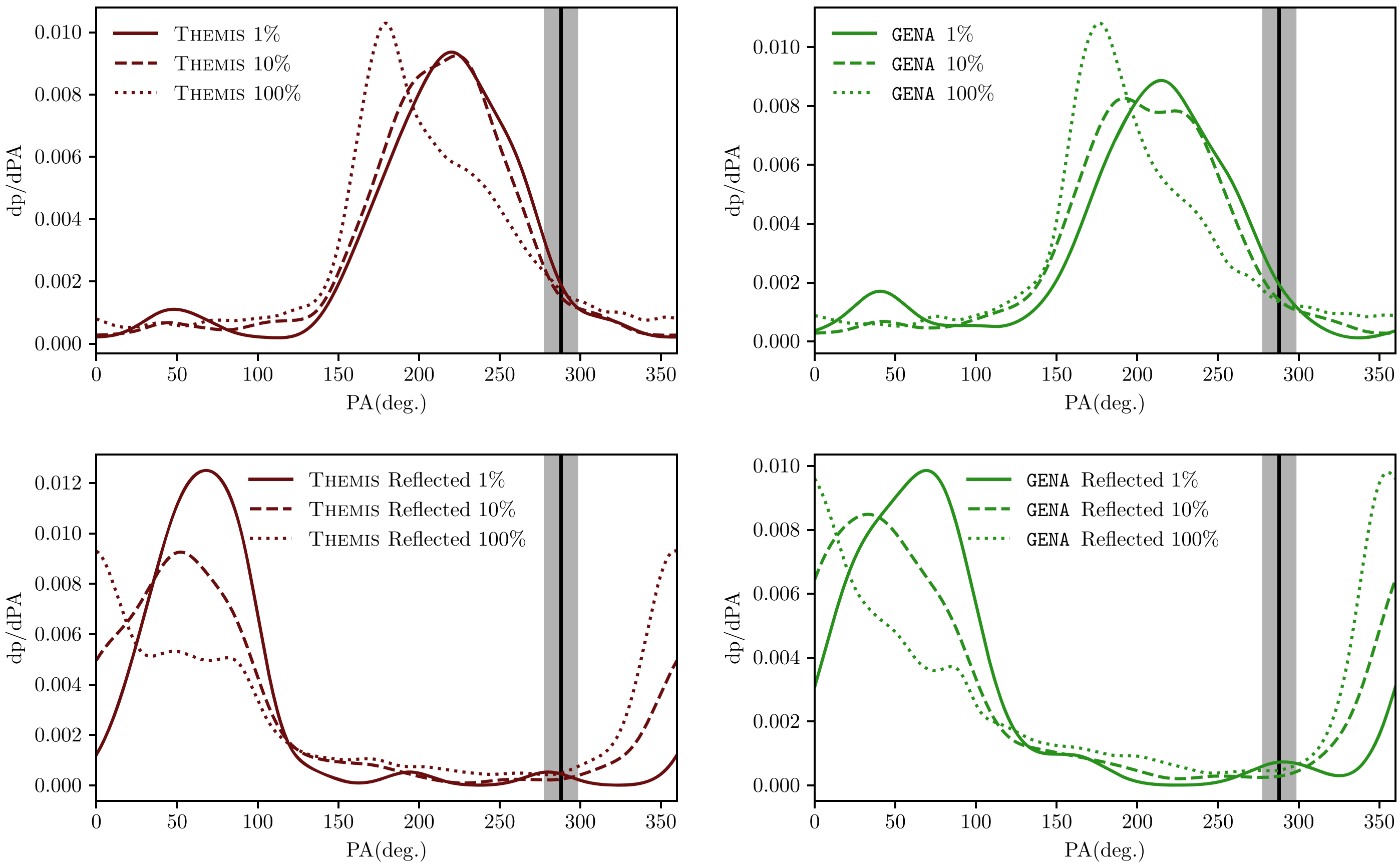

**Figure 9.** Top: distribution of best-fit PA (in degree) scored by the THEMIS (left) and GENA (right) pipelines for models with black hole spin vector pointing away from Earth ($i > 90°$ for $a_* > 0$ or $i < 90°$ for $a_* < 0$). Bottom: images with black hole spin vector pointing toward Earth ($i < 90°$ for $a_* > 0$ or $i > 90°$ for $a_* < 0$). Smooth lines were drawn with a wrapped Gaussian kernel density estimator. The three lines show (1) all images in the sample (dotted line); (2) the best-fit 10% of images within each model (dashed line); and (3) the best-fit 1% of images in each model (solid line). For reference, the vertical line shows the position angle PA $\sim 288°$ of the large-scale (mas) jet Walker et al. (2018), with the gray area from $(288 - 10)°$ to $(288 + 10)°$ indicating the observed PA variation.

We calculate $L_{\rm bol}$ with the Monte Carlo code `grmonty` (Dolence et al. 2009), which incorporates synchrotron emission, absorption, Compton scattering at all orders, and bremsstrahlung. It assumes the same thermal eDF used in generating the Image Library. We calculate $L_{\rm bol}$ for 20% of the snapshots to minimize computational cost. We then average over snapshots to find $\langle L_{\rm bol}\rangle$. The mass accretion rate $\dot{M}$ is likewise computed for each snapshot and averaged over time. We reject models with $\epsilon$ that is larger than the classical thin disk model. (Table 3 in Appendix A lists $\epsilon$ for a large set of models.) All but two of the radiatively inconsistent models are MADs with $a_* \geqslant 0$ and $R_{\rm high} = 1$. Eliminating all MAD models with $a_* \geqslant 0$ and $R_{\rm high} = 1$ does not change any of our earlier conclusions.

### 6.2. X-Ray Constraints

As part of the EHT2017 campaign, we simultaneously observed M87 with the *Chandra* X-ray observatory and the *Nuclear Spectroscopic Telescope Array* (*NuSTAR*). The best fit to simultaneous *Chandra* and *NuSTAR* observations on 2017 April 12 and 14 implies a 2–10 keV luminosity of $L_{\rm X_{obs}} = 4.4 \pm 0.1 \times 10^{40}$ erg s$^{-1}$. We used the SEDs generated from the simulations while calculating $L_{\rm bol}$ to reject models that consistently overproduce X-rays; specifically, we reject models with $\log L_{\rm X_{obs}} < \log\langle L_{\rm X}\rangle - 2\sigma(\log L_{\rm X})$. We do not reject underluminous models because the X-rays could in principle be produced by direct synchrotron emission from nonthermal electrons or by other unresolved sources. Notice that $L_{\rm X}$ is highly variable in all models so that the X-ray observations currently reject only a few models. Table 3 in Appendix A shows $\langle L_{\rm X}\rangle$ as well as upper and lower limits for a set of models that is distributed uniformly across the parameter space.

In our models the X-ray flux is produced by inverse Compton scattering of synchrotron photons. The X-ray flux is an increasing function of $\tau_T T_e^2$ where $\tau_T$ is a characteristic Thomson optical depth ($\tau_T \sim 10^{-5}$), and the characteristic amplification factor for photon energies is $\propto T_e^2$ because the X-ray band is dominated by singly scattered photons interacting with relativistic electrons (we include all scattering orders in the Monte Carlo calculation). Increasing $R_{\rm high}$ at fixed $F_\nu$(230 GHz) tends to increase $\dot{M}$ and therefore $\tau_T$ and decrease $T_e$. The increase in $T_e$ dominates in our ensemble of models, and so models with small $R_{\rm high}$ have larger $L_X$, while models with large $R_{\rm high}$ have smaller $L_X$. The effect is not strictly monotonic, however, because of noise in our sampling process and the highly variable nature of the X-ray emission.

[117] The thin disk radiative efficiency is 0.038 for $a_* = -1$, 0.057 for $a_* = 0$, and 0.42 for $a_* = 1$. See Equations (2.12) and (2.21) of Bardeen et al. (1972); the efficiency is $1 - E/\mu_p$, where $\mu_p$ is the rest mass of the particle. The rejected model list is identical if instead one simply rejects all models with $\epsilon > 0.2$.

The overluminous models are mostly SANE models with $R_{\rm high} \leqslant 20$. The model with the highest $\langle L_X \rangle = 4.2 \times 10^{42}$ erg s$^{-1}$ is a SANE, $a_* = 0$, $R_{\rm high} = 10$ model. The corresponding model with $R_{\rm high} = 1$ has $\langle L_X \rangle = 2.1 \times 10^{41}$ erg s$^{-1}$, and the difference between these two indicates the level of variability and the sensitivity of the average to the brightest snapshot. The upshot of application of the $L_X$ constraints is that $L_X$ is sensitive to $R_{\rm high}$. Very low values of $R_{\rm high}$ are disfavored. $L_X$ thus most directly constrains the electron temperature model.

### 6.3. Jet Power

Estimates of M87's jet power ($P_{\rm jet}$) have been reviewed in Reynolds et al. (1996), Li et al. (2009), de Gasperin et al. (2012), Broderick et al. (2015), and Prieto et al. (2016). The estimates range from $10^{42}$ to $10^{45}$ erg s$^{-1}$. This wide range is a consequence of both physical uncertainties in the models used to estimate $P_{\rm jet}$ and the wide range in length and timescales probed by the observations. Some estimates may sample a different epoch and thus provide little information on the state of the central engine during EHT2017. Nevertheless, observations of *HST*-1 yield $P_{\rm jet} \sim 10^{44}$ erg s$^{-1}$ (e.g., Stawarz et al. 2006). *HST*-1 is within ∼70 pc of the central engine and, taking account of relativistic time foreshortening, may be sampling the central engine $P_{\rm jet}$ over the last few decades. Furthermore, the 1.3 mm light curve of M87 as observed by SMA shows $\lesssim$50% variability over decade timescales (Bower et al. 2015). Based on these considerations it seems reasonable to adopt a very conservative lower limit on jet power $\equiv P_{\rm jet,min} = 10^{42}$ erg s$^{-1}$.

To apply this constraint we must define and measure $P_{\rm jet}$ in our models. Our procedure is discussed in detail in Appendix A. In brief, we measure the total energy flux in outflowing regions over the polar caps of the black hole in which the energy per unit rest mass exceeds $2.2\,c^2$, which corresponds to $\beta\gamma = 1$, where $\beta \equiv v/c$ and $\gamma$ is Lorentz factor. The effect of changing this cutoff is also discussed in Appendix A. Because the cutoff is somewhat arbitrary, we also calculate $P_{\rm out}$ by including the energy flux in all outflowing regions over the polar caps of the black hole; that is, it includes the energy flux in any wide-angle, low-velocity wind. $P_{\rm out}$ represents a maximal definition of jet power. Table 3 in Appendix A shows $P_{\rm jet}$ as well as a total outflow power $P_{\rm out}$.

The constraint $P_{\rm jet} > P_{\rm jet,min} = 10^{42}$ erg s$^{-1}$ rejects all $a_* = 0$ models. This conclusion is not sensitive to the definition of $P_{\rm jet}$: all $a_* = 0$ models also have total outflow power $P_{\rm out} < 10^{42}$ erg s$^{-1}$. The most powerful $a_* = 0$ model is a MAD model with $R_{\rm high} = 160$, which has $P_{\rm out} = 3.7 \times 10^{41}$ erg s$^{-1}$ and $P_{\rm jet}$ consistent with 0. We conclude that our $a_* = 0$ models are ruled out.

Can the $a_* = 0$ models be saved by changing the eDF? Probably not. There is no evidence from the GRMHD simulations that these models are capable of producing a relativistic outflow with $\beta\gamma > 1$. Suppose, however, that we are willing to identify the nonrelativistic outflow, whose power is measured by $P_{\rm out}$, with the jet. Can $P_{\rm out}$ be raised to meet our conservative threshold on jet power? Here the answer is yes, in principle, and this can be done by changing the eDF. The eDF and $P_{\rm out}$ are coupled because $P_{\rm out}$ is determined by $\dot{M}$, and $\dot{M}$ is adjusted to produce the observed compact mm flux. The relationship between $\dot{M}$ and mm flux depends upon the eDF. If the eDF is altered to produce mm photons less efficiently (for example, by lowering $T_e$ in a thermal model), then $\dot{M}$ and therefore $P_{\rm out}$ increase. A typical nonthermal eDF, by contrast, is likely to produce mm photons with greater efficiency by shifting electrons out of the thermal core and into a nonthermal tail. It will therefore lower $\dot{M}$ and thus $P_{\rm out}$. A thermal eDF with lower $T_e$ could have higher $P_{\rm out}$, as is evident in the large $R_{\rm high}$ SANE models in Table 3. There are observational and theoretical lower limits on $T_e$, however, including a lower limit provided by the observed brightness temepratue. As $T_e$ declines, $n_e$ and $B$ increase and that has implications for source linear polarization (Mościbrodzka et al. 2017; Jiménez-Rosales & Dexter 2018), which will be explored in future work. As $T_e$ declines and $n_e$ and $n_i$ increase there is also an increase in energy transfer from ions to electrons by Coulomb coupling, and this sets a floor on $T_e$.

**Table 1**
Average Image Scoring[a] Summary

| Flux[b] | $a_*$[c] | $\langle p \rangle$[d] | $N_{\rm model}$[e] | MIN($p$)[f] | MAX($p$)[g] |
|---|---|---|---|---|---|
| SANE | −0.94 | 0.33 | 24 | 0.01 | 0.88 |
| SANE | −0.5 | 0.19 | 24 | 0.01 | 0.73 |
| SANE | 0 | 0.23 | 24 | 0.01 | 0.92 |
| SANE | 0.5 | 0.51 | 30 | 0.02 | 0.97 |
| SANE | 0.75 | 0.74 | 6 | 0.48 | 0.98 |
| SANE | 0.88 | 0.65 | 6 | 0.26 | 0.94 |
| SANE | 0.94 | 0.49 | 24 | 0.01 | 0.92 |
| SANE | 0.97 | 0.12 | 6 | 0.06 | 0.40 |
| MAD | −0.94 | 0.01 | 18 | 0.01 | 0.04 |
| MAD | −0.5 | 0.75 | 18 | 0.34 | 0.98 |
| MAD | 0 | 0.22 | 18 | 0.01 | 0.62 |
| MAD | 0.5 | 0.17 | 18 | 0.02 | 0.54 |
| MAD | 0.75 | 0.28 | 18 | 0.01 | 0.72 |
| MAD | 0.94 | 0.21 | 18 | 0.02 | 0.50 |

**Notes.**
[a] The Average Image Scoring (THEMIS-AIS) is introduced in Section 4.
[b] flux: net magnetic flux on the black hole (MAD or SANE).
[c] $a_*$: dimensionless black hole spin.
[d] $\langle p \rangle$: mean of the $p$ value for the aggregated models.
[e] $N_{\rm model}$: number of aggregated models.
[f] MIN($p$): minimum $p$ value among the aggregated models.
[g] MAX($p$): maximum $p$ value among the aggregated models.

The requirement that $P_{\rm jet} > P_{\rm jet,min}$ eliminates many models other than the $a_* = 0$ models. All SANE models with $|a_*| = 0.5$ fail to produce jets with the required minimum power. Indeed, they also fail the less restrictive condition $P_{\rm out} > P_{\rm jet,min}$, so this conclusion is insensitive to the definition of the jet. We conclude that among the SANE models, only high-spin models survive.

At this point it is worth revisiting the SANE, $R_{\rm high} = 1$, $a_* = -0.94$ model that favored a low black hole mass in Section 5. These models are not rejected by a naive application of the $P_{\rm jet} > P_{\rm jet,min}$ criterion, but they are marginal. Notice, however, that we needed to assume a mass in applying the this criterion. We have consistently assumed $M = 6.2 \times 10^9\,M_\odot$. If we use the $M \sim 3 \times 10^9\,M_\odot$ implied by the best-fit $M/D$, then $\dot{M}$ drops by

**Table 2**
Rejection Table

| Flux[a] | $a_*$[b] | $R_{\rm high}$[c] | AIS[d] | $\epsilon$[e] | $L_X$[f] | $P_{\rm jet}$[g] | |
|---|---|---|---|---|---|---|---|
| SANE | −0.94 | 1 | Fail | Pass | Pass | Pass | Fail |
| SANE | −0.94 | 10 | Pass | Pass | Pass | Pass | Pass |
| SANE | −0.94 | 20 | Pass | Pass | Pass | Pass | Pass |
| SANE | −0.94 | 40 | Pass | Pass | Pass | Pass | Pass |
| SANE | −0.94 | 80 | Pass | Pass | Pass | Pass | Pass |
| SANE | −0.94 | 160 | Fail | Pass | Pass | Pass | Fail |
| SANE | −0.5 | 1 | Pass | Pass | Fail | Fail | Fail |
| SANE | −0.5 | 10 | Pass | Pass | Fail | Fail | Fail |
| SANE | −0.5 | 20 | Pass | Pass | Pass | Fail | Fail |
| SANE | −0.5 | 40 | Pass | Pass | Pass | Fail | Fail |
| SANE | −0.5 | 80 | Fail | Pass | Pass | Fail | Fail |
| SANE | −0.5 | 160 | Pass | Pass | Pass | Fail | Fail |
| SANE | 0 | 1 | Pass | Pass | Pass | Fail | Fail |
| SANE | 0 | 10 | Pass | Pass | Pass | Fail | Fail |
| SANE | 0 | 20 | Pass | Pass | Fail | Fail | Fail |
| SANE | 0 | 40 | Pass | Pass | Pass | Fail | Fail |
| SANE | 0 | 80 | Pass | Pass | Pass | Fail | Fail |
| SANE | 0 | 160 | Pass | Pass | Pass | Fail | Fail |
| SANE | +0.5 | 1 | Pass | Pass | Pass | Fail | Fail |
| SANE | +0.5 | 10 | Pass | Pass | Pass | Fail | Fail |
| SANE | +0.5 | 20 | Pass | Pass | Pass | Fail | Fail |
| SANE | +0.5 | 40 | Pass | Pass | Pass | Fail | Fail |
| SANE | +0.5 | 80 | Pass | Pass | Pass | Fail | Fail |
| SANE | +0.5 | 160 | Pass | Pass | Pass | Fail | Fail |
| SANE | +0.94 | 1 | Pass | Fail | Pass | Fail | Fail |
| SANE | +0.94 | 10 | Pass | Fail | Pass | Fail | Fail |
| SANE | +0.94 | 20 | Pass | Pass | Pass | Fail | Fail |
| SANE | +0.94 | 40 | Pass | Pass | Pass | Fail | Fail |
| SANE | +0.94 | 80 | Pass | Pass | Pass | Pass | Pass |
| SANE | +0.94 | 160 | Pass | Pass | Pass | Pass | Pass |
| MAD | −0.94 | 1 | Fail | Fail | Pass | Pass | Fail |
| MAD | −0.94 | 10 | Fail | Pass | Pass | Pass | Fail |
| MAD | −0.94 | 20 | Fail | Pass | Pass | Pass | Fail |
| MAD | −0.94 | 40 | Fail | Pass | Pass | Pass | Fail |
| MAD | −0.94 | 80 | Fail | Pass | Pass | Pass | Fail |
| MAD | −0.94 | 160 | Fail | Pass | Pass | Pass | Fail |
| MAD | −0.5 | 1 | Pass | Fail | Pass | Fail | Fail |
| MAD | −0.5 | 10 | Pass | Pass | Pass | Fail | Fail |
| MAD | −0.5 | 20 | Pass | Pass | Pass | Pass | Pass |
| MAD | −0.5 | 40 | Pass | Pass | Pass | Pass | Pass |
| MAD | −0.5 | 80 | Pass | Pass | Pass | Pass | Pass |
| MAD | −0.5 | 160 | Pass | Pass | Pass | Pass | Pass |
| MAD | 0 | 1 | Pass | Fail | Pass | Fail | Fail |
| MAD | 0 | 10 | Pass | Pass | Pass | Fail | Fail |
| MAD | 0 | 20 | Pass | Pass | Pass | Fail | Fail |
| MAD | 0 | 40 | Pass | Pass | Pass | Fail | Fail |
| MAD | 0 | 80 | Pass | Pass | Pass | Fail | Fail |
| MAD | 0 | 160 | Pass | Pass | Pass | Fail | Fail |
| MAD | +0.5 | 1 | Pass | Fail | Pass | Fail | Fail |
| MAD | +0.5 | 10 | Pass | Pass | Pass | Pass | Pass |
| MAD | +0.5 | 20 | Pass | Pass | Pass | Pass | Pass |
| MAD | +0.5 | 40 | Pass | Pass | Pass | Pass | Pass |
| MAD | +0.5 | 80 | Pass | Pass | Pass | Pass | Pass |
| MAD | +0.5 | 160 | Pass | Pass | Pass | Pass | Pass |
| MAD | +0.94 | 1 | Pass | Fail | Fail | Pass | Fail |
| MAD | +0.94 | 10 | Pass | Fail | Pass | Pass | Fail |
| MAD | +0.94 | 20 | Pass | Pass | Pass | Pass | Pass |
| MAD | +0.94 | 40 | Pass | Pass | Pass | Pass | Pass |
| MAD | +0.94 | 80 | Pass | Pass | Pass | Pass | Pass |
| MAD | +0.94 | 160 | Pass | Pass | Pass | Pass | Pass |

**Notes.**
[a] flux: net magnetic flux on the black hole (MAD, SANE).
[b] $a_*$: dimensionless black hole spin.
[c] $R_{\rm high}$: electron temperature parameter. See Equation (8).
[d] Average Image Scoring (THEMIS-AIS), models are rejected if $\langle p \rangle \leqslant 0.01$. See Section 4 and Table 1.
[e] $\epsilon$: radiative efficiency, models are rejected if $\epsilon$ is larger than the corresponding thin disk efficiency. See Section 6.1.
[f] $L_X$: X-ray luminosity; models are rejected if $\langle L_X \rangle 10^{-2\sigma} > 4.4 \times 10^{40}$ erg s$^{-1}$. See Section 6.2.
[g] $P_{\rm jet}$: jet power, models are rejected if $P_{\rm jet} \leqslant 10^{42}$ erg s$^{-1}$. See Section 6.3.

a factor of two, $P_{\rm jet}$ drops below the threshold and the model is rejected.

The lower limit on jet power $P_{\rm jet,min} = 10^{42}$ erg s$^{-1}$ is conservative and the true jet power is likely higher. If we increased $P_{\rm jet,min}$ to $3 \times 10^{42}$ erg s$^{-1}$, the only surviving models would have $|a_*| = 0.94$ and $R_{\rm high} \geqslant 10$. This conclusion is also not sensitive to the definition of the jet power: applying the same cut to $P_{\rm out}$ adds only a single model with $|a_*| < 0.94$, the $R_{\rm high} = 160$, $a_* = 0.5$ MAD model. The remainder have $a_* = 0.94$. Interestingly, the most powerful jets in our ensemble of models are produced by SANE, $a_* = -0.94$, $R_{\rm high} = 160$ models, with $P_{\rm jet} \simeq 10^{43}$ erg s$^{-1}$.

Estimates for $P_{\rm jet}$ extend to $10^{45}$ erg s$^{-1}$, but in our ensemble of models the maximum $P_{\rm jet} \sim 10^{43}$ erg s$^{-1}$. Possible explanations include: (1) $P_{\rm jet}$ is variable and the estimates probe the central engine power at earlier epochs (discussed above); (2) the $P_{\rm jet}$ estimates are too large; or (3) the models are in error. How might our models be modified to produce a larger $P_{\rm jet}$? For a given magnetic field configuration the jet power scales with $\dot{M}c^2$. To increase $P_{\rm jet}$, then, one must reduce the mm flux per accreted nucleon so that at fixed mm flux density $\dot{M}$ increases.[118] Lowering $T_e$ in a thermal model is unlikely to work because lower $T_e$ implies higher synchrotron optical depth, which increases the ring width. We have done a limited series of experiments that suggest that even a modest decrease in $T_e$ would produce a broad ring that is inconsistent with EHT2017 (Paper VI). What is required, then, is a nonthermal (or multitemperature) model with a large population of cold electrons that are invisible at mm wavelength (for a thermal subpopulation, $\Theta_{e,\rm cold} < 1$), and a population of higher-energy electrons that produces the observed mm flux (see Falcke & Biermann 1995). We have not considered such models here, but we note that they are in tension with current ideas about dissipation of turbulence because they require efficient suppression of electron heating.

The $P_{\rm jet}$ in our models is dominated by Poynting flux in the force-free region around the axis (the "funnel"), as in the Blandford & Znajek (1977) force-free magnetosphere model. The energy flux is concentrated along the walls of the funnel.[119] Tchekhovskoy et al. (2011) provided an expression for the energy flux in the funnel, the so-called Blandford–Znajek power $P_{\rm BZ}$, which becomes, in our units,

$$\begin{aligned} P_{\rm BZ} &= 2.8 f(a_*)\left(\frac{\phi}{15}\right)^2 \dot{M}c^2 \\ &= 2.2 \times 10^{43} f(a_*)\left(\frac{\phi}{15}\right)^2 \\ &\quad \times \left(\frac{\dot{M}}{10^{-6}\dot{M}_{\rm Edd}}\right)\left(\frac{M}{6.2 \times 10^9 M_\odot}\right) \text{erg s}^{-1} \end{aligned} \tag{9}$$

where $f(a_*) \approx a_*^2(1 + \sqrt{1 - a_*^2})^{-2}$ (a good approximation for $a_* < 0.95$) and $\dot{M}_{\rm Edd} = 137\, M_\odot$ yr$^{-1}$ for $M = 6.2 \times 10^9\, M_\odot$. This expression was developed for models with a thin disk in the equatorial plane. $P_{\rm BZ}$ is lower for models where the force-free region is excluded by a thicker disk around the equatorial plane. Clearly $P_{\rm BZ}$ is comparable to observational estimates of $P_{\rm jet}$.

[118] The compact mm flux density could be a factor of 2 larger than our assumed 0.5 Jy. That would raise $P_{\rm jet}$ by slightly less than a factor of 2.

[119] The total energy flux inside a cone of opening angle $\theta_0$ is proportional to $\sin^4\theta_0$ in the Blandford & Znajek (1977) monopole model if the field strength is fixed, and $\sin^2\theta_0$ if the magnetic flux is fixed.

In our models (see Table 3) $P_{\rm jet}$ follows the above scaling relation but with a smaller coefficient. The ratio of coefficients is model dependent and varies from 0.15 to 0.83. This is likely because the force-free region is restricted to a cone around the poles of the black hole, and the width of the cone varies by model. Indeed, the coefficient is larger for MAD than for SANE models, which is consistent with this idea because MAD models have a wide funnel and SANE models have a narrow funnel. This also suggests that future comparison of synthetic 43 and 86 GHz images from our models with lower-frequency VLBI data may further constrain the magnetic flux on the black hole.

The connection between the Poynting flux in the funnel and black hole spin has been discussed for some time in the simulation literature, beginning with McKinney & Gammie (2004; see also McKinney 2006; McKinney & Narayan 2007). The structure of the funnel magnetic field can be time-averaged and shown to match the analytic solution of Blandford & Znajek (1977). Furthermore, the energy flux density can be time-averaged and traced back to the event horizon. Is the energy contained in black hole spin sufficient to drive the observed jet over the jet lifetime? The spindown timescale is $\tau = (M - M_{\rm irr})c^2/P_{\rm jet}$, where $M_{\rm irr} \equiv M((1 + \sqrt{1 - a_*^2})/2)^{1/2}$ is the irreducible mass of the black hole. For the $a_* = 0.94$ MAD model with $R_{\rm high} = 160$, $\tau = 7.3 \times 10^{12}$ yr, which is long compared to a Hubble time ($\sim 10^{10}$ yr). Indeed, the spindown time for all models is long compared to the Hubble time.

We conclude that for models that have sufficiently powerful jets and are consistent with EHT2017, $P_{\rm jet}$ is driven by extraction of black hole spin energy through the Blandford–Znajek process.

### 6.4. Constraint Summary

We have applied constraints from AIS, a radiative self-consistency constraint, a constraint on maximum X-ray luminosity, and a constraint on minimum jet power. Which models survive? Here we consider only models for which we have calculated $L_X$ and $L_{\rm bol}$. Table 2 summarizes the results. Here we consider only $i = 163°$ (for $a_* \geqslant 0$) and $i = 17°$ (for $a_* < 0$). The first three columns give the model parameters. The next four columns show the result of application of each constraint: THEMIS-AIS (here broken out by individual model rather than groups of models), radiative efficiency ($\epsilon < \epsilon_{\rm thin\ \ disk}$), $L_X$, and $P_{\rm jet}$.

The final column gives the logical AND of the previous four columns, and allows a model to pass only if it passes all tests. Evidently most of the SANE models fail, with the exception of some $a_* = -0.94$ models and a few $a_* = 0.94$ models with large $R_{\rm high}$. A much larger fraction of the MAD models pass, although $a_* = 0$ models all fail because of inadequate jet power. MAD models with small $R_{\rm high}$ also fail. It is the jet power constraint that rejects the largest number of models.

## 7. Discussion

We have interpreted the EHT2017 data using a limited library of models with attendant limitations. Many of the limitations stem from the GRMHD model, which treats the plasma as an ideal fluid governed by equations that encode conservation laws for particle number, momentum, and energy. The eDF, in particular, is described by a number density and temperature, rather than a full distribution function, and the electron temperature $T_e$ is assumed to be a function of the local ion temperature and plasma $\beta_{\rm p}$. Furthermore, all models assume a Kerr black hole spacetime, but there are alternatives. Here we consider some of the model limitations and possible extensions, including to models beyond general relativity.

### 7.1. Radiative Effects

Post-processed GRMHD simulations that are consistent with EHT data and the flux density of 1.3 mm emission in M87 can yield unphysically large radiative efficiencies (see Section 6). This implies that the radiative cooling timescale is comparable to or less than the advection timescale. As a consequence, including radiative cooling in simulations may be necessary to recover self-consistent models (see Mościbrodzka et al. 2011; Dibi et al. 2012). In our models we use a single parameter, $R_{\rm high}$, to adjust $T_e$ and account for all effects that might influence the electron energy density. How good is this approximation?

The importance of radiative cooling can be assessed using newly developed, state-of-the-art general relativistic radiation GRMHD ("radiation GRMHD") codes. Sądowski et al. (2013b; see also Sądowski et al. 2014, 2017; McKinney et al. 2014) applied the M1 closure (Levermore 1984), which treats the radiation as a relativistic fluid. Ryan et al. (2015) introduced a Monte Carlo radiation GRMHD method, allowing for full frequency-dependent radiation transport. Models for turbulent dissipation into the electrons and ions, as well as heating and cooling physics that sets the temperature ratio $T_i/T_e$, have been added to GRMHD and radiative GRMHD codes and used in simulations of Sgr A* (Ressler et al. 2015, 2017; Chael et al. 2018) and M87 (Ryan et al. 2018; Chael et al. 2019). While the radiative cooling and Coulomb coupling physics in these simulations is well understood, the particle heating process, especially the relative heating rates of ions and electrons, remains uncertain.

Radiation GRMHD models are computationally expensive per run and do not have the same scaling freedom as the GRMHD models, so they need to be repeatedly re-run with different initial conditions until they produce the correct 1.3 mm flux density. It is therefore impractical to survey the parameter space using radiation GRMHD. It is possible, however, to check individual GRMHD models against existing radiation GRMHD models of M87 (Ryan et al. 2018; Chael et al. 2019).

The SANE radiation GRMHD models of Ryan et al. (2018) with $a_* = 0.94$ and $M = 6 \times 10^9\,M_\odot$ can be compared to GRMHD SANE $a_* = 0.94$ models at various values of $R_{\rm high}$. The radiative models have $\dot{M}/\dot{M}_{\rm Edd} = 5.2 \times 10^{-6}$ and $P_{\rm jet} = 5.1 \times 10^{41}$ erg s$^{-1}$. The GRMHD models in this work have, for $1 \leqslant R_{\rm high} \leqslant 160$, $0.36 \times 10^{-6} \leqslant \dot{M}/\dot{M}_{\rm Edd} \leqslant 20 \times 10^{-6}$, and $0.22 \leqslant P_{\rm jet}/(10^{41}$ erg s$^{-1}) \leqslant 12$ (Table 3). Evidently the mass accretion rates and jet powers in the GRMHD models span a wide range that depends on $R_{\rm high}$, but when we choose $R_{\rm high} = 10 - 20$ they are similar to what is found in the radiative GRMHD model when using the turbulent electron heating model (Howes 2010).

We have also directly compared the $T_e$ distribution in the emitting region, and the radiation GRMHD model is quite close to the $R_{\rm high} = 10$ model. The resulting images are qualitatively similar, with an asymmetric photon ring that is brighter in the south and a weak inner ring associated with the funnel wall emission as in Figure 2. The radiation GRMHD SANE model, like all our nonradiative GRMHD SANE models (except the $R_{\rm high} = 160$ model), would be ruled out by the condition $P_{\rm jet} > 10^{42}$ erg s$^{-1}$.

The MAD radiation GRMHD models of Chael et al. (2019) with $a_* = 0.94$ and $M = 6.2 \times 10^9 M_\odot$ can be compared to GRMHD MAD $a_* = 0.94$ models at various values of $R_{\rm high}$. Chael et al. (2019) uses two dissipation models: the Howes (2010, hereafter H10) model of heating from a Landau-damped turbulent cascade, and the Rowan et al. (2017, hereafter R17) model of heating based on simulations of transrelativistic magnetic reconnection. The (H10, R17) models have $\dot{M}/\dot{M}_{\rm Edd} = (3.6, 2.3) \times 10^{-6}$ and $P_{\rm jet} = (6.6, 13) \times 10^{42}$ erg s$^{-1}$. The GRMHD models have, for $1 \leqslant R_{\rm high} \leqslant 160$, $0.13 \times 10^{-6} \leqslant \dot{M}/\dot{M}_{\rm Edd} \leqslant 1.4 \times 10^{-6}$ and $2.3 \leqslant P_{\rm jet}/(10^{42}$ erg s$^{-1}) \leqslant 8.8$ (Table 3). In the radiation GRMHD MAD models $\dot{M}$ lies in the middle of the range spanned by the nonradiative GRMHD models, and jet power lies at the upper end of the range spanned by the nonradiative GRMHD models. The $T_e$ distributions in the radiative and nonradiative MAD models differ: the mode of the radiation GRMHD model $T_e$ distribution is about a factor of 3 below the mode of the $T_e$ distribution in the $R_{\rm high} = 20$ GRMHD model, and the GRMHD model has many more zones at $\Theta_e \sim 100$ that contribute to the final image than the radiation GRMHD models. This difference is a consequence of the $R_{\rm high}$ model for $T_e$: in MAD models almost all the emission emerges at $\beta_{\rm p} \lesssim 1$, so $R_{\rm high}$, which changes $T_e$ in the $\beta_{\rm p} > 1$ region, offers little control over $T_e$ in the emission region. Nevertheless, the jet power and accretion rates are similar in the radiative and nonradiative MAD models, and the time-averaged radiative and nonradiative images are qualitatively indistinguishable. This suggests that the image is determined mainly by the spacetime geometry and is insensitive to the details of the plasma evolution.

This review of radiative effects is encouraging but incomplete: it only considers a limited selection of models and a narrow set of observational constraints. Future studies of time dependence and polarization are likely to sharpen the contrast between radiative and nonradiative models.

### 7.2. Nonthermal Electrons

Throughout this Letter we have considered only a thermal eDF. While a thermal eDF can account for the observed emission at mm wavelengths in M87 (e.g., Mościbrodzka et al. 2016; Prieto et al. 2016; Ryan et al. 2018; Chael et al. 2019), eDFs that include a nonthermal tail can also explain the observed SED (Broderick & Loeb 2009; Yu et al. 2011; Dexter et al. 2012; Li et al. 2016; Davelaar et al. 2018; J. Davelaar et al. 2019, in preparation).

The role of nonthermal electrons (and positrons) in producing the observed compact emission is not a settled question, and cannot be settled in this first investigation of EHT2017 models, but there are constraints. The number density, mean velocity, and energy density of the eDF are fixed or limited by the GRMHD models. In addition, the eDF cannot on average sustain features that would be erased by kinetic instabilities on timescales short compared to $r_g c^{-1}$. Some nonthermal eDFs increase $F_\nu/\dot{M}$ in comparison to a thermal eDF, implying lower values of $\dot{M}$ than quoted above (Ball et al. 2018; Davelaar et al. 2018; J. Davelaar et al. 2019, in preparation). These lower values of $\dot{M}$ can slightly change the source morphology, e.g., by decreasing the visibility of the approaching jet (e.g., Dexter et al. 2012).

One can evaluate the influence of nonthermal eDFs in several ways. For example, it is possible to study simplified, phenomenological models. Emission features due to the cooling of nonthermal electrons may then reveal how and where the nonthermal electrons are produced (Pu et al. 2017). Emission features created by the injection of nonthermal electrons within GRMHD models of the jet and their subsequent cooling will be studied separately (T. Kawashima et al. 2019, in preparation). The effect of nonthermal eDFs can also be studied by post-processing of ideal GRMHD models if one assumes that the electrons have a fixed, parameterized form such as a power-law distribution (Dexter et al. 2012) or a $\kappa$-distribution (Davelaar et al. 2018; J. Davelaar et al. 2019, in preparation). These parameterized models produce SEDs that agree with radio to near-infrared data, but they are approximations to the underlying physics and do not resolve the microscopic processes that accelerate particles. One can also include dissipative processes explicitly in the GRMHD models, including scalar resistivity (Palenzuela et al. 2009; Dionysopoulou et al. 2013; Del Zanna et al. 2016; Qian et al. 2017; Ripperda et al. 2019), heat fluxes and viscosities (pressure anisotropies; Chandra et al. 2015; Ressler et al. 2015; Foucart et al. 2017), and particle acceleration (e.g., Chael et al. 2017). Ultimately special and general relativistic particle-in-cell codes (Watson & Nishikawa 2010; Chen et al. 2018; Levinson & Cerutti 2018; Parfrey et al. 2019) will enable direct investigations of kinetic processes.

### 7.3. Other Models and Analysis Limitations

We have used a number of other approximations in generating our models. Among the most serious ones are as follows.

(1) Fast Light Approximation. A GRMHD simulation produces a set of dump files containing the model state at a single global (Kerr–Schild) coordinate time. Because the dynamical time is only slightly longer than the light-crossing time, in principle one needs to trace rays through a range of coordinate times, i.e., by interpolation between multiple closely spaced dump files. In practice this is difficult because a high cadence of output files is required, limiting the speed of the GRMHD simulations and requiring prohibitively large data storage. In addition, the cost of ray tracing through multiple output files is high. Because of this, we adopt the commonly used fast light approximation in which GRMHD variables are read from a single dump file and held steady during the ray tracing. Including light-travel time delays produces minor changes to the small-scale image structure and to light curves (e.g., Dexter et al. 2010; Bronzwaer et al. 2018; Z. Younsi et al. 2019b, in preparation), although it is essential for the study of variability on the light-crossing timescale.

(2) Untilted Disks. We have assumed that the disk angular momentum vector and black hole spin vector are (anti-) aligned. There is no reason for the angular momentum vector of the accretion flow on large scales to align with the black hole spin vector, and there is abundant evidence for misaligned disks in AGNs (e.g., Miyoshi et al. 1995). How might disk tilt affect our results?

Tilting the disk by as little as $\sim 15^\circ$ is enough to set up a standing, two-armed spiral shock close to the ISCO (Fragile & Blaes 2008). This shock directly affects the morphology of mm wavelength images, especially at low inclination, in models of Sgr A* (Dexter & Fragile 2013, especially Figure 5), producing an obvious two-armed spiral pattern on the sky. If this structure were also present in images of tilted models of M87, then it is possible that even a modest tilt could be ruled out.

If a modest tilt is present in M87 it is unlikely to affect our conclusion regarding the sign of black hole spin. That conclusion depends on emission from funnel wall plasma in counter-rotating ($a_* < 0$) disks. The funnel wall plasma is loaded onto funnel plasma field lines by local instabilities at the wall and then rotates with the funnel and therefore the black hole (Wong et al. 2019). The funnel wall is already unsteady, fluctuating by tens of degrees in azimuth and in time, so a modest tilt seems unlikely to dramatically alter the funnel wall structure.

Is there observational evidence for tilt in M87? In numerical studies of tilted disks the jet emerges perpendicular to the disk (Liska et al. 2018), and tilted disks are expected to precess. One might then expect that a tilted source would produce a jet that exhibits periodic variations, or periodic changes in jet direction with distance from the source, as seen in other sources. There is little evidence of this in M87 (see Park et al. 2019 for a discussion of possible misaligned structure in the jet). Indeed, Walker et al. (2018) saw at most small displacements of the jet with time and distance from the source at mas scales. In sum, there is therefore little observational motivation for considering tilted disk models.

Tilted disk models of M87 are an interesting area for future study. It is possible that the inner disk may align with the black hole via a thick-disk variant of the Bardeen & Petterson (1975) effect. Existing tilted thick-disk GRMHD simulations (e.g., Fragile et al. 2007; McKinney et al. 2013; Shiokawa 2013; Liska et al. 2018) show some evidence for alignment and precession (McKinney et al. 2013; Shiokawa 2013; Liska et al. 2018), but understanding of the precession and alignment timescales is incomplete. It will be challenging to extend the Image Library to include a survey of tilted disk models, however, because with tilted disks there are two new parameters: the two angles that describe the orientation of the outer disk with respect to the black hole spin vector and the line of sight.

(3) Pair Production. In some models of M87 the mm emission is dominated by electron-positron pairs within the funnel, even close to the horizon scale (see Beskin et al. 1992; Levinson & Rieger 2011; Mościbrodzka et al. 2011; Broderick & Tchekhovskoy 2015; Hirotani & Pu 2016). The pairs are produced from the background radiation field or from a pair-cascade process following particle acceleration by unscreened electric fields, which we cannot evaluate using ideal GRMHD models. We leave it to future work to assess whether or not these models can plausibly suppress emission from the disk and funnel wall, and simultaneously produce a sufficiently powerful jet.

(4) Numerical Treatment of Low-density Regions. Virtually all MHD simulations, including ours, use a “floor” procedure that resets the density if it falls below a minimum value. If this is not done, then truncation error accumulates dramatically in the low-density regions and the solution is corrupted. If the volume where floors are activated contains only a small fraction of the simulation mass, momentum, and energy, then most aspects of the solution are unaffected by this procedure (e.g., McKinney & Gammie 2004).

In regions where the floors are activated the temperature of the plasma is no longer reliable. This is why we cut off emission from regions with $B^2/\rho > 1$, where floors are commonly activated. In models where floors are only activated in the funnel (e.g., most SANE models), the resulting images are insensitive to the choice of cutoff $B^2/\rho$. In MAD models the regions of low and high density are mixed because lightly loaded magnetic field lines that are trapped in the hole bubbles outward through the disk. In this case emission at $\nu > 230$ GHz can be sensitive to the choice of cutoff $B^2/\rho$ Chael et al. (2019). The sense of the effect is that greater cutoff $B^2/\rho$ implies more emission at high frequency. Our use of a cutoff $B^2/\rho = 1$ is therefore likely to underestimate mm emission and therefore overestimate $\dot{M}$ and $P_{\rm jet}$. Accurate treatment of the dynamics and thermodynamics of low-density regions and especially sharp boundaries between low- and high-density regions is a fundamental numerical problem in black hole accretion flow modeling that merits further attention.

### *7.4. Alternatives to Kerr Black Holes*

Although our working hypothesis has been that M87 contains a Kerr black hole, it is interesting to consider whether or not the data is also consistent with alternative models for the central object. These alternatives can be grouped into three main categories: (i) black holes within general relativity that include additional fields; (ii) black hole solutions from alternative theories of gravity or incorporating quantum effects; (iii) black hole “mimickers,” i.e., compact objects, both within general relativity or in alternative theories, whose properties could be fine-tuned to resemble those of black holes.

The first category includes, for example, black holes in Einstein–Maxwell–dilaton-axion gravity (e.g., García et al. 1995; Mizuno et al. 2018), black holes with electromagnetic or Newman–Unti-Tamburino (NUT) charges (e.g., Grenzebach et al. 2014), regular black holes in nonlinear electrodynamics (e.g., Abdujabbarov et al. 2016), black hole metrics affected by a cosmological constant (e.g., Dymnikova 1992) or a dark matter halo (e.g., Hou et al. 2018), and black holes with scalar wigs (e.g., Barranco et al. 2017) or hair (e.g., Herdeiro & Radu 2014). While the shadows of this class of compact objects are expected to be similar to Kerr and therefore cannot be ruled out immediately by current observations (Mizuno et al. 2018), the most extreme examples of black holes surrounded by massive scalar field configurations should produce additional lobes in the shadow or disconnected dark regions (Cunha et al. 2015). As these features are not found in the EHT2017 image, these alternatives are not viable models for M87.

The second category comprises black hole solutions with classical modifications to general relativity, as well as effects coming from approaches to quantum gravity (see, e.g., Moffat 2015; Dastan et al. 2016; Younsi et al. 2016; Amir et al. 2018; Eiroa & Sendra 2018; Giddings & Psaltis 2018). These alternatives have shadows that are qualitatively very similar to those of Kerr black holes and are not distinguishable with present EHT capabilities. However, higher-frequency observations, together with the degree of polarization of the emitted radiation or the variability of the accretion flow, can be used to assess their viability.

Finally, the third category comprises compact objects such as spherically symmetric naked singularities (e.g., Joshi et al. 2014), superspinars (Kerr with $|a_*| > 1$, which are axisymmetric spacetime with naked singularities), and regular horizonless objects, either with or without a surface. Examples of regular surfaceless objects are: boson stars (Kaup 1968), traversable wormholes, and clumps of self-interacting dark matter (Saxton et al. 2016), while examples of black hole mimickers with a surface are gravastars (Mazur & Mottola 2004) and collapsed polymers (Brustein & Medved 2017), to cite only a few. Because the exotic genesis of these black hole mimickers is essentially unknown, their physical properties are essentially unconstrained, thus making the distinction from black holes rather challenging (see, however, Chirenti & Rezzolla 2007, 2016). Nevertheless, some conclusions can drawn already. For instance, the shadow of a superspinar is very different from that of a black hole (Bambi & Freese 2009), and the EHT2017 observations rule out any superspinar model for M87. Similarly, for certain parameter ranges, the shadows of spherically symmetric naked singularities have been found to consist of a filled disk with no dark region[120] in the center (Shaikh et al. 2019); clearly, this class of models is ruled out. In the same vein, because the shadows of wormholes can exhibit large deviations from those of black holes (see, e.g., Bambi 2013; Nedkova et al. 2013; Shaikh 2018), a large portion of the corresponding space of parameters can be constrained with the present observations.

A comparison of EHT2017 data with the boson star model, as a representative horizonless and surfaceless black hole mimicker, and a gravastar model as a representative horizonless black hole mimicker, will be presented in Olivares et al. (2019a). Both models produce images with ring-like features similar to those observed by EHT2017, which are consistent with the results of Broderick & Narayan (2006), who also consider black hole alternatives with a surface. The boson star generically requires masses that are substantially different from that expected for M87 (H. Olivares et al. 2019b, in preparation), while the gravastar has accretion variability that is considerably different from that onto a black hole.

In summary, because each of the many exotic alternatives to Kerr black holes can span an enormous space of parameters that is only poorly constrained, the comparisons carried out here must be considered preliminary. Nevertheless, they show that the EHT2017 observations are not consistent with several of the alternatives to Kerr black holes, and that some of those models that produce similar images show rather different dynamics in the accretion flow and in its variability. Future observations and more detailed theoretical modeling, combined with multiwavelength campaigns and polarimetric measurements, will further constrain alternatives to Kerr black holes.

## 8. Conclusion

In this Letter we have made a first attempt at understanding the physical implications of a single, high-quality EHT data set for M87. We have compared the data to a library of mock images produced from GRMHD simulations by GRRT calculations. The library covers a parameter space that is substantially larger than earlier model surveys. The results of this comparison are consistent with the hypothesis that the compact 1.3 mm emission in M87 arises within a few $r_g$ of a Kerr black hole, and that the ring-like structure of the image is generated by strong gravitational lensing and Doppler beaming. The models predict that the asymmetry of the image depends on the sense of black hole spin. If this interpretation is accurate, then the spin vector of the black hole in M87 points away from Earth (the black hole spins clockwise on the sky). The models also predict that there is a strong energy flux directed away from the poles of the black hole, and that this energy flux is electromagnetically dominated. If the models are correct, then the central engine for the M87 jet is powered by the electromagnetic extraction of free energy associated with black hole spin via the Blandford–Znajek process.

In our models, M87's compact mm emission is generated by the synchrotron mechanism. Our ability to make physical inferences based on the models is therefore intimately tied to the quality of our understanding of the eDF. We have used a thermal model with a single free parameter that adjusts the ratio of ion to electron temperature in regions with plasma $\beta_p > 1$ (i.e., regions where magnetic pressure is less than gas pressure). This simple model does not span the range of possible plasma behavior. The theory of high temperature, collisionless plasmas must be better understood if this core physical uncertainty of sub-Eddington black hole accretion is to be eliminated. At present our understanding is inadequate, and alternative eDF models occupy a large, difficult-to-explore parameter space with the potential to surprise. Despite these uncertainties, many of the models produce images with similar morphology that is consistent with EHT2017 data. This suggests that the image shape is controlled mainly by gravitational lensing and the spacetime geometry, rather than details of the plasma physics.

Although the EHT2017 images are consistent with the vast majority of our models, parts of the parameter space can be rejected on physical grounds or by comparison with contemporaneous data at other wavelengths. We reject some models because, even though all models are variable, some models are *too variable* to be consistent with the data. We can also reject models based on a radiative efficiency cut (the models are not self-consistent and would cool quickly if radiative effects were included), an X-ray luminosity cut using contemporaneous *Chandra* and *NuSTAR* data, and on a jet-power cut. The requirement that the jet power exceed a conservative lower limit of $10^{42}$ erg s$^{-1}$ turns out to eliminate many models, including all models with $a_* = 0$.

We have examined the astrophysical implications of only a subset of EHT2017 data; much remains to be done, and there are significant opportunities for further constraining the models. EHT2017 data includes tracks from four separate days of observing; each day is 2.8 $r_g c^{-1}$ (see Paper IV). This timescale is short compared to the decorrelation timescale of simulated images, which is $\sim$50 $r_g c^{-1}$, and smaller than the light-crossing time of the source plasma. Analysis techniques that use short-timescale variations in the data will need to be developed and are likely to recover new, more stringent constraints on the model from the EHT2017 data set. EHT2017 took polarized data as well. Our simulations already predict full polarization maps, albeit for our simple eDF model. Comparison of model polarization maps of the source with EHT2017 data are likely to sharply limit the space of allowed models (Mościbrodzka et al. 2017). Finally, in this Letter the only multiwavelength companion data that we consider are X-ray observations. Simultaneous data are available at many other wavelengths, from the radio to the gamma-rays, and is likely to further limit the range of acceptable models and guide the implementation of predictive electron physics models.

[120] The width of the ring, the central flux depression, and a quantitative discussion of the black hole shadow can be found in Paper VI.

In this Letter we have focused on the time-dependent ideal GRMHD model. Physically motivated, semi-analytic models including nonthermal emission have not been applied yet and will be discussed in future papers (A. E. Broderick et al. 2019b, in preparation; T. Kawashima et al. 2019, in preparation; H.-Y. Pu et al. 2019, in preparation).

We have also not yet considered how the physical properties of the jet are constrained by lower-frequency VLBI observations, which constrain jet kinematics (Mertens et al. 2016; Britzen et al. 2017; Hada et al. 2017; Kim et al. 2018; Walker et al. 2018), the jet width profile (Asada & Nakamura 2012; Hada et al. 2013; Nakamura et al. 2018), the total jet power at kilo-parsec scale (Owen et al. 2000; Stawarz et al. 2006), the jet power (e.g., Kino et al. 2014, 2015), the core shift (Hada et al. 2011), and the symmetric limb-brightening structure (Takahashi et al. 2018; Kim et al. 2018). The jet width profile is potentially very interesting because it depends on the magnetic flux $\phi$: the jet internal magnetic pressure $\propto\phi^2$. We therefore expect (and see in our numerical simulations; see Figure 4) that MAD jets are wider at the base than SANE jets. Future theoretical work will help connect the ring-like structure seen in EHT2017 to the large-scale jet (M. Nakamura et al. 2019, in preparation).

A second epoch of observations ($\gtrsim 50\ r_{\rm g} c^{-1} \sim 2$ weeks after EHT2017, when the models suggest that source structure will decorrelate) will increase the power of the average image analysis to reject models. The EHT2017 data were able to reject one entire category of models with confidence: high magnetic flux (MAD), retrograde, high-spin models. Other categories of models, such as the low magnetic flux, high-spin models, are assigned comparatively low probabilities by the average image scoring scheme. Data taken later, more than a decorrelation time after EHT2017 (model decorrelation times are of order two weeks), will provide an independent realization of the source. The probabilities attached to individual models by average image scoring will then multiply. For example, a model with probability 0.05 that is assigned probability 0.05 in comparison to a second epoch of observation would then have probability $0.05^2 = 2.5 \times 10^{-3}$, and would be strongly disfavored by the average image scoring criterion (see Section 4).

Future EHT 345 GHz campaigns (Paper II) will provide excellent constraints, particularly on the width of the ring. The optical depth on every line of sight through the source is expected to decrease (the drop is model and location dependent). In our models this makes the ring narrower, better defined, easier to measure accurately from VLBI data, and less dependent on details of the source plasma model.

Certain features of the model are geometric and should be present in future EHT observations. The photon ring is a persistent feature of the model related to the mass and distance to the black hole. It should be present in the next EHT campaign unless there is a dramatic change in $\dot{M}$, which would be evident in the SED. The asymmetry in the photon ring is also a persistent feature of the model because, we have argued, it is controlled by the black hole spin. The asymmetry should therefore remain in the southern half of the ring for the next EHT campaign, unless there is a dramatic tilt of the inner accretion flow. If the small-scale and large-scale jet are aligned, then EHT2017 saw the brightest region at unusually small PA, and future campaigns are likely (but not certain) to see the peak brightness shift further to the west. Future 230 GHz EHT campaigns (Paper II) will thus sharply test the GRMHD source models.

Together with complementary studies that are presently targeting either the supermassive black hole candidate at the Galactic Center (Eckart & Genzel 1997; Ghez et al. 1998; Gravity Collaboration et al. 2018a, 2018b) or stellar-mass binary black holes whose gravitational-wave emission is recorded by the LIGO and Virgo detectors (Abbott et al. 2016), the results provided here are consistent with the existence of astrophysical black holes. More importantly, they clearly indicate that their phenomenology, despite being observed on mass scales that differ by eight orders of magnitude, follows very closely the one predicted by general relativity. This demonstrates the complementarity of experiments studying black holes on all scales, promising much imp roved tests of gravity in its most extreme regimes.

The authors of this Letter thank the following organizations and programs: the Academy of Finland (projects 274477, 284495, 312496); the Advanced European Network of E-infrastructures for Astronomy with the SKA (AENEAS) project, supported by the European Commission Framework Programme Horizon 2020 Research and Innovation action under grant agreement 731016; the Alexander von Humboldt Stiftung; the Black Hole Initiative at Harvard University, through a grant (60477) from the John Templeton Foundation; the China Scholarship Council; Comisión Nacional de Investigación Científica y Tecnológica (CONICYT, Chile, via PIA ACT172033, Fondecyt 1171506, BASAL AFB-170002, ALMA-conicyt 31140007); Consejo Nacional de Ciencia y Tecnológica (CONACYT, Mexico, projects 104497, 275201, 279006, 281692); the Delaney Family via the Delaney Family John A. Wheeler Chair at Perimeter Institute; Dirección General de Asuntos del Personal Académico—Universidad Nacional Autónoma de México (DGAPA—UNAM, project IN112417); the European Research Council Synergy Grant "BlackHoleCam: Imaging the Event Horizon of Black Holes" (grant 610058); the Generalitat Valenciana postdoctoral grant APOSTD/2018/177; the Gordon and Betty Moore Foundation (grants GBMF-3561, GBMF-5278); the Istituto Nazionale di Fisica Nucleare (INFN) sezione di Napoli, iniziative specifiche TEONGRAV; the International Max Planck Research School for Astronomy and Astrophysics at the Universities of Bonn and Cologne; the Jansky Fellowship program of the National Radio Astronomy Observatory (NRAO); the Japanese Government (Monbukagakusho: MEXT) Scholarship; the Japan Society for the Promotion of Science (JSPS) Grant-in-Aid for JSPS Research Fellowship (JP17J08829); JSPS Overseas Research Fellowships; the Key Research Program of Frontier Sciences, Chinese Academy of Sciences (CAS, grants QYZDJ-SSW-SLH057, QYZDJ-SSW-SYS008); the Leverhulme Trust Early Career Research Fellowship; the Max-Planck-Gesellschaft (MPG); the Max Planck Partner Group of the MPG and the CAS; the MEXT/JSPS KAKENHI (grants 18KK0090, JP18K13594, JP18K03656, JP18H03721, 18K03709, 18H01245, 25120007); the MIT International Science and Technology Initiatives (MISTI) Funds; the Ministry of Science and Technology (MOST) of Taiwan (105-2112-M-001-025-MY3, 106-2112-M-001-011, 106-2119-M-001-027, 107-2119-M-001-017, 107-2119-M-001-020, and 107-2119-M-110-005);

the National Aeronautics and Space Administration (NASA, Fermi Guest Investigator grant 80NSSC17K0649); the National Institute of Natural Sciences (NINS) of Japan; the National Key Research and Development Program of China (grant 2016YFA0400704, 2016YFA0400702); the National Science Foundation (NSF, grants AST-0096454, AST-0352953, AST-0521233, AST-0705062, AST-0905844, AST-0922984, AST-1126433, AST-1140030, DGE-1144085, AST-1207704, AST-1207730, AST-1207752, MRI-1228509, OPP-1248097, AST-1310896, AST-1312651, AST-1337663, AST-1440254, AST-1555365, AST-1715061, AST-1614868, AST-1615796, AST-1716327, OISE-1743747, AST-1816420); the Natural Science Foundation of China (grants 11573051, 11633006, 11650110427, 10625314, 11721303, 11725312, 11873028, 11873073, U1531245, 11473010); the Natural Sciences and Engineering Research Council of Canada (NSERC, including a Discovery Grant and the NSERC Alexander Graham Bell Canada Graduate Scholarships-Doctoral Program); the National Youth Thousand Talents Program of China; the National Research Foundation of Korea (the Global PhD Fellowship Grant: grants NRF-2015H1A2A1033752, 2015-R1D1A1A01056807, the Korea Research Fellowship Program: NRF-2015H1D3A1066561); the Netherlands Organization for Scientific Research (NWO) VICI award (grant 639.043.513) and Spinoza Prize SPI 78-409; the New Scientific Frontiers with Precision Radio Interferometry Fellowship awarded by the South African Radio Astronomy Observatory (SARAO), which is a facility of the National Research Foundation (NRF), an agency of the Department of Science and Technology (DST) of South Africa; the Onsala Space Observatory (OSO) national infrastructure, for the provisioning of its facilities/observational support (OSO receives funding through the Swedish Research Council under grant 2017-00648) the Perimeter Institute for Theoretical Physics (research at Perimeter Institute is supported by the Government of Canada through the Department of Innovation, Science and Economic Development Canada and by the Province of Ontario through the Ministry of Economic Development, Job Creation and Trade); the Russian Science Foundation (grant 17-12-01029); the Spanish Ministerio de Economía y Competitividad (grants AYA2015-63939-C2-1-P, AYA2016-80889-P); the State Agency for Research of the Spanish MCIU through the "Center of Excellence Severo Ochoa" award for the Instituto de Astrofísica de Andalucía (SEV-2017-0709); the Toray Science Foundation; the US Department of Energy (USDOE) through the Los Alamos National Laboratory (operated by Triad National Security, LLC, for the National Nuclear Security Administration of the USDOE (Contract 89233218CNA000001)); the Italian Ministero dell'Istruzione Università e Ricerca through the grant Progetti Premiali 2012-iALMA (CUP C52I13000140001); ALMA North America Development Fund; Chandra TM6-17006X and DD7-18089X; a Sprows Family VURF Fellowship; a NSERC Discovery Grant; the FRQNT Nouveaux Chercheurs program; CIFAR; the NINS program of Promoting Research by Networking among Institutions(Grant Number 01421701); MEXT as a priority issue (Elucidation of the fundamental laws and evolution of the universe) to be tackled by using post-K Computer and JICFuS; part of this work used XC50 at the Center for Computational Astrophysics, National Astronomical Observatory of Japan.This work used the Extreme Science and Engineering Discovery Environment (XSEDE), supported by NSF grant ACI-1548562, and CyVerse, supported by NSF grants DBI-0735191, DBI-1265383, and DBI-1743442. XSEDE Stampede2 resource at TACC was allocated through TG-AST170024 and TG-AST080026N. XSEDE JetStream resource at PTI and TACC was allocated through AST170028. The simulations were performed in part on the SuperMUC cluster at the LRZ in Garching, on the LOEWE cluster in CSC in Frankfurt, and on the HazelHen cluster at the HLRS in Stuttgart. This research was enabled in part by support provided by Compute Ontario (http://computeontario.ca), Calcul Quebec (http://www.calculquebec.ca) and Compute Canada (http://www.computecanada.ca). Results in this Letter are based in part on observations made by the Chandra X-ray Observatory (Observation IDs 20034, 20035, 19457, 19458, 00352, 02707, 03717) and the Nuclear Spectroscopic Telescope Array (NuSTAR; Observation IDs 90202052002, 90202052004, 60201016002, 60466002002). The authors thank Belinda Wilkes, Fiona Harrison, Pat Slane, Joshua Wing, Karl Forster, and the *Chandra* and *NuSTAR* scheduling, data processing, and archive teams for making these challenging simultaneous observations possible. We thank the staff at the participating observatories, correlation centers, and institutions for their enthusiastic support. This Letter makes use of the following ALMA data: ADS/JAO.ALMA#2016.1.01154.V. ALMA is a partnership of the European Southern Observatory (ESO; Europe, representing its member states), NSF, and National Institutes of Natural Sciences of Japan, together with National Research Council (Canada), Ministry of Science and Technology (MOST; Taiwan), Academia Sinica Institute of Astronomy and Astrophysics (ASIAA; Taiwan), and Korea Astronomy and Space Science Institute (KASI; Republic of Korea), in cooperation with the Republic of Chile. The Joint ALMA Observatory is operated by ESO, Associated Universities, Inc. (AUI)/NRAO, and the National Astronomical Observatory of Japan (NAOJ). The NRAO is a facility of the NSF operated under cooperative agreement by AUI. APEX is a collaboration between the Max-Planck-Institut für Radioastronomie (Germany), ESO, and the Onsala Space Observatory (Sweden). The SMA is a joint project between the SAO and ASIAA and is funded by the Smithsonian Institution and the Academia Sinica. The JCMT is operated by the East Asian Observatory on behalf of the NAOJ, ASIAA, and KASI, as well as the Ministry of Finance of China, Chinese Academy of Sciences, and the National Key R&D Program (No. 2017YFA0402700) of China. Additional funding support for the JCMT is provided by the Science and Technologies Facility Council (UK) and participating universities in the UK and Canada. The LMT project is a joint effort of the Instituto Nacional de Astrofísica, Óptica y Electrónica (Mexico) and the University of Massachusetts at Amherst (USA). The IRAM 30-m telescope on Pico Veleta, Spain is operated by IRAM and supported by CNRS (Centre National de la Recherche Scientifique, France), MPG (Max-Planck-Gesellschaft, Germany) and IGN (Instituto Geográfico Nacional, Spain). The SMT is operated by the Arizona Radio Observatory, a part of the Steward Observatory of the University of Arizona, with financial support of operations from the State of Arizona and financial support for instrumentation development from the NSF. Partial SPT support is provided by the NSF Physics Frontier Center award (PHY-0114422) to the Kavli Institute of Cosmological Physics at the University of Chicago (USA), the Kavli Foundation, and the GBMF (GBMF-947). The SPT hydrogen maser was provided on loan from the GLT, courtesy of ASIAA. The SPT is supported by the National Science Foundation through grant PLR- 1248097. Partial support is also provided by the NSF Physics Frontier Center grant PHY-1125897 to the Kavli Institute of Cosmological Physics at the University of Chicago, the Kavli Foundation and the Gordon and Betty Moore Foundation grant GBMF-947.The EHTC has received generous donations of FPGA chips from Xilinx Inc., under the Xilinx University Program. The EHTC has benefited from technology shared under open-source license by the Collaboration for Astronomy Signal Processing and Electronics Research (CASPER). The EHT project is grateful to T4Science and Microsemi for their assistance with Hydrogen Masers. This research has made use of NASA's Astrophysics Data System. We gratefully acknowledge the support provided by the extended staff of the ALMA, both from the inception of the ALMA Phasing Project through the observational campaigns of 2017 and 2018. We would like to thank A. Deller and W. Brisken for EHT-specific support with the use of DiFX. We acknowledge the significance that Maunakea, where the SMA and JCMT EHT stations are located, has for the indigenous Hawai'ian people.

## Appendix A
## Table of Simulation Results

Below we provide a table of simulation results for models with a standard inclination of 17° between the approaching jet and the line of sight. In the notation of this Letter this corresponds to $i = 17°$ for $a_* < 0$ or $i = 163°$ for $a_* \geqslant 0$. The table shows models for which we were able to calculate $L_{\rm bol}$ and $L_{\rm X}$. When $M$ is needed to calculate, e.g., $P_{\rm jet}$, we assume $M = 6.2 \times 10^9\, M_\odot$.

The first, third, and fourth columns in the table identify the model parameters: SANE or MAD based on dimensionless flux, $a_*$, and $R_{\rm high}$. Once these parameters are specified, an average value of $\dot{M}$ for the model, which is shown in last column, can be found from the requirement that the average flux density of 1.3 mm emission is ~0.5 Jy (see Paper IV). This $\dot{M}$ is shown in units of the Eddington accretion rate $\dot{M}_{\rm Edd} = 137 M_\odot\ {\rm yr}^{-1}$. The measured average dimensionless magnetic flux $\phi$ is shown in the second column. Notice that $\phi$ is determined solely from the GRMHD simulation and is independent of the mass scaling $\mathscr{M}$ and the mass $M$ used to fix the flux density. It is also independent of the electron thermodynamics ($R_{\rm high}$).

The fifth column shows the radiative efficiency, which is the bolometric luminosity $L_{\rm bol}$ over $\dot{M}c^2$. Here $L_{\rm bol}$ was found from a relativistic Monte Carlo radiative transport model that includes synchrotron emission, Compton scattering (all orders), and bremsstrahlung. The Monte Carlo calculation makes no approximations in treating the Compton scattering (see Dolence et al. 2009). Bremsstrahlung is negligible in all models.

The sixth column shows predicted X-ray luminosity $L_{\rm X}$ in the 2–10 keV band. This was calculated using the same relativistic Monte Carlo radiative transport model as for $L_{\rm bol}$. There are three numbers in this column: the average $\langle L_{\rm X}\rangle$ (left) of the 20 sample spectra used in the calculation, and a maximum and minimum value. The maximum and minimum are obtained by taking the standard deviation $\sigma(\log_{10} L_{\rm X})$ and setting the maximum (minimum) to $10^{+2\sigma}\langle L_{\rm X}\rangle$ ($10^{-2\sigma}\langle L_{\rm X}\rangle$).

The seventh column shows the jet power

$$P_{\rm jet} \equiv \int_{\beta\gamma > (\beta\gamma)_{\rm cut}} d\theta\ \frac{1}{\Delta t} \int dt d\phi\ \sqrt{-g}\,(-T^r{}_t - \rho u^r). \qquad (10)$$

The integral is evaluated at $r = 40\ r_{\rm g}$ for SANE models and $r = 100\ r_{\rm g}$ for MAD models. These radii were chosen because they are close to the outer boundary of the computational domain. Here $\Delta t$ is the duration of the time-average, $-T^r{}_t$ is a component of the stress-energy tensor representing outward radial energy flux, $g$ is the determinant of the (covariant) metric, $\rho$ is the rest-mass density, and $u^r$ is the radial component of the four-velocity. Here we use Kerr–Schild $t$, $r$, $\theta$, $\phi$ for clarity; in practice, the integral is evaluated in simulation coordinates. The quantity in parentheses is the outward energy flux with the rest-mass energy flux subtracted off. The $\theta$ integral is done after time averaging and azimuthal integration over the region where

$$(\beta\gamma)^2 \equiv \left(\frac{-T^r{}_t}{\rho u^r}\right)^2 - 1 \ > (\beta\gamma)^2_{\rm cut}. \qquad (11)$$

Here $\beta\gamma$ would be the radial four-velocity as $r \to \infty$ if the flow were steady and all internal magnetic and internal energy were converted to kinetic energy. In Table 3 we use $(\beta\gamma)^2_{\rm cut} = 1$ to define the jet. This is equivalent to restricting the jet to regions where the total energy per unit rest-mass (including the rest-mass energy) exceeds $\sqrt{5}\,c^2 \simeq 2.2c^2$.

The ninth column shows the total outflow power $P_{\rm out}$, defined using the same integral as in Equation (10), but with the $\theta$ integral carried out over the entire region around the poles where there is steady outflow (and $\theta < 1$, although the result is insensitive to this condition). $P_{\rm out}$ thus includes both the narrow, fast, relativistic jet and any wide-angle, slow, or nonrelativistic outflow. It is the maximal $P_{\rm jet}$ under any definition of jet power.

Finally, the tenth column shows the ratio of the electromagnetic to total energy flux in the jet. In most cases this number is close to 1; i.e., the jet is Poynting dominated. This measurement is sensitive to the numerical treatment of low-density regions in the jet where the jet can be artificially loaded with plasma by numerical "floors" in the GRMHD evolution. More accurate treatment of the funnel would raise values in this column.

Our choice of $(\beta\gamma)^2_{\rm cut}$, and therefore $P_{\rm jet}$, is somewhat arbitrary. To probe the sensitivity of $P_{\rm jet}$ to $(\beta\gamma)^2_{\rm cut}$, Figure 10 shows the ratio $P_{\rm jet}/P_{\rm out}$ (which is determined by the GRMHD model and is thus independent of the electron thermodynamics, i.e., $R_{\rm high}$) as a function of $(\beta\gamma)^2_{\rm cut}$.

The eighth and tenth columns show the jet and outflow efficiency. This is determined by the GRMHD evolution, i.e., it is independent of electron thermodynamics ($R_{\rm high}$). It is >0.1 only for MAD models with $a_* \geqslant 0.5$.

The eleventh column shows the fraction of $P_{\rm jet}$ in Poynting flux. This fraction is large for all models, and meaningless for the $a_* = 0$ models, which have $P_{\rm jet}$ that is so small that it is difficult to measure accurately.

The problem of defining $P_{\rm jet}$ and $P_{\rm out}$ has been discussed extensively in the literature (e.g., Narayan et al. 2012; Yuan et al. 2015; Mościbrodzka et al. 2016), where alternative definitions of unbound regions and of the jet have been used, some based on a fluid Bernoulli parameter $B_e \equiv$

**Table 3**
Model Table

| Flux | $\phi$ | Spin | $R_{\rm high}$ | $L_{\rm bol}/(\dot{M}\ c^2)$ | $L_{\rm X}$ (cgs) | $P_{\rm jet}$ (cgs) | $P_{\rm jet}/(\dot{M}\ c^2)$ | $P_{\rm out}$ (cgs) | $P_{\rm out}/(\dot{M}\ c^2)$ | $P_{\rm jet,em}/P_{\rm jet}$ | $\dot{M}/\dot{M}_{\rm Edd}$ |
|---|---|---|---|---|---|---|---|---|---|---|---|
| SANE | 1.02 | −0.94 | 1 | $1.27 \times 10^{-2}$ | $3.18^{<49.55}_{>0.20} \times 10^{41}$ | $1.16 \times 10^{42}$ | $5.34 \times 10^{-3}$ | $1.19 \times 10^{42}$ | $5.48 \times 10^{-3}$ | 0.84 | $2.77 \times 10^{-5}$ |
| SANE | 1.02 | −0.94 | 10 | $1.6 \times 10^{-3}$ | $9.62^{<64.42}_{>1.44} \times 10^{40}$ | $4.94 \times 10^{42}$ | $5.34 \times 10^{-3}$ | $5.07 \times 10^{42}$ | $5.48 \times 10^{-3}$ | 0.84 | $1.19 \times 10^{-4}$ |
| SANE | 1.02 | −0.94 | 20 | $6.09 \times 10^{-4}$ | $3.26^{<11.86}_{>0.90} \times 10^{40}$ | $5.8 \times 10^{42}$ | $5.34 \times 10^{-3}$ | $5.96 \times 10^{42}$ | $5.48 \times 10^{-3}$ | 0.84 | $1.39 \times 10^{-4}$ |
| SANE | 1.02 | −0.94 | 40 | $2.45 \times 10^{-4}$ | $8.89^{<50.53}_{>1.56} \times 10^{39}$ | $7.02 \times 10^{42}$ | $5.34 \times 10^{-3}$ | $7.21 \times 10^{42}$ | $5.48 \times 10^{-3}$ | 0.84 | $1.69 \times 10^{-4}$ |
| SANE | 1.02 | −0.94 | 80 | $1.33 \times 10^{-4}$ | $2.65^{<18.26}_{>0.39} \times 10^{39}$ | $8.89 \times 10^{42}$ | $5.34 \times 10^{-3}$ | $9.13 \times 10^{42}$ | $5.48 \times 10^{-3}$ | 0.84 | $2.13 \times 10^{-4}$ |
| SANE | 1.02 | −0.94 | 160 | $7.12 \times 10^{-5}$ | $6.36^{<55.27}_{>0.73} \times 10^{38}$ | $1.2 \times 10^{43}$ | $5.34 \times 10^{-3}$ | $1.23 \times 10^{43}$ | $5.48 \times 10^{-3}$ | 0.84 | $2.87 \times 10^{-4}$ |
| SANE | 1.11 | −0.5 | 1 | $1.62 \times 10^{-2}$ | $1.97^{<3.94}_{>0.98} \times 10^{41}$ | $2.62 \times 10^{40}$ | $1.86 \times 10^{-4}$ | $3.84 \times 10^{40}$ | $2.72 \times 10^{-4}$ | 0.88 | $1.81 \times 10^{-5}$ |
| SANE | 1.11 | −0.5 | 10 | $2.17 \times 10^{-3}$ | $1.94^{<5.40}_{>0.69} \times 10^{41}$ | $1.95 \times 10^{41}$ | $1.86 \times 10^{-4}$ | $2.85 \times 10^{41}$ | $2.72 \times 10^{-4}$ | 0.88 | $1.34 \times 10^{-4}$ |
| SANE | 1.11 | −0.5 | 20 | $6.69 \times 10^{-4}$ | $3.72^{<7.72}_{>1.80} \times 10^{40}$ | $2.26 \times 10^{41}$ | $1.86 \times 10^{-4}$ | $3.31 \times 10^{41}$ | $2.72 \times 10^{-4}$ | 0.88 | $1.56 \times 10^{-4}$ |
| SANE | 1.11 | −0.5 | 40 | $2.47 \times 10^{-4}$ | $9.44^{<13.37}_{>6.67} \times 10^{39}$ | $2.62 \times 10^{41}$ | $1.86 \times 10^{-4}$ | $3.83 \times 10^{41}$ | $2.72 \times 10^{-4}$ | 0.88 | $1.81 \times 10^{-4}$ |
| SANE | 1.11 | −0.5 | 80 | $1.26 \times 10^{-4}$ | $1.23^{<4.58}_{>0.33} \times 10^{39}$ | $3.2 \times 10^{41}$ | $1.86 \times 10^{-4}$ | $4.68 \times 10^{41}$ | $2.72 \times 10^{-4}$ | 0.88 | $2.21 \times 10^{-4}$ |
| SANE | 1.11 | −0.5 | 160 | $7.86 \times 10^{-5}$ | $3.72^{<16.68}_{>0.83} \times 10^{38}$ | $4.21 \times 10^{41}$ | $1.86 \times 10^{-4}$ | $6.16 \times 10^{41}$ | $2.72 \times 10^{-4}$ | 0.88 | $2.9 \times 10^{-4}$ |
| SANE | 0.99 | 0 | 1 | $3.17 \times 10^{-2}$ | $2.08^{<194.22}_{>0.02} \times 10^{41}$ | $2.24 \times 10^{36}$ | $4.4 \times 10^{-8}$ | $5.22 \times 10^{39}$ | $1.03 \times 10^{-4}$ | 1.01 | $6.5 \times 10^{-6}$ |
| SANE | 0.99 | 0 | 10 | $1.88 \times 10^{-2}$ | $4.2^{<425.40}_{>0.04} \times 10^{42}$ | $4.38 \times 10^{37}$ | $4.4 \times 10^{-8}$ | $1.02 \times 10^{41}$ | $1.03 \times 10^{-4}$ | 1.01 | $1.27 \times 10^{-4}$ |
| SANE | 0.99 | 0 | 20 | $5.83 \times 10^{-3}$ | $1.57^{<39.69}_{>0.06} \times 10^{42}$ | $8.02 \times 10^{37}$ | $4.4 \times 10^{-8}$ | $1.87 \times 10^{41}$ | $1.03 \times 10^{-4}$ | 1.01 | $2.33 \times 10^{-4}$ |
| SANE | 0.99 | 0 | 40 | $7.8 \times 10^{-4}$ | $8.92^{<41.45}_{>1.92} \times 10^{40}$ | $9.16 \times 10^{37}$ | $4.4 \times 10^{-8}$ | $2.14 \times 10^{41}$ | $1.03 \times 10^{-4}$ | 1.01 | $2.66 \times 10^{-4}$ |
| SANE | 0.99 | 0 | 80 | $1.69 \times 10^{-4}$ | $2.5^{<19.17}_{>0.33} \times 10^{39}$ | $1.03 \times 10^{38}$ | $4.4 \times 10^{-8}$ | $2.41 \times 10^{41}$ | $1.03 \times 10^{-4}$ | 1.01 | $3 \times 10^{-4}$ |
| SANE | 0.99 | 0 | 160 | $1.08 \times 10^{-4}$ | $3.44^{<13.32}_{>0.89} \times 10^{38}$ | $1.23 \times 10^{38}$ | $4.4 \times 10^{-8}$ | $2.87 \times 10^{41}$ | $1.03 \times 10^{-4}$ | 1.01 | $3.57 \times 10^{-4}$ |
| SANE | 1.10 | 0.5 | 1 | $4.97 \times 10^{-2}$ | $5.5^{<34.41}_{>0.88} \times 10^{40}$ | $2.57 \times 10^{39}$ | $1.63 \times 10^{-4}$ | $9.19 \times 10^{39}$ | $5.86 \times 10^{-4}$ | 0.88 | $2.01 \times 10^{-6}$ |
| SANE | 1.10 | 0.5 | 10 | $5.98 \times 10^{-3}$ | $4.73^{<88.59}_{>0.25} \times 10^{40}$ | $1.91 \times 10^{40}$ | $1.64 \times 10^{-4}$ | $6.84 \times 10^{40}$ | $5.86 \times 10^{-4}$ | 0.88 | $1.5 \times 10^{-5}$ |
| SANE | 1.10 | 0.5 | 20 | $3.33 \times 10^{-3}$ | $3.83^{<49.18}_{>0.30} \times 10^{40}$ | $4.09 \times 10^{40}$ | $1.64 \times 10^{-4}$ | $1.47 \times 10^{41}$ | $5.86 \times 10^{-4}$ | 0.88 | $3.2 \times 10^{-5}$ |
| SANE | 1.10 | 0.5 | 40 | $1.74 \times 10^{-3}$ | $2.52^{<22.73}_{>0.28} \times 10^{40}$ | $8.02 \times 10^{40}$ | $1.64 \times 10^{-4}$ | $2.87 \times 10^{41}$ | $5.86 \times 10^{-4}$ | 0.88 | $6.28 \times 10^{-5}$ |
| SANE | 1.10 | 0.5 | 80 | $6.95 \times 10^{-4}$ | $7.84^{<91.92}_{>0.67} \times 10^{39}$ | $1.27 \times 10^{41}$ | $1.64 \times 10^{-4}$ | $4.55 \times 10^{41}$ | $5.86 \times 10^{-4}$ | 0.88 | $9.95 \times 10^{-5}$ |
| SANE | 1.10 | 0.5 | 160 | $2.78 \times 10^{-4}$ | $1.37^{<22.85}_{>0.08} \times 10^{39}$ | $1.69 \times 10^{41}$ | $1.63 \times 10^{-4}$ | $6.06 \times 10^{41}$ | $5.86 \times 10^{-4}$ | 0.88 | $1.33 \times 10^{-4}$ |
| SANE | 1.64 | 0.94 | 1 | 1.4 | $2.38^{<359.03}_{>0.02} \times 10^{41}$ | $2.2 \times 10^{40}$ | $7.76 \times 10^{-3}$ | $3.38 \times 10^{40}$ | $1.19 \times 10^{-2}$ | 0.82 | $3.63 \times 10^{-7}$ |
| SANE | 1.64 | 0.94 | 10 | $2.7 \times 10^{-1}$ | $2.79^{<508.99}_{>0.02} \times 10^{41}$ | $1.4 \times 10^{41}$ | $7.76 \times 10^{-3}$ | $2.15 \times 10^{41}$ | $1.19 \times 10^{-2}$ | 0.82 | $2.31 \times 10^{-6}$ |
| SANE | 1.64 | 0.94 | 20 | $1.74 \times 10^{-1}$ | $5.75^{<1685.98}_{>0.02} \times 10^{41}$ | $3.22 \times 10^{41}$ | $7.76 \times 10^{-3}$ | $4.94 \times 10^{41}$ | $1.19 \times 10^{-2}$ | 0.82 | $5.31 \times 10^{-6}$ |
| SANE | 1.64 | 0.94 | 40 | $7.2 \times 10^{-2}$ | $4.71^{<2490.36}_{>0.01} \times 10^{41}$ | $5.97 \times 10^{41}$ | $7.76 \times 10^{-3}$ | $9.17 \times 10^{41}$ | $1.19 \times 10^{-2}$ | 0.82 | $9.84 \times 10^{-6}$ |
| SANE | 1.64 | 0.94 | 80 | $2.38 \times 10^{-2}$ | $1.42^{<860.83}_{>0.00} \times 10^{41}$ | $8.87 \times 10^{41}$ | $7.76 \times 10^{-3}$ | $1.36 \times 10^{42}$ | $1.19 \times 10^{-2}$ | 0.82 | $1.46 \times 10^{-5}$ |
| SANE | 1.64 | 0.94 | 160 | $8.45 \times 10^{-3}$ | $3.22^{<1687.88}_{>0.01} \times 10^{40}$ | $1.23 \times 10^{42}$ | $7.76 \times 10^{-3}$ | $1.89 \times 10^{42}$ | $1.19 \times 10^{-2}$ | 0.82 | $2.03 \times 10^{-5}$ |
| MAD | 8.04 | −0.94 | 1 | $7.61 \times 10^{-1}$ | $2.12^{<17.74}_{>0.25} \times 10^{41}$ | $1.36 \times 10^{42}$ | $2.09 \times 10^{-1}$ | $1.6 \times 10^{42}$ | $2.46 \times 10^{-1}$ | 0.75 | $8.32 \times 10^{-7}$ |
| MAD | 8.04 | −0.94 | 10 | $7.54 \times 10^{-2}$ | $5.76^{<68.06}_{>0.49} \times 10^{40}$ | $1.97 \times 10^{42}$ | $2.09 \times 10^{-1}$ | $2.32 \times 10^{42}$ | $2.46 \times 10^{-1}$ | 0.75 | $1.21 \times 10^{-6}$ |
| MAD | 8.04 | −0.94 | 20 | $3.76 \times 10^{-2}$ | $2.27^{<29.09}_{>0.18} \times 10^{40}$ | $2.38 \times 10^{42}$ | $2.09 \times 10^{-1}$ | $2.8 \times 10^{42}$ | $2.46 \times 10^{-1}$ | 0.75 | $1.46 \times 10^{-6}$ |
| MAD | 8.04 | −0.94 | 40 | $2.07 \times 10^{-2}$ | $6.18^{<77.36}_{>0.49} \times 10^{39}$ | $3 \times 10^{42}$ | $2.09 \times 10^{-1}$ | $3.54 \times 10^{42}$ | $2.46 \times 10^{-1}$ | 0.75 | $1.84 \times 10^{-6}$ |
| MAD | 8.04 | −0.94 | 80 | $1.17 \times 10^{-2}$ | $1.32^{<26.36}_{>0.07} \times 10^{39}$ | $3.99 \times 10^{42}$ | $2.09 \times 10^{-1}$ | $4.71 \times 10^{42}$ | $2.46 \times 10^{-1}$ | 0.75 | $2.45 \times 10^{-6}$ |
| MAD | 8.04 | −0.94 | 160 | $6.52 \times 10^{-3}$ | $2.57^{<46.76}_{>0.14} \times 10^{38}$ | $5.7 \times 10^{42}$ | $2.09 \times 10^{-1}$ | $6.73 \times 10^{42}$ | $2.46 \times 10^{-1}$ | 0.75 | $3.5 \times 10^{-6}$ |
| MAD | 12.25 | −0.5 | 1 | $2.96 \times 10^{-1}$ | $1.39^{<11.56}_{>0.17} \times 10^{41}$ | $3.43 \times 10^{41}$ | $4.91 \times 10^{-2}$ | $6.04 \times 10^{41}$ | $8.64 \times 10^{-2}$ | 0.82 | $8.95 \times 10^{-7}$ |
| MAD | 12.25 | −0.5 | 10 | $4.53 \times 10^{-2}$ | $2.43^{<19.86}_{>0.30} \times 10^{40}$ | $5.31 \times 10^{41}$ | $4.92 \times 10^{-2}$ | $9.33 \times 10^{41}$ | $8.64 \times 10^{-2}$ | 0.82 | $1.38 \times 10^{-6}$ |
| MAD | 12.25 | −0.5 | 20 | $2.67 \times 10^{-2}$ | $8.18^{<77.51}_{>0.86} \times 10^{39}$ | $6.45 \times 10^{41}$ | $4.92 \times 10^{-2}$ | $1.13 \times 10^{42}$ | $8.64 \times 10^{-2}$ | 0.82 | $1.68 \times 10^{-6}$ |
| MAD | 12.25 | −0.5 | 40 | $1.69 \times 10^{-2}$ | $2.17^{<22.33}_{>0.21} \times 10^{39}$ | $8.07 \times 10^{41}$ | $4.92 \times 10^{-2}$ | $1.42 \times 10^{42}$ | $8.64 \times 10^{-2}$ | 0.82 | $2.1 \times 10^{-6}$ |
| MAD | 12.25 | −0.5 | 80 | $1.07 \times 10^{-2}$ | $4.87^{<50.76}_{>0.47} \times 10^{38}$ | $1.05 \times 10^{42}$ | $4.92 \times 10^{-2}$ | $1.85 \times 10^{42}$ | $8.64 \times 10^{-2}$ | 0.82 | $2.74 \times 10^{-6}$ |
| MAD | 12.25 | −0.5 | 160 | $6.43 \times 10^{-3}$ | $1.09^{<7.06}_{>0.17} \times 10^{38}$ | $1.46 \times 10^{42}$ | $4.92 \times 10^{-2}$ | $2.57 \times 10^{42}$ | $8.64 \times 10^{-2}$ | 0.82 | $3.81 \times 10^{-6}$ |
| MAD | 15.44 | 0 | 1 | $2.67 \times 10^{-1}$ | $1.22^{<14.60}_{>0.10} \times 10^{41}$ | 0.0 | 0.0 | $8.39 \times 10^{40}$ | $1.51 \times 10^{-2}$ | 0.00 | $7.12 \times 10^{-7}$ |

**Table 3**
(Continued)

| Flux | $\phi$ | Spin | $R_{\rm high}$ | $L_{\rm bol}/(\dot{M}\ c^2)$ | $L_{\rm X}$ (cgs) | $P_{\rm jet}$ (cgs) | $P_{\rm jet}/(\dot{M}\ c^2)$ | $P_{\rm out}$ (cgs) | $P_{\rm out}/(\dot{M}\ c^2)$ | $P_{\rm jet,em}/P_{\rm jet}$ | $\dot{M}/\dot{M}_{\rm Edd}$ |
|---|---|---|---|---|---|---|---|---|---|---|---|
| MAD | 15.44 | 0 | 10 | $4.53 \times 10^{-2}$ | $1.86^{<31.55}_{>0.11} \times 10^{40}$ | 0.0 | 0.0 | $1.39 \times 10^{41}$ | $1.51 \times 10^{-2}$ | 0.00 | $1.18 \times 10^{-6}$ |
| MAD | 15.44 | 0 | 20 | $2.81 \times 10^{-2}$ | $5.98^{<101.81}_{>0.35} \times 10^{39}$ | 0.0 | 0.0 | $1.71 \times 10^{41}$ | $1.51 \times 10^{-2}$ | 0.00 | $1.46 \times 10^{-6}$ |
| MAD | 15.44 | 0 | 40 | $1.85 \times 10^{-2}$ | $1.63^{<27.75}_{>0.10} \times 10^{39}$ | 0.0 | 0.0 | $2.15 \times 10^{41}$ | $1.51 \times 10^{-2}$ | 0.00 | $1.82 \times 10^{-6}$ |
| MAD | 15.44 | 0 | 80 | $1.21 \times 10^{-2}$ | $3.51^{<61.34}_{>0.20} \times 10^{38}$ | 0.0 | 0.0 | $2.77 \times 10^{41}$ | $1.51 \times 10^{-2}$ | 0.00 | $2.35 \times 10^{-6}$ |
| MAD | 15.44 | 0 | 160 | $7.63 \times 10^{-3}$ | $8.06^{<80.62}_{>0.81} \times 10^{37}$ | 0.0 | 0.0 | $3.73 \times 10^{41}$ | $1.51 \times 10^{-2}$ | 0.00 | $3.17 \times 10^{-6}$ |
| MAD | 15.95 | 0.5 | 1 | $5.45 \times 10^{-1}$ | $1.57^{<11.98}_{>0.21} \times 10^{41}$ | $4.64 \times 10^{41}$ | $1.16 \times 10^{-1}$ | $6.74 \times 10^{41}$ | $1.69 \times 10^{-1}$ | 0.85 | $5.11 \times 10^{-7}$ |
| MAD | 15.95 | 0.5 | 10 | $9.45 \times 10^{-2}$ | $2.71^{<36.30}_{>0.20} \times 10^{40}$ | $8.07 \times 10^{41}$ | $1.16 \times 10^{-1}$ | $1.17 \times 10^{42}$ | $1.69 \times 10^{-1}$ | 0.85 | $8.89 \times 10^{-7}$ |
| MAD | 15.95 | 0.5 | 20 | $5.54 \times 10^{-2}$ | $9.67^{<126.69}_{>0.74} \times 10^{39}$ | $1.02 \times 10^{42}$ | $1.16 \times 10^{-1}$ | $1.49 \times 10^{42}$ | $1.69 \times 10^{-1}$ | 0.85 | $1.13 \times 10^{-6}$ |
| MAD | 15.95 | 0.5 | 40 | $3.5 \times 10^{-2}$ | $3.3^{<39.01}_{>0.28} \times 10^{39}$ | $1.32 \times 10^{42}$ | $1.16 \times 10^{-1}$ | $1.92 \times 10^{42}$ | $1.69 \times 10^{-1}$ | 0.85 | $1.45 \times 10^{-6}$ |
| MAD | 15.95 | 0.5 | 80 | $2.22 \times 10^{-2}$ | $8^{<91.84}_{>0.70} \times 10^{38}$ | $1.74 \times 10^{42}$ | $1.16 \times 10^{-1}$ | $2.52 \times 10^{42}$ | $1.69 \times 10^{-1}$ | 0.85 | $1.92 \times 10^{-6}$ |
| MAD | 15.95 | 0.5 | 160 | $1.35 \times 10^{-2}$ | $1.79^{<8.44}_{>0.38} \times 10^{38}$ | $2.38 \times 10^{42}$ | $1.16 \times 10^{-1}$ | $3.46 \times 10^{42}$ | $1.69 \times 10^{-1}$ | 0.85 | $2.62 \times 10^{-6}$ |
| MAD | 12.78 | 0.94 | 1 | 3.65 | $5.19^{<43.60}_{>0.62} \times 10^{41}$ | $1.97 \times 10^{42}$ | $8.23 \times 10^{-1}$ | $2.29 \times 10^{42}$ | $9.55 \times 10^{-1}$ | 0.80 | $3.07 \times 10^{-7}$ |
| MAD | 12.78 | 0.94 | 10 | $3.68 \times 10^{-1}$ | $1.3^{<13.22}_{>0.13} \times 10^{41}$ | $3.04 \times 10^{42}$ | $8.23 \times 10^{-1}$ | $3.52 \times 10^{42}$ | $9.55 \times 10^{-1}$ | 0.80 | $4.73 \times 10^{-7}$ |
| MAD | 12.78 | 0.94 | 20 | $1.79 \times 10^{-1}$ | $5^{<56.22}_{>0.44} \times 10^{40}$ | $3.73 \times 10^{42}$ | $8.23 \times 10^{-1}$ | $4.33 \times 10^{42}$ | $9.55 \times 10^{-1}$ | 0.80 | $5.81 \times 10^{-7}$ |
| MAD | 12.78 | 0.94 | 40 | $9.43 \times 10^{-2}$ | $1.54^{<22.13}_{>0.11} \times 10^{40}$ | $4.74 \times 10^{42}$ | $8.23 \times 10^{-1}$ | $5.5 \times 10^{42}$ | $9.55 \times 10^{-1}$ | 0.80 | $7.38 \times 10^{-7}$ |
| MAD | 12.78 | 0.94 | 80 | $5.19 \times 10^{-2}$ | $3.74^{<80.85}_{>0.17} \times 10^{39}$ | $6.26 \times 10^{42}$ | $8.23 \times 10^{-1}$ | $7.27 \times 10^{42}$ | $9.55 \times 10^{-1}$ | 0.80 | $9.75 \times 10^{-7}$ |
| MAD | 12.78 | 0.94 | 160 | $2.82 \times 10^{-2}$ | $6.97^{<186.48}_{>0.26} \times 10^{38}$ | $8.75 \times 10^{42}$ | $8.23 \times 10^{-1}$ | $1.02 \times 10^{43}$ | $9.55 \times 10^{-1}$ | 0.80 | $1.36 \times 10^{-6}$ |

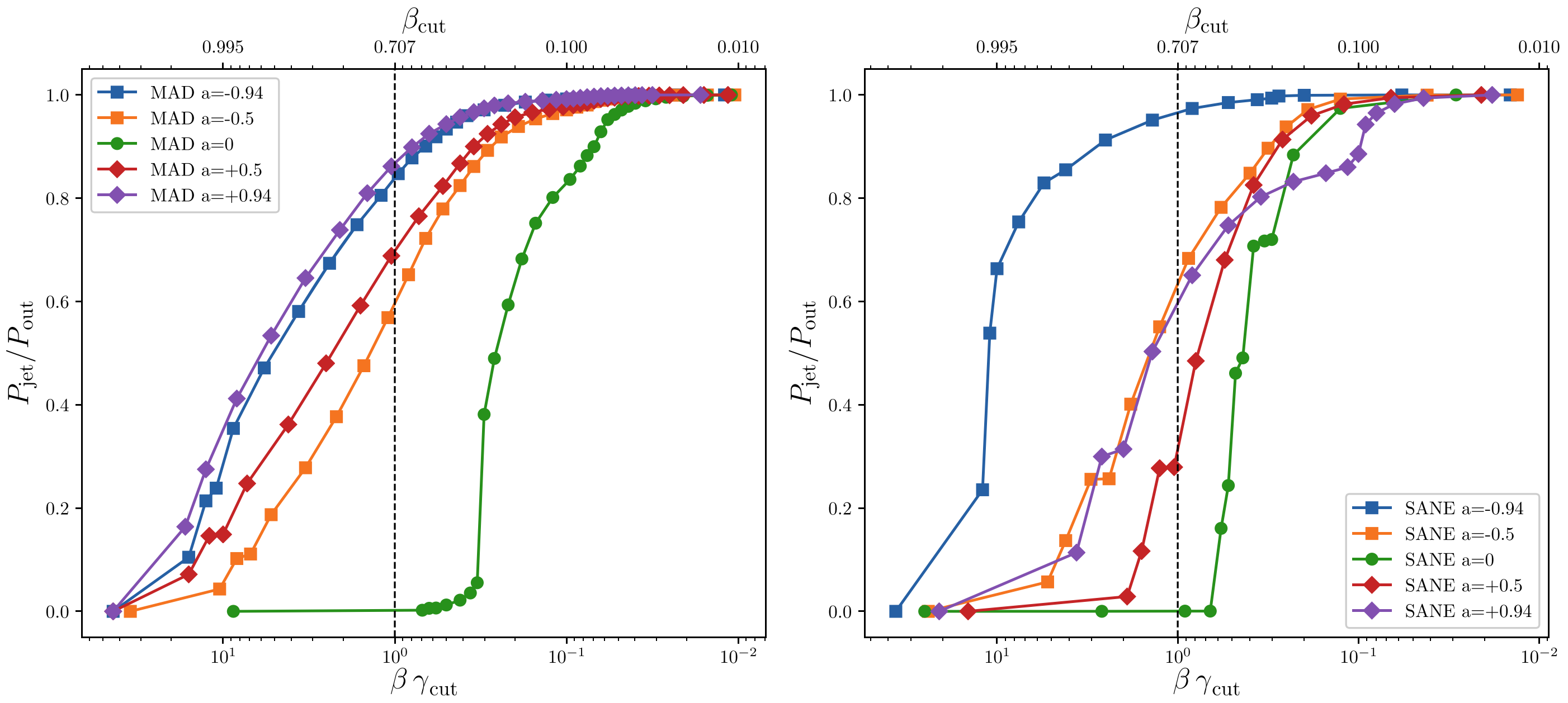

**Figure 10.** Ratio $P_{\rm jet}/P_{\rm out}$ as a function of the outflow velocity cutoff parameter $\beta\gamma_{\rm cut}$. Evidently, as the cut is decreased, so that the maximum asymptotic speed of the jet flow is decreased, an increasing fraction of $P_{\rm out}$ is classified as $P_{\rm jet}$. Our nominal cutoff is $\beta\gamma = 1$, which corresponds to $\beta \equiv v^r/c = 1/\sqrt{2}$. Using this definition, $P_{\rm jet}$ for $a_* = 0$ models is small because the energy flux in the relativistic outflow is small.

$-u_t(\rho + u + p)/\rho - 1$, while others use $\mu$ (the ratio of energy flux to rest mass flux), which is directly related to our $\beta\gamma$.

## Appendix B
## Image Decomposition into Midplane, Nearside, and Farside Components

In Section 3.3 we presented representative images from the Image Library spanning a broad range of values in both $a_*$ and $R_{\rm high}$. It was noted that for SANEs with low values of $R_{\rm high}$ the emission is concentrated more in the midplane, whereas for larger values of $R_{\rm high}$ this emission is concentrated in the funnel wall. In particular, Figure 4 presented temporal- and azimuthal-averaged images of the point of origin of photons comprising images from $a_* = 0.94$ MAD and SANE simulations with $R_{\rm high} = 10$ and 160.

Figure 11 presents the decomposition of the four images in Figure 4 into components that we refer to as: midplane (material within 32°.7 of the midplane), nearside (material within 1 radian, or 57°.3, of the polar axis nearest to the observer), and farside (material within 1 radian of the polar axis furthest from the observer). From inspection of the first three models (rows) in Figure 11, the ratio of nearside to farside flux in the simulations is small (compared to the midplane) and of order unity and the midplane emission is dominant, as in Figure 4.

However, for the SANE, $R_{\rm high} = 160$ model the farside emission contributes a flux that exceeds that produced from the midplane, and is significantly brighter than the nearside emission. This is in agreement with what is seen in the bottom-right panel of Figures 4, and can be understood to arise from the SANE model possessing an optically thin disk and bright funnel wall in the $R_{\rm high} = 160$ case, compared to SANE, $R_{\rm high} = 10$, as also seen in Figures 2 and 3. Due to the reduced opacity along the line of sight in this case, mm photons can pass through both the intervening nearside material and the midplane without significant attenuation, before reaching the photospheric boundary in the farside component (where $\tau \sim 1$), where they originate. The image decomposition and its application to M87's image structure will be explored further in Z. Younsi et al. 2019a (in preparation).

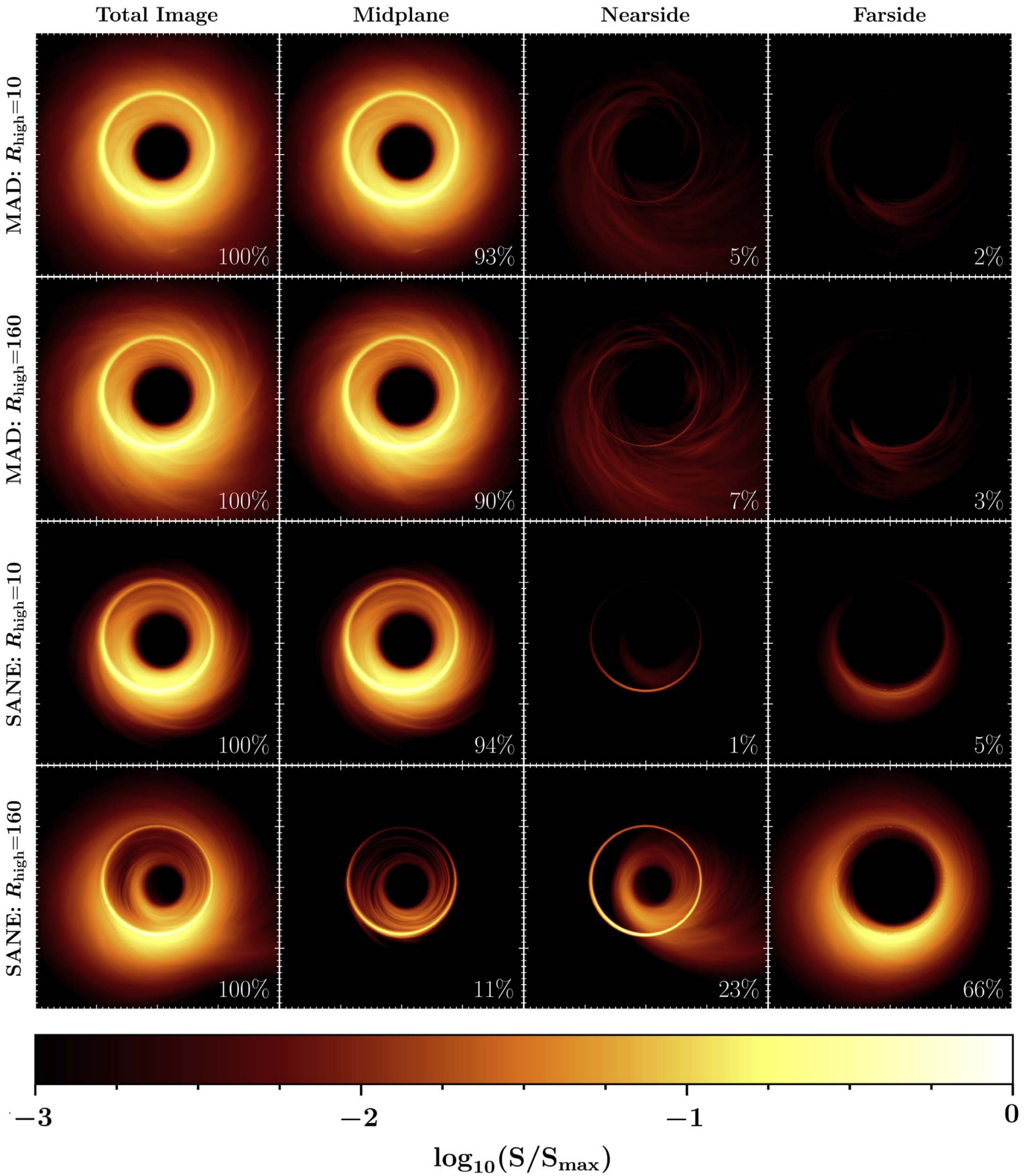

**Figure 11.** Decomposition of time-averaged 1.3 mm images from Figure 4 into midplane, nearside, and farside components (MAD and SANE models with $a_* = 0.94$). Each model (row) of the figure corresponds to a simulation in Figure 4. The percentage of the total image flux from each component is indicated in the bottom right of each panel. The color scale is logarithmic and spans three decades in total flux with respect to the total image from each model, chosen in order to emphasize both nearside and farside components, which are nearly invisible when shown in a linear scale. The field of view is 80 $\mu$as.

## ORCID iDs

Kazunori Akiyama https://orcid.org/0000-0002-9475-4254
Antxon Alberdi https://orcid.org/0000-0002-9371-1033
Rebecca Azulay https://orcid.org/0000-0002-2200-5393
Anne-Kathrin Baczko https://orcid.org/0000-0003-3090-3975
Mislav Baloković https://orcid.org/0000-0003-0476-6647
John Barrett https://orcid.org/0000-0002-9290-0764
Lindy Blackburn https://orcid.org/0000-0002-9030-642X
Katherine L. Bouman https://orcid.org/0000-0003-0077-4367
Geoffrey C. Bower https://orcid.org/0000-0003-4056-9982
Christiaan D. Brinkerink https://orcid.org/0000-0002-2322-0749
Roger Brissenden https://orcid.org/0000-0002-2556-0894
Silke Britzen https://orcid.org/0000-0001-9240-6734
Avery E. Broderick https://orcid.org/0000-0002-3351-760X
Do-Young Byun https://orcid.org/0000-0003-1157-4109
Andrew Chael https://orcid.org/0000-0003-2966-6220
Chi-kwan Chan https://orcid.org/0000-0001-6337-6126
Shami Chatterjee https://orcid.org/0000-0002-2878-1502
Ilje Cho https://orcid.org/0000-0001-6083-7521
Pierre Christian https://orcid.org/0000-0001-6820-9941
John E. Conway https://orcid.org/0000-0003-2448-9181
Geoffrey B. Crew https://orcid.org/0000-0002-2079-3189
Yuzhu Cui https://orcid.org/0000-0001-6311-4345
Jordy Davelaar https://orcid.org/0000-0002-2685-2434
Mariafelicia De Laurentis https://orcid.org/0000-0002-9945-682X
Roger Deane https://orcid.org/0000-0003-1027-5043
Jessica Dempsey https://orcid.org/0000-0003-1269-9667
Gregory Desvignes https://orcid.org/0000-0003-3922-4055
Jason Dexter https://orcid.org/0000-0003-3903-0373
Sheperd S. Doeleman https://orcid.org/0000-0002-9031-0904
Ralph P. Eatough https://orcid.org/0000-0001-6196-4135
Heino Falcke https://orcid.org/0000-0002-2526-6724
Vincent L. Fish https://orcid.org/0000-0002-7128-9345
Raquel Fraga-Encinas https://orcid.org/0000-0002-5222-1361
José L. Gómez https://orcid.org/0000-0003-4190-7613
Peter Galison https://orcid.org/0000-0002-6429-3872
Charles F. Gammie https://orcid.org/0000-0001-7451-8935
Boris Georgiev https://orcid.org/0000-0002-3586-6424
Roman Gold https://orcid.org/0000-0003-2492-1966
Minfeng Gu (顾敏峰) https://orcid.org/0000-0002-4455-6946
Mark Gurwell https://orcid.org/0000-0003-0685-3621
Kazuhiro Hada https://orcid.org/0000-0001-6906-772X
Ronald Hesper https://orcid.org/0000-0003-1918-6098
Luis C. Ho (何子山) https://orcid.org/0000-0001-6947-5846
Mareki Honma https://orcid.org/0000-0003-4058-9000
Chih-Wei L. Huang https://orcid.org/0000-0001-5641-3953
Shiro Ikeda https://orcid.org/0000-0002-2462-1448
Sara Issaoun https://orcid.org/0000-0002-5297-921X
David J. James https://orcid.org/0000-0001-5160-4486
Michael Janssen https://orcid.org/0000-0001-8685-6544
Britton Jeter https://orcid.org/0000-0003-2847-1712
Wu Jiang (江悟) https://orcid.org/0000-0001-7369-3539
Michael D. Johnson https://orcid.org/0000-0002-4120-3029
Svetlana Jorstad https://orcid.org/0000-0001-6158-1708
Taehyun Jung https://orcid.org/0000-0001-7003-8643
Mansour Karami https://orcid.org/0000-0001-7387-9333
Ramesh Karuppusamy https://orcid.org/0000-0002-5307-2919
Tomohisa Kawashima https://orcid.org/0000-0001-8527-0496
Garrett K. Keating https://orcid.org/0000-0002-3490-146X
Mark Kettenis https://orcid.org/0000-0002-6156-5617
Jae-Young Kim https://orcid.org/0000-0001-8229-7183
Junhan Kim https://orcid.org/0000-0002-4274-9373
Motoki Kino https://orcid.org/0000-0002-2709-7338
Jun Yi Koay https://orcid.org/0000-0002-7029-6658
Patrick M. Koch https://orcid.org/0000-0003-2777-5861
Shoko Koyama https://orcid.org/0000-0002-3723-3372
Michael Kramer https://orcid.org/0000-0002-4175-2271
Carsten Kramer https://orcid.org/0000-0002-4908-4925
Thomas P. Krichbaum https://orcid.org/0000-0002-4892-9586
Tod R. Lauer https://orcid.org/0000-0003-3234-7247
Sang-Sung Lee https://orcid.org/0000-0002-6269-594X
Yan-Rong Li (李彦荣) https://orcid.org/0000-0001-5841-9179
Zhiyuan Li (李志远) https://orcid.org/0000-0003-0355-6437
Michael Lindqvist https://orcid.org/0000-0002-3669-0715
Kuo Liu https://orcid.org/0000-0002-2953-7376
Elisabetta Liuzzo https://orcid.org/0000-0003-0995-5201
Laurent Loinard https://orcid.org/0000-0002-5635-3345
Ru-Sen Lu (路如森) https://orcid.org/0000-0002-7692-7967
Nicholas R. MacDonald https://orcid.org/0000-0002-6684-8691
Jirong Mao (毛基荣) https://orcid.org/0000-0002-7077-7195
Sera Markoff https://orcid.org/0000-0001-9564-0876
Daniel P. Marrone https://orcid.org/0000-0002-2367-1080
Alan P. Marscher https://orcid.org/0000-0001-7396-3332
Iván Martí-Vidal https://orcid.org/0000-0003-3708-9611
Lynn D. Matthews https://orcid.org/0000-0002-3728-8082
Lia Medeiros https://orcid.org/0000-0003-2342-6728
Karl M. Menten https://orcid.org/0000-0001-6459-0669
Yosuke Mizuno https://orcid.org/0000-0002-8131-6730
Izumi Mizuno https://orcid.org/0000-0002-7210-6264
James M. Moran https://orcid.org/0000-0002-3882-4414
Kotaro Moriyama https://orcid.org/0000-0003-1364-3761
Monika Moscibrodzka https://orcid.org/0000-0002-4661-6332
Cornelia Müller https://orcid.org/0000-0002-2739-2994
Hiroshi Nagai https://orcid.org/0000-0003-0292-3645
Neil M. Nagar https://orcid.org/0000-0001-6920-662X
Masanori Nakamura https://orcid.org/0000-0001-6081-2420
Ramesh Narayan https://orcid.org/0000-0002-1919-2730
Iniyan Natarajan https://orcid.org/0000-0001-8242-4373
Chunchong Ni https://orcid.org/0000-0003-1361-5699
Aristeidis Noutsos https://orcid.org/0000-0002-4151-3860
Héctor Olivares https://orcid.org/0000-0001-6833-7580
Daniel C. M. Palumbo https://orcid.org/0000-0002-7179-3816
Ue-Li Pen https://orcid.org/0000-0003-2155-9578
Dominic W. Pesce https://orcid.org/0000-0002-5278-9221
Oliver Porth https://orcid.org/0000-0002-4584-2557
Ben Prather https://orcid.org/0000-0002-0393-7734
Jorge A. Preciado-López https://orcid.org/0000-0002-4146-0113
Hung-Yi Pu https://orcid.org/0000-0001-9270-8812
Venkatessh Ramakrishnan https://orcid.org/0000-0002-9248-086X

Ramprasad Rao https://orcid.org/0000-0002-1407-7944
Alexander W. Raymond https://orcid.org/0000-0002-5779-4767
Bart Ripperda https://orcid.org/0000-0002-7301-3908
Freek Roelofs https://orcid.org/0000-0001-5461-3687
Eduardo Ros https://orcid.org/0000-0001-9503-4892
Mel Rose https://orcid.org/0000-0002-2016-8746
Alan L. Roy https://orcid.org/0000-0002-1931-0135
Chet Ruszczyk https://orcid.org/0000-0001-7278-9707
Benjamin R. Ryan https://orcid.org/0000-0001-8939-4461
Kazi L. J. Rygl https://orcid.org/0000-0003-4146-9043
David Sánchez-Arguelles https://orcid.org/0000-0002-7344-9920
Mahito Sasada https://orcid.org/0000-0001-5946-9960
Tuomas Savolainen https://orcid.org/0000-0001-6214-1085
Lijing Shao https://orcid.org/0000-0002-1334-8853
Zhiqiang Shen (沈志强) https://orcid.org/0000-0003-3540-8746
Des Small https://orcid.org/0000-0003-3723-5404
Bong Won Sohn https://orcid.org/0000-0002-4148-8378
Jason SooHoo https://orcid.org/0000-0003-1938-0720
Fumie Tazaki https://orcid.org/0000-0003-0236-0600
Paul Tiede https://orcid.org/0000-0003-3826-5648
Remo P. J. Tilanus https://orcid.org/0000-0002-6514-553X
Michael Titus https://orcid.org/0000-0002-3423-4505
Kenji Toma https://orcid.org/0000-0002-7114-6010
Pablo Torne https://orcid.org/0000-0001-8700-6058
Sascha Trippe https://orcid.org/0000-0003-0465-1559
Ilse van Bemmel https://orcid.org/0000-0001-5473-2950
Huib Jan van Langevelde https://orcid.org/0000-0002-0230-5946
Daniel R. van Rossum https://orcid.org/0000-0001-7772-6131
John Wardle https://orcid.org/0000-0002-8960-2942
Jonathan Weintroub https://orcid.org/0000-0002-4603-5204
Norbert Wex https://orcid.org/0000-0003-4058-2837
Robert Wharton https://orcid.org/0000-0002-7416-5209
Maciek Wielgus https://orcid.org/0000-0002-8635-4242
George N. Wong https://orcid.org/0000-0001-6952-2147
Qingwen Wu (吴庆文) https://orcid.org/0000-0003-4773-4987
André Young https://orcid.org/0000-0003-0000-2682
Ken Young https://orcid.org/0000-0002-3666-4920
Ziri Younsi https://orcid.org/0000-0001-9283-1191
Feng Yuan (袁峰) https://orcid.org/0000-0003-3564-6437
J. Anton Zensus https://orcid.org/0000-0001-7470-3321
Guangyao Zhao https://orcid.org/0000-0002-4417-1659
Shan-Shan Zhao https://orcid.org/0000-0002-9774-3606
Jadyn Anczarski https://orcid.org/0000-0003-4317-3385
Frederick K. Baganoff https://orcid.org/0000-0003-3852-6545
Andreas Eckart https://orcid.org/0000-0001-6049-3132
Joseph R. Farah https://orcid.org/0000-0003-4914-5625
Daryl Haggard https://orcid.org/0000-0001-6803-2138
Daniel Michalik https://orcid.org/0000-0002-7618-6556
Andrew Nadolski https://orcid.org/0000-0001-9479-9957
Joseph Neilsen https://orcid.org/0000-0002-8247-786X
Michael A. Nowak https://orcid.org/0000-0001-6923-1315
Rurik A. Primiani https://orcid.org/0000-0003-3910-7529
Paul Yamaguchi https://orcid.org/0000-0002-6017-8199
Shuo Zhang https://orcid.org/0000-0002-2967-790X

The Event Horizon Telescope Collaboration

Kazunori Akiyama[1,2,3,4], Antxon Alberdi[5], Walter Alef[6], Keiichi Asada[7], Rebecca Azulay[8,9,6], Anne-Kathrin Baczko[6], David Ball[10], Mislav Baloković[4,11], John Barrett[2], Dan Bintley[12], Lindy Blackburn[4,11], Wilfred Boland[13], Katherine L. Bouman[4,11,14], Geoffrey C. Bower[15], Michael Bremer[16], Christiaan D. Brinkerink[17], Roger Brissenden[4,11], Silke Britzen[6], Avery E. Broderick[18,19,20], Dominique Broguiere[16], Thomas Bronzwaer[17], Do-Young Byun[21,22], John E. Carlstrom[23,24,25,26], Andrew Chael[4,11], Chi-kwan Chan[10,27], Shami Chatterjee[28], Koushik Chatterjee[29], Ming-Tang Chen[15], Yongjun Chen (陈永军)[30,31], Ilje Cho[21,22], Pierre Christian[10,11], John E. Conway[32], James M. Cordes[28], Geoffrey B. Crew[2], Yuzhu Cui[33,34], Jordy Davelaar[17], Mariafelicia De Laurentis[35,36,37], Roger Deane[38,39], Jessica Dempsey[12], Gregory Desvignes[6], Jason Dexter[40], Sheperd S. Doeleman[4,11], Ralph P. Eatough[6], Heino Falcke[17], Vincent L. Fish[2], Ed Fomalont[1], Raquel Fraga-Encinas[17], Per Friberg[12], Christian M. Fromm[36], José L. Gómez[5], Peter Galison[1,41,42], Charles F. Gammie[43,44], Roberto García[16], Olivier Gentaz[16], Boris Georgiev[19,20], Ciriaco Goddi[17,45], Roman Gold[36], Minfeng Gu (顾敏峰)[30,46], Mark Gurwell[11], Kazuhiro Hada[33,34], Michael H. Hecht[2], Ronald Hesper[47], Luis C. Ho (何子山)[48,49], Paul Ho[7], Mareki Honma[33,34], Chih-Wei L. Huang[7], Lei Huang (黄磊)[30,46], David H. Hughes[50], Shiro Ikeda[3,51,52,53], Makoto Inoue[7], Sara Issaoun[17], David J. James[4,11], Buell T. Jannuzi[10], Michael Janssen[17], Britton Jeter[19,20], Wu Jiang (江悟)[30], Michael D. Johnson[4,11], Svetlana Jorstad[54,55], Taehyun Jung[21,22], Mansour Karami[18,19], Ramesh Karuppusamy[6], Tomohisa Kawashima[3], Garrett K. Keating[11], Mark Kettenis[56], Jae-Young Kim[6], Junhan Kim[10], Jongsoo Kim[21], Motoki Kino[3,57], Jun Yi Koay[7], Patrick M. Koch[7], Shoko Koyama[7], Michael Kramer[6], Carsten Kramer[16], Thomas P. Krichbaum[6], Cheng-Yu Kuo[58], Tod R. Lauer[59], Sang-Sung Lee[21], Yan-Rong Li (李彦荣)[60], Zhiyuan Li (李志远)[61,62], Michael Lindqvist[32], Kuo Liu[6], Elisabetta Liuzzo[63], Wen-Ping Lo[7,64], Andrei P. Lobanov[6], Laurent Loinard[65,66], Colin Lonsdale[2], Ru-Sen Lu (路如森)[30,6], Nicholas R. MacDonald[6], Jirong Mao (毛基荣)[67,68,69], Sera Markoff[29,70], Daniel P. Marrone[10], Alan P. Marscher[54], Iván Martí-Vidal[32,71], Satoki Matsushita[7], Lynn D. Matthews[2], Lia Medeiros[10,72], Karl M. Menten[6], Yosuke Mizuno[36], Izumi Mizuno[12], James M. Moran[4,11], Kotaro Moriyama[33,2], Monika Moscibrodzka[17], Cornelia Müller[6,17], Hiroshi Nagai[3,34], Neil M. Nagar[73], Masanori Nakamura[7], Ramesh Narayan[4,11], Gopal Narayanan[74], Iniyan Natarajan[39], Roberto Neri[16], Chunchong Ni[19,20], Aristeidis Noutsos[6], Hiroki Okino[33,75], Héctor Olivares[36], Tomoaki Oyama[33], Feryal Özel[10], Daniel C. M. Palumbo[4,11], Nimesh Patel[11], Ue-Li Pen[76,77,78,18], Dominic W. Pesce[4,11], Vincent Piétu[16], Richard Plambeck[79], Aleksandar PopStefanija[74], Oliver Porth[36,29], Ben Prather[43], Jorge A. Preciado-López[18], Dimitrios Psaltis[10], Hung-Yi Pu[18], Venkatessh Ramakrishnan[73], Ramprasad Rao[15], Mark G. Rawlings[12], Alexander W. Raymond[4,11], Luciano Rezzolla[36], Bart Ripperda[36], Freek Roelofs[17], Alan Rogers[2], Eduardo Ros[6], Mel Rose[10], Arash Roshanineshat[10], Helge Rottmann[6], Alan L. Roy[6], Chet Ruszczyk[2], Benjamin R. Ryan[80,81], Kazi L. J. Rygl[63], Salvador Sánchez[82], David Sánchez-Arguelles[50,83], Mahito Sasada[33,84], Tuomas Savolainen[6,85,86], F. Peter Schloerb[74], Karl-Friedrich Schuster[16], Lijing Shao[6,49], Zhiqiang Shen (沈志强)[30,31], Des Small[56], Bong Won Sohn[21,22,87], Jason SooHoo[2], Fumie Tazaki[33], Paul Tiede[19,20], Remo P. J. Tilanus[17,45,88], Michael Titus[2], Kenji Toma[89,90], Pablo Torne[6,82], Tyler Trent[10], Sascha Trippe[91], Shuichiro Tsuda[33], Ilse van Bemmel[56], Huib Jan van Langevelde[56,92], Daniel R. van Rossum[17], Jan Wagner[6], John Wardle[93], Jonathan Weintroub[4,11], Norbert Wex[6], Robert Wharton[6], Maciek Wielgus[4,11], George N. Wong[43], Qingwen Wu (吴庆文)[94], André Young[17], Ken Young[11], Ziri Younsi[95,36], Feng Yuan (袁峰)[30,46,96], Ye-Fei Yuan (袁业飞)[97], J. Anton Zensus[6], Guangyao Zhao[21], Shan-Shan Zhao[17,61], Ziyan Zhu[42], Jadyn Anczarski[98], Frederick K. Baganoff[99], Andreas Eckart[6,100], Joseph R. Farah[11,101,4], Daryl Haggard[102,103,104], Zheng Meyer-Zhao[7,105], Daniel Michalik[106,107], Andrew Nadolski[44], Joseph Neilsen[98], Hiroaki Nishioka[7], Michael A. Nowak[108], Nicolas Pradel[7], Rurik A. Primiani[109], Kamal Souccar[74], Laura Vertatschitsch[11,109], Paul Yamaguchi[11], and Shuo Zhang[99]

[1] National Radio Astronomy Observatory, 520 Edgemont Rd, Charlottesville, VA 22903, USA
[2] Massachusetts Institute of Technology Haystack Observatory, 99 Millstone Road, Westford, MA 01886, USA
[3] National Astronomical Observatory of Japan, 2-21-1 Osawa, Mitaka, Tokyo 181-8588, Japan
[4] Black Hole Initiative at Harvard University, 20 Garden Street, Cambridge, MA 02138, USA
[5] Instituto de Astrofísica de Andalucía-CSIC, Glorieta de la Astronomía s/n, E-18008 Granada, Spain
[6] Max-Planck-Institut für Radioastronomie, Auf dem Hügel 69, D-53121 Bonn, Germany
[7] Institute of Astronomy and Astrophysics, Academia Sinica, 11F of Astronomy-Mathematics Building, AS/NTU No. 1, Sec. 4, Roosevelt Rd, Taipei 10617, Taiwan, R.O.C.
[8] Departament d'Astronomia i Astrofísica, Universitat de València, C. Dr. Moliner 50, E-46100 Burjassot, València, Spain
[9] Observatori Astronòmic, Universitat de València, C. Catedrático José Beltrán 2, E-46980 Paterna, València, Spain
[10] Steward Observatory and Department of Astronomy, University of Arizona, 933 N. Cherry Ave., Tucson, AZ 85721, USA
[11] Center for Astrophysics | Harvard & Smithsonian, 60 Garden Street, Cambridge, MA 02138, USA
[12] East Asian Observatory, 660 N. A'ohoku Pl., Hilo, HI 96720, USA
[13] Nederlandse Onderzoekschool voor Astronomie (NOVA), PO Box 9513, 2300 RA Leiden, The Netherlands
[14] California Institute of Technology, 1200 East California Boulevard, Pasadena, CA 91125, USA
[15] Institute of Astronomy and Astrophysics, Academia Sinica, 645 N. A'ohoku Place, Hilo, HI 96720, USA
[16] Institut de Radioastronomie Millimétrique, 300 rue de la Piscine, 38406 Saint Martin d'Hères, France

[17] Department of Astrophysics, Institute for Mathematics, Astrophysics and Particle Physics (IMAPP), Radboud University, P.O. Box 9010, 6500 GL Nijmegen, The Netherlands
[18] Perimeter Institute for Theoretical Physics, 31 Caroline Street North, Waterloo, ON, N2L 2Y5, Canada
[19] Department of Physics and Astronomy, University of Waterloo, 200 University Avenue West, Waterloo, ON, N2L 3G1, Canada
[20] Waterloo Centre for Astrophysics, University of Waterloo, Waterloo, ON N2L 3G1 Canada
[21] Korea Astronomy and Space Science Institute, Daedeok-daero 776, Yuseong-gu, Daejeon 34055, Republic of Korea
[22] University of Science and Technology, Gajeong-ro 217, Yuseong-gu, Daejeon 34113, Republic of Korea
[23] Kavli Institute for Cosmological Physics, University of Chicago, Chicago, IL 60637, USA
[24] Department of Astronomy and Astrophysics, University of Chicago, 5640 South Ellis Avenue, Chicago, IL 60637, USA
[25] Department of Physics, University of Chicago, 5720 South Ellis Avenue, Chicago, IL 60637, USA
[26] Enrico Fermi Institute, University of Chicago, 5640 South Ellis Avenue, Chicago, IL 60637, USA
[27] Data Science Institute, University of Arizona, 1230 N. Cherry Ave., Tucson, AZ 85721, USA
[28] Cornell Center for Astrophysics and Planetary Science, Cornell University, Ithaca, NY 14853, USA
[29] Anton Pannekoek Institute for Astronomy, University of Amsterdam, Science Park 904, 1098 XH, Amsterdam, The Netherlands
[30] Shanghai Astronomical Observatory, Chinese Academy of Sciences, 80 Nandan Road, Shanghai 200030, People's Republic of China
[31] Key Laboratory of Radio Astronomy, Chinese Academy of Sciences, Nanjing 210008, People's Republic of China
[32] Department of Space, Earth and Environment, Chalmers University of Technology, Onsala Space Observatory, SE-439 92 Onsala, Sweden
[33] Mizusawa VLBI Observatory, National Astronomical Observatory of Japan, 2-12 Hoshigaoka, Mizusawa, Oshu, Iwate 023-0861, Japan
[34] Department of Astronomical Science, The Graduate University for Advanced Studies (SOKENDAI), 2-21-1 Osawa, Mitaka, Tokyo 181-8588, Japan
[35] Dipartimento di Fisica "E. Pancini", Universitá di Napoli "Federico II", Compl. Univ. di Monte S. Angelo, Edificio G, Via Cinthia, I-80126, Napoli, Italy
[36] Institut für Theoretische Physik, Goethe-Universität Frankfurt, Max-von-Laue-Straße 1, D-60438 Frankfurt am Main, Germany
[37] INFN Sez. di Napoli, Compl. Univ. di Monte S. Angelo, Edificio G, Via Cinthia, I-80126, Napoli, Italy
[38] Department of Physics, University of Pretoria, Lynnwood Road, Hatfield, Pretoria 0083, South Africa
[39] Centre for Radio Astronomy Techniques and Technologies, Department of Physics and Electronics, Rhodes University, Grahamstown 6140, South Africa
[40] Max-Planck-Institut für Extraterrestrische Physik, Giessenbachstr. 1, D-85748 Garching, Germany
[41] Department of History of Science, Harvard University, Cambridge, MA 02138, USA
[42] Department of Physics, Harvard University, Cambridge, MA 02138, USA
[43] Department of Physics, University of Illinois, 1110 West Green St, Urbana, IL 61801, USA
[44] Department of Astronomy, University of Illinois at Urbana-Champaign, 1002 West Green Street, Urbana, Illinois 61801, USA
[45] Leiden Observatory—Allegro, Leiden University, P.O. Box 9513, 2300 RA Leiden, The Netherlands
[46] Key Laboratory for Research in Galaxies and Cosmology, Chinese Academy of Sciences, Shanghai 200030, People's Republic of China
[47] NOVA Sub-mm Instrumentation Group, Kapteyn Astronomical Institute, University of Groningen, Landleven 12, 9747 AD Groningen, The Netherlands
[48] Department of Astronomy, School of Physics, Peking University, Beijing 100871, People's Republic of China
[49] Kavli Institute for Astronomy and Astrophysics, Peking University, Beijing 100871, People's Republic of China
[50] Instituto Nacional de Astrofísica, Óptica y Electrónica. Apartado Postal 51 y 216, 72000. Puebla Pue., México
[51] The Institute of Statistical Mathematics, 10-3 Midori-cho, Tachikawa, Tokyo, 190-8562, Japan
[52] Department of Statistical Science, The Graduate University for Advanced Studies (SOKENDAI), 10-3 Midori-cho, Tachikawa, Tokyo 190-8562, Japan
[53] Kavli Institute for the Physics and Mathematics of the Universe, The University of Tokyo, 5-1-5 Kashiwanoha, Kashiwa, 277-8583, Japan
[54] Institute for Astrophysical Research, Boston University, 725 Commonwealth Ave., Boston, MA 02215, USA
[55] Astronomical Institute, St. Petersburg University, Universitetskij pr., 28, Petrodvorets,198504 St.Petersburg, Russia
[56] Joint Institute for VLBI ERIC (JIVE), Oude Hoogeveensedijk 4, 7991 PD Dwingeloo, The Netherlands
[57] Kogakuin University of Technology and Engineering, Academic Support Center, 2665-1 Nakano, Hachioji, Tokyo 192-0015, Japan
[58] Physics Department, National Sun Yat-Sen University, No. 70, Lien-Hai Rd, Kaosiung City 80424, Taiwan, R.O.C
[59] National Optical Astronomy Observatory, 950 North Cherry Ave., Tucson, AZ 85719, USA
[60] Key Laboratory for Particle Astrophysics, Institute of High Energy Physics, Chinese Academy of Sciences, 19B Yuquan Road, Shijingshan District, Beijing, People's Republic of China
[61] School of Astronomy and Space Science, Nanjing University, Nanjing 210023, People's Republic of China
[62] Key Laboratory of Modern Astronomy and Astrophysics, Nanjing University, Nanjing 210023, People's Republic of China
[63] Italian ALMA Regional Centre, INAF-Istituto di Radioastronomia, Via P. Gobetti 101, 40129 Bologna, Italy
[64] Department of Physics, National Taiwan University, No.1, Sect.4, Roosevelt Rd., Taipei 10617, Taiwan, R.O.C
[65] Instituto de Radioastronomía y Astrofísica, Universidad Nacional Autónoma de México, Morelia 58089, México
[66] Instituto de Astronomía, Universidad Nacional Autónoma de México, CdMx 04510, México
[67] Yunnan Observatories, Chinese Academy of Sciences, 650011 Kunming, Yunnan Province, People's Republic of China
[68] Center for Astronomical Mega-Science, Chinese Academy of Sciences, 20A Datun Road, Chaoyang District, Beijing, 100012, People's Republic of China
[69] Key Laboratory for the Structure and Evolution of Celestial Objects, Chinese Academy of Sciences, 650011 Kunming, People's Republic of China
[70] Gravitation Astroparticle Physics Amsterdam (GRAPPA) Institute, University of Amsterdam, Science Park 904, 1098 XH Amsterdam, The Netherlands
[71] Centro Astronómico de Yebes (IGN), Apartado 148, 19180 Yebes, Spain
[72] Department of Physics, Broida Hall, University of California Santa Barbara, Santa Barbara, CA 93106, USA
[73] Astronomy Department, Universidad de Concepción, Casilla 160-C, Concepción, Chile
[74] Department of Astronomy, University of Massachusetts, 01003, Amherst, MA, USA
[75] Department of Astronomy, Graduate School of Science, The University of Tokyo, 7-3-1 Hongo, Bunkyo-ku, Tokyo 113-0033, Japan
[76] Canadian Institute for Theoretical Astrophysics, University of Toronto, 60 St. George Street, Toronto, ON M5S 3H8, Canada
[77] Dunlap Institute for Astronomy and Astrophysics, University of Toronto, 50 St. George Street, Toronto, ON M5S 3H4, Canada
[78] Canadian Institute for Advanced Research, 180 Dundas St West, Toronto, ON M5G 1Z8, Canada
[79] Radio Astronomy Laboratory, University of California, Berkeley, CA 94720, USA
[80] CCS-2, Los Alamos National Laboratory, P.O. Box 1663, Los Alamos, NM 87545, USA
[81] Center for Theoretical Astrophysics, Los Alamos National Laboratory, Los Alamos, NM, 87545, USA
[82] Instituto de Radioastronomía Milimétrica, IRAM, Avenida Divina Pastora 7, Local 20, 18012, Granada, Spain
[83] Consejo Nacional de Ciencia y Tecnología, Av. Insurgentes Sur 1582, 03940, Ciudad de México, México
[84] Hiroshima Astrophysical Science Center, Hiroshima University, 1-3-1 Kagamiyama, Higashi-Hiroshima, Hiroshima 739-8526, Japan
[85] Aalto University Department of Electronics and Nanoengineering, PL 15500, 00076 Aalto, Finland
[86] Aalto University Metsähovi Radio Observatory, Metsähovintie 114, 02540 Kylmälä, Finland
[87] Department of Astronomy, Yonsei University, Yonsei-ro 50, Seodaemun-gu, 03722 Seoul, Republic of Korea
[88] Netherlands Organisation for Scientific Research (NWO), Postbus 93138, 2509 AC Den Haag, The Netherlands

[89] Frontier Research Institute for Interdisciplinary Sciences, Tohoku University, Sendai 980-8578, Japan
[90] Astronomical Institute, Tohoku University, Sendai 980-8578, Japan
[91] Department of Physics and Astronomy, Seoul National University, Gwanak-gu, Seoul 08826, Republic of Korea
[92] Leiden Observatory, Leiden University, Postbus 2300, 9513 RA Leiden, The Netherlands
[93] Physics Department, Brandeis University, 415 South Street, Waltham, MA 02453, USA
[94] School of Physics, Huazhong University of Science and Technology, Wuhan, Hubei, 430074, People's Republic of China
[95] Mullard Space Science Laboratory, University College London, Holmbury St. Mary, Dorking, Surrey, RH5 6NT, UK
[96] School of Astronomy and Space Sciences, University of Chinese Academy of Sciences, No. 19A Yuquan Road, Beijing 100049, People's Republic of China
[97] Astronomy Department, University of Science and Technology of China, Hefei 230026, People's Republic of China
[98] Department of Physics,Villanova University, 800 E. Lancaster Ave, Villanova, PA, 19085, USA
[99] Kavli Institute for Astrophysics and Space Research, Massachusetts Institute of Technology, Cambridge, MA 02139, USA
[100] Physikalisches Institut der Universität zu Köln, Zülpicher Str. 77, D-50937 Köln, Germany
[101] University of Massachusetts Boston, 100 William T, Morrissey Blvd, Boston, MA 02125, USA
[102] Department of Physics, McGill University, 3600 University Street, Montréal, QC H3A 2T8, Canada
[103] McGill Space Institute, McGill University, 3550 University Street, Montréal, QC H3A 2A7, Canada
[104] CIFAR Azrieli Global Scholar, Gravity and the Extreme Universe Program, Canadian Institute for Advanced Research, 661 University Avenue, Suite 505, Toronto, ON M5G 1M1, Canada
[105] ASTRON, Oude Hoogeveensedijk 4, 7991 PD Dwingeloo, The Netherlands
[106] Science Support Office, Directorate of Science, European Space Research and Technology Centre (ESA/ESTEC), Keplerlaan 1, 2201 AZ Noordwijk, The Netherlands
[107] University of Chicago, 5640 South Ellis Avenue, Chicago, IL 60637, USA
[108] Physics Dept., CB 1105, Washington University, One Brookings Drive, St. Louis, MO 63130-4899, USA
[109] Systems and Technology Research, 600 West Cummings Park, Woburn, MA 01801, USA